\documentclass[referee]{raa}   
\usepackage{graphicx,times}             
\usepackage{natbib}
\usepackage{amssymb,amsmath}
\bibpunct{(}{)}{;}{a}{}{,}

\usepackage[a4paper=true,pagebackref=true]{hyperref}
\hypersetup{colorlinks = true, linkcolor = green, anchorcolor = red, citecolor = blue, filecolor = red, pagecolor = red, urlcolor = red}

\usepackage{graphicx}
\usepackage{textcomp}
\usepackage{tabularx}
\usepackage{adjustbox}
\usepackage{xcolor}
\usepackage{afterpage}
\hypersetup{colorlinks=true,citecolor=cyan,urlcolor=cyan,linkcolor=blue}

\usepackage{tabularx}
\usepackage{adjustbox}
\usepackage{rotating}

\usepackage{lscape}

\begin{document}
\title{Comparative case study of two methods to assess the eruptive potential of selected active regions}

\volnopage{Vol.0 (20xx) No.0, 000--000}      
\setcounter{page}{1}     

\author{Francesca Zuccarello\inst{1,2}
\and Ilaria Ermolli\inst{3}
\and Marianna B. Kors\'os\inst{4,5}
\and Fabrizio Giorgi\inst{3}
\and Salvo L. Guglielmino\inst{2}
\and Robertus Erd\'elyi \inst{4,6,7}
\and Paolo Romano\inst{2}}

\institute{Dipartimento di Fisica e Astronomia ``Ettore Majorana'', Universit\`a di Catania, Via S.~Sofia 78, I 95123 Catania, Italy\\
\and
INAF Osservatorio Astrofisico di Catania, Via S.~Sofia 78, I-95123 Catania, Italy\\
\and
INAF Osservatorio Astronomico di Roma, Via Frascati 33, I-00078 Monte Porzio Catone, Italy\\
\and
Department of Physics, Aberystwyth University, Ceredigion, Cymru, SY23 3BZ, UK,\\ 
{\it Corresponding author:} mak102@aber.ac.uk\\
\and
Department of Astronomy, E\"otv\"os Lor\'and University, P\'azm\'any P. s\'et\'any 1/A, Budapest, H-1117, Hungary\\
\and
Solar Physics \& Space Plasma Research Center (SP2RC), School of Mathematics and Statistics, University of Sheffield, Hicks Building, Hounsfield Road, S3 7RH, UK\\
\and 
Gyula Bay Zolt\'an Solar Observatory (GSO), Hungarian Solar Physics Foundation (HSPF), Pet\H{o}fi t\'er 3., Gyula, H-5700, Hungary\\
\vs\no}

 {\small Received~~2021 month day; accepted~~20xx~~month day}

\abstract{
Solar eruptive events, like flares and coronal mass ejections, are characterized by the rapid release of energy that can give rise to emission of radiation across the entire electromagnetic spectrum and to an abrupt significant increase in the kinetic energy of particles. These energetic phenomena can have important effects on the Space Weather conditions and therefore it is necessary to understand their origin, in particular, what is the eruptive potential of an active region (AR). 
In these case studies, we compare two distinct methods that were used in previous works to investigate the variations of some characteristic physical parameters during the pre-flare states of flaring ARs. These methods consider: i) the magnetic flux evolution and the magnetic helicity accumulation, and ii) the fractal and multi-fractal properties of  flux concentrations in ARs.
Our comparative analysis is based on time series of photospheric data obtained by the Solar Dynamics Observatory between March 2011 and June 2013. We selected two distinct samples of ARs: one is distinguished by the occurrence of more energetic M- and X-class flare events, that may have a rapid effect on not just the near-Earth space, but also on the terrestrial environment; the second is characterized by no-flares or having just few C- and B-class flares. 
We found that the two tested methods complement each other in their ability to assess the eruptive potentials of ARs and could be employed to identify ARs prone to flaring activity.
Based on the presented case study, we suggest that using a combination of  different methods may aid  to identify more reliably the eruptive potentials of ARs and help to better understand the pre-flare states.
}

\titlerunning{Comparison of two methods to assess ARs eruptive potential}
\authorrunning{Zuccarello et al.}
\maketitle

%

\section{Introduction}

Solar flares occur in and around active regions (ARs) when magnetic energy, which has built up in regions of the solar atmosphere, is suddenly released by magnetic reconnection processes. These processes can also 
drive solar plasma into the heliosphere by giving rise to coronal mass ejections (CMEs) that often occur concurrently with flares \citep[see, e.g.,][and references therein]{Priest_2002,Shibata_2011}.  
Observations collected over many years have shown that key aspects of evolution leading to eruptive events in ARs are the size and topology of the magnetic region \citep[see, e.g.,][]{Romano2014, Romano2019}, the non-potential component of the magnetic field in the AR, the occurrence of destabilization conditions, and the associated energies \citep[for a review, see, e.g.,][]{Benz_2017}.

Prediction of solar eruptive events, such as flares and CMEs, has recently evolved from  a tool to test our understanding of the processes leading the evolution of solar magnetic regions into a societal need to both prevent and mitigate the potentials of damage to modern technologies due to Space Weather events \citep[see, e.g.,][and references therein]{Bonadonna2017,Mann2018,Bingham2019,Opgenoorth2019,Plainaki2020}. Therefore, the literature is rich in flare-prediction methods applied to a wide range of observations of ARs \citep[see, e.g.][]{McCloskey2018,Leka2018,Campi2019,Falco2019,Goodman2020,Lin2020}, as well as a number of studies devoted to the comparison between different prediction methods \citep[see, e.g.,][and references therein]{Barnes_2016,Leka2019a,Leka2019b,Park2020}.

In this work, our aim is to exploit the diverse information obtained by two individual methods on the pre-flare conditions, which could be useful and practical for a future operational service to assess the eruptive potentials of an AR.

\subsection*{First technique -- Magnetic flux and magnetic helicity trend}

The first technique employs methods that analyze  the magnetic flux ($\Phi$) and magnetic helicity ($H$) of ARs \citep{smyrli}. They assume that the excess energy with respect to the potential field is injected into the source region of a flare by its Poynting flux: 
\begin{equation}\label{eq:00}
 P=(c/4\pi) \int (\vec E \times \vec B) ~ dS,  
\end{equation}
where the electric vector \textbf{E} is expressed by Ohm's law: 
\begin{equation}\label{eq:0}
\vec{E} =  - (1/c) ~ \vec v \times \vec B  
\end{equation}
and $c$, \textbf{v}, and \textbf{B} indicate the speed of electromagnetic radiation in the vacuum, the plasma velocity, and the magnetic field vector, respectively.  Helicity injection can occur by the emergence of new magnetic flux and by the presence of a significant velocity component ($v_{t} $) perpendicular to the magnetic field. \citet{kusano2002} described a method for the determination of the tangential velocity field $v_{t} $  to estimate the injected energy, while \citet{schrijver2005} argued that the helicity carried by the emerging field has a stronger impact than the field shearing at the surface \citep{li}. 

By comparing the energy and helicity budgets of quiet- and flare-productive ARs, \citet{georg2007} concluded that the estimated free energy and the helicity are both reliable quantities for their distinction. \citet{labonte2007} found that a sufficient helicity injection rate for an X-class flare to occur is at least $6 \cdot 10^{36}$~Mx$^{2}$~s$^{-1}$, where this threshold was proposed as a necessary dynamical condition for flare eruption. Therefore helicity injection rate may be an excellent flare-risk signature. Considering dynamical properties in connection with halo CMEs, \citet{smyrli} examined the temporal variation of the helicity flux in a sample of ARs. They found that no typical pre-CME behavioural pattern can be identified, but in some cases the post-CME state showed that a significant amount of helicity had indeed been carried away during such events.  More recently, \citet{elmhamdi} reported characteristic flare-related patterns of the  AR tilt angle variation, which is another property related to helicity.

\subsection*{Second technique -- Fractal and multi-fractal parameters}
The second technique focuses on the morphology of flux concentrations in ARs, in particular on the level of intermittency in surface magnetic field patterns \citep{ermolli2014}. High level of intermittency means strong tangential discontinuities in the magnetic field, which  may initiate reconnection events. \citet{Abramenko2003} introduced a number of parameters to describe the structural complexity of  magnetic regions and they found characteristic patterns in the pre-flare behaviour, but only on a time scale of a few tens of minutes prior the flaring events. The parameters inferred from photospheric magnetograms reported in  \citet{Abramenko2003} were evaluated by considering variations of the structure function for ARs; the results obtained  indicate the enhancement of plasma turbulence in connection with the flares. 

Following \citet{Abramenko2003}, multiple studies have reported distinct values of the fractal and multi-fractal parameters measured for ARs with different flare-classes \citep[see, e.g.][]{Mcateer_2005,georg2012}. However, such measurements are not very efficient for distinguishing ARs with flare activity, because the large measured dispersion values of an AR could produce the same class flare events. Further investigating the sensitivity of fractal and multifractal measurements on flare activity of ARs, 
\citet{ermolli2014} showed that the dispersion of results and temporal evolution of measured values are all affected by the spatial resolution and cadence of the analyzed observations. In particular, the analysis of data from the Solar Dynamic Observatory/Helioseismic and Magnetic Imager \citep[SDO/HMI,][]{Scherrer_etal2012, Schou_etal2012,Wachter_etal2012} with a higher resolution produced less noisy results than those obtained from the relatively low resolution of Solar and Heliospheric Observatory/Michelson Doppler Imager \citep[SOHO/MDI,][]{Scherrer_1995} data, employed largely in earlier studies.
\citet{giorgi2014} reported that measurements of fractal and multifractal parameters carried out  on a large sample of  SDO/HMI observations, which are the highest-resolution full-disk synoptic magnetograms available to date, allow distinguishing ARs that host more energetic events from relatively flare-quiet ARs. Nevertheless, these measurements do not allow us to distinguish between the C- and M-class flaring
ARs, nor between the  M- and X-class ARs. 
Furthermore, 
\citet{giorgi2014} reported consistent changes on the time series of the measured parameters at flare occurrence only on $\simeq 50\%$
of the analyzed  ARs and $\simeq 50\%$ of the M- and X-class events.

\subsection*{Paper organization}

 In Sections~\ref{2} and ~\ref{3}, we describe the data set of five flare-productive and five flare-quiet ARs and the applied methods in detail, respectively. In Section~\ref{4}, we summarize the results obtained from the two applied methods. Next, in Section~\ref{5}, we discuss the obtained results and in Section~\ref{6} we give our conclusions.

\section{Data} \label{2}

We analyzed the time series of ten ARs observed between 2011 March 6 and 2013 Jun 24 by SDO/HMI. We selected from the NOAA catalog the following regions: i) five {\it flare-productive} ARs: 11166, 11283, 11429, 11515, 11520 and (ii) five {\it flare-quiet} ARs: 11267, 11512, 11589, 11635 and 11775. All these ARs had $\delta$-spot(s).

The selected flare-productive regions were cradles of M- and X-class flares during their passage over the solar disk, with peak soft X-ray flux greater than or equal to $10^{-5}~$ W m$^{-2}$ as reported in the soft X-ray (SXR) flux catalog of the Geostationary Orbiting Environmental Satellites (GOES), available at the NOAA Space Weather Prediction Center\footnote{http://www.swpc.noaa.gov/products/goes-x-ray-flux}. The selected five flare-quiet ARs hosted only very low energetic B- and C-class flares with $10^{-7}$-$10^{-6}~$ W m$^{-2}$ peak in the soft X-ray flux. 

For each selected AR, we analyzed time series of Level 1.5 SDO/HMI full-disk photospheric LOS magnetograms. 
The time interval of the analyzed data series was chosen in such a way that each AR was within $\approx \pm 30^{\circ}$ from the central meridian, in order to avoid significant uncertainties due to projection effects. All the magnetogram data were corrected for the angle between the magnetic field direction and the observer's LOS \citep{Woo03} and were co-aligned by applying the standard differential rotation rate reported by \citet{How90}.

We extracted sub-arrays centered on the selected ARs from the SDO/HMI full-disk observations. The data set restricted to the time interval with the AR position within $\approx \pm 30^{\circ}$ from the central meridian consist of LOS magnetograms, each of 4096 $\times$4096 pixels,  with a pixel size of 0.505 arcsec and cadence  from 12 to 96 minutes. Tables \ref{table1} and \ref{table1bis} summarize the details of the analyzed regions, the time interval considered for each AR and the dimension of the analyzed sub-arrays.

\begin{table*}
\caption{Selected time intervals and subfield of the flare-productive ARs for the helicity and fractal analysis.\label{table1}}
\centering
 \begin{tabular}{cccccc}
\hline\hline
AR & Start date & Start time & End date  & End time & Subfield\\
 &  & UT & & UT & (arcsec) \\
\hline

 AR 11166	& Mar 6, 2011	& 22:00     & Mar 10, 2011 	& 22:00    & 512 $\times$ 512 	\\
AR 11283	& Sep 3, 2011   & 22:00     & Sep 7, 2011 	& 22:00    & 512 $\times$ 512	\\
AR 11429	& Mar 6, 2012 	& 21:00     & Mar 10, 2012  & 22:00    & 440 $\times$ 440 \\
AR 11515    & Jul  1, 2012 & 01:00   & Jul 5, 2012  		& 04:00 	&  400 $\times$ 400\\

AR 11520   & Jul 10, 2012 & 08:00   & Jul 14, 2012 		& 16:00 	&  240 $\times$ 240\\
 \hline
 \end{tabular}
\end{table*}

\begin{table*}
\caption{Selected time intervals and subfield of the flare-quiet ARs for the helicity and fractal analysis.\label{table1bis}}
\centering
 \begin{tabular}{cccccc}
\hline\hline
AR & Start date & Start time & End date  & End time & Subfield\\
 &  & UT & & UT & (arcsec) \\
\hline

AR 11267	& Aug 6, 2011	& 02:00     & Aug 10, 2011 	& 02:00    & 240 $\times$ 240 	\\
AR 11512	& Jun 26, 2012   & 24:00     & Jun 30, 2012 	& 24:00    & 240 $\times$ 240	\\
AR 11589	& Oct 13, 2012 	& 14:00     & Oct 17, 2012  & 14:00    & 512 $\times$ 512 \\
AR 11635    & Dec 22, 2012 & 21:00   & Dec 26, 2012  		& 21:00 	&  512 $\times$ 512\\
AR 11775    & Jun 19, 2013 & 10:00   & Jun 23, 2013 		& 10:00 	&  240 $\times$ 240\\
 \hline
 \end{tabular}
\end{table*}

\section{ Methods} \label{3}
\subsection{Magnetic flux and magnetic helicity trend}
This method is based on the analysis of the magnetic flux ($\Phi$) and magnetic helicity ($H$) of ARs \citep{smyrli}. Considering SDO/HMI LOS magnetograms with the subfield (centered on each AR) specified in the last column of Tables \ref{table1} and \ref{table1bis}, we measured the evolution of the positive ($\Phi_+$), negative ($\Phi_-$), and unsigned ($\Phi$) magnetic fluxes. We estimated the uncertainty of the magnetic flux by propagating the experimental errors and considering the SDO/HMI sensitivity of 10~G \citep{Schou_etal2012}. 

The magnetic helicity flux values were determined by the mean magnetograms corresponding to the average between two sequential ones in the analyzed series. We, then, measured the horizontal velocity fields by means of the Differential Affine Velocity Estimator technique \citep[DAVE,][]{Sch05,Sch06}, by using a full-width-at-half-maximum (FWHM) of the apodizing window of 11 pixels (5.5 arcsec), as suggested by \citet{Sch08}. 

The rate of change of the magnetic helicity flux  $dH/dt$  was estimated using the following equation (see \citealt{Par05}):

 \begin{equation}  \label{eq22}
 \frac{dH}{dt} = -\frac{1}{2\pi} \int_{S} \int_{S'} \frac{d\theta(r)}{dt} B_{n} B'_{n} dS dS', 
   \end{equation}

\noindent where $S$ = $S'$ is the integration surface,  $d\theta(r)/dt$ is the relative rotation rate of pairs of photospheric positions defined by $x$ and $x'$, where the condition $\vec{r} = \vec{x} - \vec{x'}$ holds.  Moreover, $B_{n}$ indicates the component of the magnetic field normal to the surface $S$.

Once the magnetic helicity flux is obtained, we could estimate the corresponding accumulation of magnetic helicity ($H$) for the selected ARs, using the equation:

 \begin{equation}  \label{eq23}
H = \int_{0}^{\Delta t}  \frac{dH}{dt} dt, 
   \end{equation}
   
\noindent where $\Delta t$ indicates the total observational time interval for each AR. 

 As far as the calculation of $H$ is concerned, it is important to stress that helicity is a signed quantity, resulting from the algebraic sum of right-handed (conventionally positive) and left-handed (conventionally negative) values. This means that a helicity increase could be related to an increasing dominance of one of the two signs, because the injection of the prevailing sign becomes stronger, or the injection of the deficient sign becomes weaker, or both, along with conceivable combinations. Conversely, a helicity decrease could be related to a weakening injection of the
prevailing sign, a strengthening injection of the deficient sign, or both, along with combinations. Therefore, in order to investigate the contribution of the right-handed (in the following indicated as $H^+$) and left-handed ($H^-$) helicity, we have also taken into account the contributions of these quantities separately.

\subsection{Fractal and multi-fractal parameters}

This method is based on the study of the level of intermittency in surface magnetic field patterns \citep{ermolli2014}. We estimated the generalized fractal dimensions $D_0$ and $D_8$ and the multifractal Contribution Diversity $C_{div}$ and Dimensional Diversity $D_{div}$ on each subarray extracted from the SDO/HMI LOS magnetograms series considered in our study. We recall that these measures describe self-similar properties of the solar magnetic field, which are a signature of the turbulent physical processes that govern the evolution of the solar magnetic regions from the interplay between plasma flows and magnetic field. We derived $D_0$, $D_8$, $C_{div}$, and $D_{div}$  following previous studies in the literature by, e.g., \citet{Mcateer_2005}, \citet{Criscuoli_etal2009}, \citet{georg2012}, and \citet{ermolli2014}. We analysed $D_0$ and $D_8$, as in previous studies, because they are less sensitive to pixel scale and pixelization \citep{Lawrence1996}. The  theory behind the computed quantities, as well as the accuracy of the methods and algorithms employed on solar data, are extensively described in, e.g., \citet{Abramenko2003}, \citet{Abramenko2005}, \citet{Sen2007}, and in the above papers. However, we summarize here the main characteristics of the methods applied in this study in what follows. 

Given a measure of an observable $P$, the Generalized Fractal Dimension is 
defined as:

 \begin{equation}  \label{eq2}
 D_q = \frac{1}{q-1} \lim_{\epsilon \to  0}\frac{\ln I_q(\epsilon)}{\ln \epsilon} , 
   \end{equation}
   
\noindent 
where $q$ is a real number, 
 \begin{equation}  \label{eq3}
I_q(\epsilon) = \sum_{i=1}^{N_{\epsilon}} P_i (\epsilon)^q , 
   \end{equation}

\noindent 
and we have introduced the more compact notation:  
$N_{\epsilon}$ instead of $N(\epsilon)$. In our study, $P_i ( \epsilon )$ is  the  normalized magnetic flux  
$P_i(\epsilon) =  \frac{\left| {\sum_j \Phi_j } \right|}{\Phi_{tot}} $, where $j$ is the image pixel that runs in a box of size $\epsilon$, $\Phi_j$ is the magnetic flux at image pixel $j$, and $\Phi_{tot}$ is the magnetic flux in the image. 
We analyzed the temporal variation of the generalized dimension $D_0$ and $D_8$, where, according to Equation (\ref{eq2}):
 \begin{equation}  \label{eq4a}
 D_0 = -\lim_{\epsilon \to  0}\frac{\ln I_0 \left( \epsilon \right) }{\ln \epsilon} ,
   \end{equation}
 \begin{equation}  \label{eq4}
 D_8 = \frac{1}{7} \lim_{\epsilon \to  0}\frac{\ln I_8 \left( \epsilon \right) }{\ln \epsilon} .
   \end{equation} 
  
The multifractal spectrum $F(\alpha_q)$  is defined as:
 \begin{equation}  \label{eq6}
F(\alpha _q) = \lim_{\epsilon \to 0} \frac{\sum _{i=1}^{N_{\epsilon}} \eta_{i,q} (\epsilon) \ln \eta _{i,q} (\epsilon)}{\ln \epsilon}  ,   \end{equation} 
where:  
 \begin{equation}  \label{eq62}
\eta _{i,q} (\epsilon) = \frac{P_i (\epsilon) ^q}{\sum _{i=1}^{N_{\epsilon}} P_i ( \epsilon ) ^q} , 
 \end{equation} 
 and 
  \begin{equation}  \label{eq63}
\alpha _q = \lim _{\epsilon \to 0} \frac{\sum _{i=1}^{N _ {\epsilon}} \eta _{i,q} (\epsilon) \ln P_i (\epsilon)}{\ln \epsilon} .
 \end{equation}
$\alpha$ represents the density of the quantity analysed in the study, here normalized magnetic flux, F($\alpha$) is the fractal dimension of the points of the domain with the same value of  $\alpha$.
Following \citet{Conlon_etal2008}, the $C_{div}$ and $D_{div}$ considered in our study are defined as: 
 \begin{equation}  \label{eq7}
C_{div} = (\alpha _q) _{max} - (\alpha _q)_{min} \; ; \; D_{div} = F (\alpha _q)_{max}- F (\alpha _q)_{min} 
\end{equation}

\noindent
respectively. These measures are a signature of the turbulent processes involving plasma flows and magnetic field in and around solar magnetic regions, as recalled above. Following \citet{Conlon_etal2008}, we restricted our analysis to $q>0$, since numerical errors are larger for negative values of the exponent $q$.

We inferred the above fractal and multifractal parameters by using the 
box-counting technique \citep{Mandelbrot_1983}, which is usually employed for  fractal and multifractal analysis  (see, e.g., \citealp{Evertsz_Mandelbrot_1992}). The method  consists of covering the image of the analysed region with boxes of different sizes $\epsilon$  and then estimating the slope of the linear relation: 

   \begin{equation}  \label{eq1}
 \ln N (\epsilon) = D \lim_{\epsilon \to  0} \ln \frac{1}{\epsilon} + C ,  
   \end{equation}

\noindent 
where $N(\epsilon)$ is the number of boxes that cover the studied region, C is a constant and D is the fractal dimension. 

For each analysed sub-array, we computed $D_0$, $D_8$, $C_{\mathrm{div}}$, and $D_{\mathrm{div}}$ and then investigated their variation with respect to the evolution of the AR. We take into account the results derived from the unsigned and signed flux measurements of the magnetic field in the data. 

Next, for each sub-array, we evaluated the above parameters on the set of pixels characterized by a  LOS magnetic flux larger than $\pm$ 25 G that is $\approx \pm 3$ times  the standard deviation of quiet Sun magnetic flux distribution on the sub-array. We derived $D_0$ and $D_8$  from the least-squares best fit of the scaling relation of Equation (\ref{eq4}), and $C_{\mathrm{div}}$, and $D_{\mathrm{div}}$   from Equations (\ref{eq6})  and (\ref{eq7}). The uncertainty associated with the measured values was assumed to be equal to the 2-sigma uncertainty for the parameters returned by the regression fit of Equation (\ref{eq4}) and 
from error propagation.

\section{Results} \label{4}

\subsection{Magnetic flux and magnetic helicity }
\subsubsection{Flare-productive ARs}
Figure~\ref{fig1}  (left panels) shows the evolution of the positive-polarity ($\Phi_{+}$), negative-polarity ($\Phi_{-}$), and unsigned ($\Phi$) magnetic flux measured using the time series of each analyzed flare-productive AR. The panels in the middle column of Figure~\ref{fig1} display the magnetic helicity accumulation $H$ in the studied ARs. The positive (right-handed) $H^+$ and the absolute value of the negative (left-handed) $H^-$ helicity accumulation are reported in the right panels of Figure~\ref{fig1}.  We plot the measured  fluxes with their corresponding error bars that were derived as outlined in Sect.~\ref{3}. 
Vertical lines in each plot denote the  time of occurrence of M- (red), and X- (green) class flares; when the flare was associated with a CME, the thickness of the vertical line is enhanced.

In the following, we summarize the main findings derived from the analysis of Fig.~\ref{fig1}, concerning  the evolution of $\Phi$, $H$, $H^+$ and $H^-$ for each flaring AR. 
 
\begin{itemize}
	\item \textbf{AR 11166}:  the magnetic flux increases during the entire analyzed time interval. $\Phi_{+}$ and $\Phi_{-}$ are quite balanced (initially $\Phi_{+}$ is slightly higher, but after $\sim 62$ hr the situation is reversed).
The accumulated magnetic helicity is positive and shows a persistent increase from the beginning and during all the selected time interval; there is not any measured change after the M-class flare associated with the CME. The situation is similar after the X-class flare. We only report a flattening of the $H$ trend after the second M-class flare, followed by a new increase after the third M-class flare. Both the right- and the left-handed helicity show a persistent increase, even if $H^+$ shows a more significant increase during the analyzed time interval. It is worthwhile to note that, taking into account that this AR was on the northern hemisphere, it did not follow the general pattern cycle-invariant hemispheric helicity rule \citep{See90,Pev95}. 
	\item \textbf{AR 11283}: after an initial rising phase, $\Phi_{-}$ is almost constant, while $\Phi_{+}$ decreases during the investigated time interval. The positive and negative magnetic fluxes show an imbalance, with a higher $\Phi_{-}$, increasing in time. 
The values of $H$ are lower than for AR 11166 (see Table \ref{table2}) and this property might be related to the fact that the magnetic flux in AR 11166 was generally increasing, while in this AR it is characterized by lower values and it is decreasing. Similarly to AR 11166, both the right- and left-handed helicity show a persistent increase, with the $H^+$ exhibiting a more significant increase during the analyzed time interval.
Finally, we note that, also in this case, the AR did not follow the hemispheric helicity rule. For more details about the behaviour of helicity in this AR, see, e.g., \citet{Romano2015}.
\item \textbf {AR 11429}:  $\Phi_{-}$ decreases during the studied time interval, while  $\Phi_{+}$, after an initial increase,  remains almost constant. There is a flux imbalance such that $\Phi_{-} \sim 1.66$ $\Phi_{+}$, at the beginning, but the imbalance decreases with time.
 The magnetic helicity, characterized by negative values,  shows a very rapid increase until the X-class flare, that is associated with a CME, takes place. For almost 50 hr later, $H$ shows an alternation between positive and negative values and, after an M-class flare associated with a follow-up CME, there is a steeper increase of negative $H$ accumulation. The right- and left-handed helicity accumulation have similar values and increasing trend till the occurrence of the M-class flare associated with the CME, and later on  $H^-$ increases more rapidly than $H^+$. Interestingly, this AR did not follow the Hale Law: in fact, even though being on the northern hemisphere, its leading polarity was positive, appearing as a possible rogue AR \citep[see, e.g.,][]{rogueAR}. However, taking into account that the $H$ sign is negative, the AR followed the helicity hemispheric rule \citep[see, e.g.,][for a more elaborated analysis of this AR]{elmhamdi}.
	\item \textbf{AR 11515}:  $\Phi_{-}$ is almost constant during the analysed time interval, while $\Phi_{+}$ increases continuously. Initially, $\Phi_{-}$ $>$ $\Phi_{+}$, but after an elapsed time of t = 42 hr the situation is reversed. It is worth noting that at this time an M-class flare occurred with a CME and that the slope of the magnetic helicity accumulation changes, as described in the following. The accumulated magnetic helicity of this AR remains close to zero for the first $\sim 8$ hr (note that during the same time interval the magnetic flux also remains constant). Next, $H$ starts to increase (with negative values) and shows a phase of flattening few hr before the second M-class flare with a CME. Later on, $H$ increases again, without any clear variation signature after three M-class flares, which occurred between t  = 75 hr and t = 82 hr. The analysis of the right- and left-handed $H$ shows that both are characterized by a continuous increase, with $H^-$ showing a more significant increase during the analyzed time interval. This AR did not follow the helicity hemispheric rule.
\item \textbf{AR 11520}:  $\Phi_{-}$ decreases continuously, while $\Phi_{+}$ is almost constant. Initially $\Phi_{-}$ $>$ $\Phi_{+}$, but at time t =55 hr, the situation is reversed. Note that the time of reversal occurs  when $H$  changes its trend (from increasing to decreasing). 
	In fact, $H$ increases very rapidly until a maximum value of $8 \cdot 10^{42}$~Mx$^{2}$ is reached. $H$ is almost flat for $\sim 2$ hr and then an X-class flare with a CME, takes place. Immediately before and after this event, the trend of $H$ changes as it starts to decrease. Both the right- and the left-handed helicity show a continuous increase, but it is possible to distinguish an almost specular change in the trend of the relevant curves after the occurrence of the X-class flare. The occurrence of an M-class flare associated with the CME at time t = 93 hr does not seem to be related to any changes in the trend of $H$. This AR followed the helicity hemispheric rule. 
 \end{itemize}

\begin{table*}
\centering
\caption{Main characteristics of the flaring ARs. CM passage indicates the day on which the AR was on the central meridian; Yes or No in the fourth row indicate whether or not the magnetic polarity of the leading spot followed the Hale's Law; $\Phi_{max}$ indicates the maximum value reached by $\Phi$, the unsigned magnetic flux during the period analyzed;
$\Phi$ imbalance indicates the maximum imbalance between the two magnetic polarities and $\Phi_{-}$ and $\Phi_{+}$ 
indicate the negative and positive magnetic flux, respectively;
$\left| H \right|$ indicates the unsigned magnetic helicity accumulation.\label{table2}}

 \begin{tabular}{cccccc}
\hline\hline
Parameter & AR 11166 & AR 11283 & AR 11429 & AR 11515 & AR 11520 \\
\hline
{\small AR classification}	& $\beta \gamma \delta$ 	& $\beta \gamma \delta $     & $\beta \gamma \delta$ 	& $\beta \gamma \delta $    & $\beta \gamma \delta$ 	\\
{\small CM passage}	& Mar 8, 2011	& Sept 5, 2011     & Mar 8, 2012 	& Jul 3, 2012    &  Jul 12, 2012	\\
{\small Average Lat}	& N11	& N13     & N17		& S17    & S17	\\
{\small Hale Law} &	Yes	& Yes   & No 	& Yes & Yes\\
{\small $\Phi_{max}$ (Mx)} &	$2.8 \cdot 10^{22}$ 	& $1.6 \cdot 10^{22}$ & $3.7 \cdot 10^{22}$ 	& $3.3 \cdot 10^{22}$ & $5.0 \cdot 10^{22}$\\
{\small Max $\Phi$ imbalance} & {\small $\Phi_{+}\sim 1.1\cdot \Phi_{-}$ }	& {\small $\Phi_{-} \sim 1.8\cdot \Phi_{+}$} &  {\small $\Phi_{-} \sim 1.6\cdot \Phi_{+}$} 	& {\small $\Phi_{+} \sim 1.5\cdot \Phi_{-}$} & {\small $\Phi_{-} \sim 1.5\cdot \Phi_{+}$}\\
{\small $\mid{}H\mid$ max (Mx$^{2}$)}	& $6.6 \cdot 10^{42}$	& $1.9 \cdot 10^{42}$    &$ 3.0 \cdot 10^{42}$ 	& $ 7.7 \cdot 10^{42}$ & $8.0 \cdot 10^{42}$\\
{\small $H$ Sign} &	 Positive	& Positive & Negative 	& Negative & Positive\\
{\small Hemispheric rule} & No	& No   & Yes 	& No & Yes\\
\hline
\end{tabular}
\\
\vspace{0.2cm}
\end{table*}

Table \ref{table2} summarizes the  results of our measurements of the magnetic flux and helicity accumulation for the five analyzed flaring ARs.
 
\subsubsection{Flare-quiet ARs}
In the following, we summarize the results derived from the analysis of Fig.~\ref{fig1_bis}, showing the evolution of $\Phi$ and $H$, as well as the behaviour of the left-handed and right-handed helicity accumulation for each flare-quiet AR.
 
\begin{itemize}
	\item \textbf{AR 11267}: the magnetic flux shows a general decreasing trend during the analyzed time interval. However, while $\Phi_{+}$ is smoothly decreasing, $\Phi_{-}$ shows after about 10 hr from the beginning of the analyzed interval, an increasing phase, which later becomes constant and then decreases.
The accumulated magnetic helicity is negative and shows a continuous increase from the beginning till $\sim 50$ hr, when it starts to show an opposite (i.e., decreasing) behaviour. Both left-handed and right-handed $H$ show a persistent increase, but while the increase of $H^+$ is constant, the increase of $H^-$ is steeper during the first $\sim 50$ hr and smoother in the remaining time interval. This AR did not follow the helicity hemispheric rule. Another analysis of the helicity for this AR can be found in \citet{Guglielmino2016}. 
 \item \textbf{AR 11512}: the total magnetic flux is characterized by a decreasing trend during the analyzed time interval. This behaviour is mainly due to the decreasing $\Phi_{-}$, while $\Phi_{+}$ remains constant during the analyzed interval. 
The accumulated magnetic helicity is negative and shows a persistent increase (with different steepness) from the beginning till $\sim 60$ hr, when it starts to show an opposite (i.e., decreasing) behaviour, followed after about 10 hr by a new increase. Both left-handed and right-handed $H$ show a persistent increase, but similarly to AR 11267, even if less evident, the increase of $H^+$ is constant, while the increase of $H^-$ is steeper during the first $\sim 60$ hr and smoother in the remaining time interval. Like AR 11267, this AR did not follow the helicity hemispheric rule.
	\item \textbf{AR 11589}: the magnetic flux shows a decreasing trend during the analyzed time interval. This is more evident in the $\Phi_{+}$ and less in the $\Phi_{-}$. The accumulated magnetic helicity is characterized by negative values and shows a continuous increase from the beginning till $\sim 80$ hr, when it becomes almost constant. The left- and right-handed $H$ are both linearly increasing during the analyzed time interval.
	\item \textbf{AR 11635}: the magnetic flux shows an alternate behaviour: initially decreasing, then increasing and finally decreasing again. $\Phi_{-}$ is always higher than $\Phi_{+}$.
The accumulated magnetic helicity has positive values and shows an increasing trend, initially quite smooth, but steeper after $\sim 30$ hr. Successively, we can notice a phase of constant value of $H$ and finally a decrease. The right- and left-handed $H$, both increasing with time, are characterized by an initial similar trend, but later on $H^-$ increases more slowly. This AR did not follow the helicity hemispheric rule.
	\item \textbf{AR 11775}: the magnetic flux (total and negative) shows a decreasing trend during the analyzed time interval, while $\Phi_{+}$ is almost constant. The accumulated magnetic helicity has positive values and shows a continuous and steep increase from the beginning till $\sim 30$ hr, when it becomes almost constant. $H^-$ is characterized by a linear increasing trend, while $H^+$ shows initially a similar trend, but after $\sim 30$ hr the increase is smoother.
	\end{itemize}

Table~\ref{table2bis} summarises the  results of the magnetic flux and helicity accumulation for the five analysed flare-quiet ARs.

From the comparison of the helicity accumulation trend between the flare-producing and flare-quiet ARs, we can conclude that for the first class of ARs $H$ is generally characterised by a persistent accumulation of higher magnitudes and senses of helicity, with the only exception of AR 11520. In this respect, it should be noted that the change of the trend in this AR takes place in correspondence  to the occurrence of a CME: this behaviour has already been reported in previous studies (see, e.g. \citet{smyrli}) and has been interpreted as due to a process by which a significant amount of helicity can be carried away during the CME. Differently, the flare-quiet ARs do not show a persistent accumulation of higher magnitudes and senses of helicity: $H$ can be initially increasing and later on decreasing (ARs 11267 and 11635), initially increasing and later on (almost) constant (ARs 11512, 11589, and 11775).

Another element that seems to differentiate the two groups of ARs is related to the different values of the right- and left-handed helicity accumulation, being about 3 - 4 times larger for the flare-productive ARs with respect to the flare-quiet ARs (as already noticed, the lower values of the $H$ accumulation for AR 11283 might be related to a decreasing and lower magnetic flux).

\begin{table*}
\caption{Main characteristics of the flare-quiet ARs. Same Parameters as in Table \ref{table2}.\label{table2bis}}
\centering
 \begin{tabular}{cccccc}
\hline\hline

Parameter &AR 11267 & AR 11512 & AR 11589 & AR 11635 & AR 11775 \\
\hline
{\small AR classification}	& $\beta \gamma \delta$ 	& $\beta \gamma \delta $     & $\beta \gamma \delta$ 	& $\beta \gamma \delta $    & $\beta \gamma \delta$ 	\\
{\small CM passage}	& Aug 7, 2011	& Jun 28, 2012     & Oct 10, 2012 	& Dec 24, 2012    &  Jun 21, 2013	\\
{\small Average Lat}	& S16	& S16     & N13		& N11    & S26	\\
{\small Hale Law} &	Yes	& Yes   & Yes 	& Yes & Yes\\
{\small $\Phi_{max}$ (Mx)} &	$3.8 \cdot 10^{21}$ 	& $1.0 \cdot 10^{22}$ & $1.7 \cdot 10^{22}$ 	& $1.3 \cdot 10^{22}$ & $1.5 \cdot 10^{22}$\\
{\small Max $\Phi$ imbalance} & {\small $\Phi_{-}\sim 2.0\cdot \Phi_{+}$ }	& {\small $\Phi_{+} \sim 2.0\cdot \Phi_{-}$} &  {\small $\Phi_{-} \sim 1.8\cdot \Phi_{+}$} 	& {\small $\Phi_{-} \sim 1.9\cdot \Phi_{+}$} & {\small $\Phi_{-} \sim 1.7\cdot \Phi_{+}$}\\
{\small $\mid{}H\mid$ max (Mx$^{2}$)}	& $1.9 \cdot 10^{41}$	& $1.2 \cdot 10^{42}$    &$ 2.0 \cdot 10^{42}$ 	& $ 1.1 \cdot 10^{42}$ & $1.2 \cdot 10^{42}$\\
{\small $H$ Sign} &	 Negative	& Negative & Negative 	& Positive & Positive\\
{\small Hemispheric rule} & No	& No   & Yes 	& No & Yes\\
\hline
\end{tabular}
\\
\vspace{0.2cm}
\end{table*}

\afterpage{
\begin{figure*}
	\centerline{\includegraphics[scale=0.125,trim={90 285 10 215},clip,width=5.15cm]{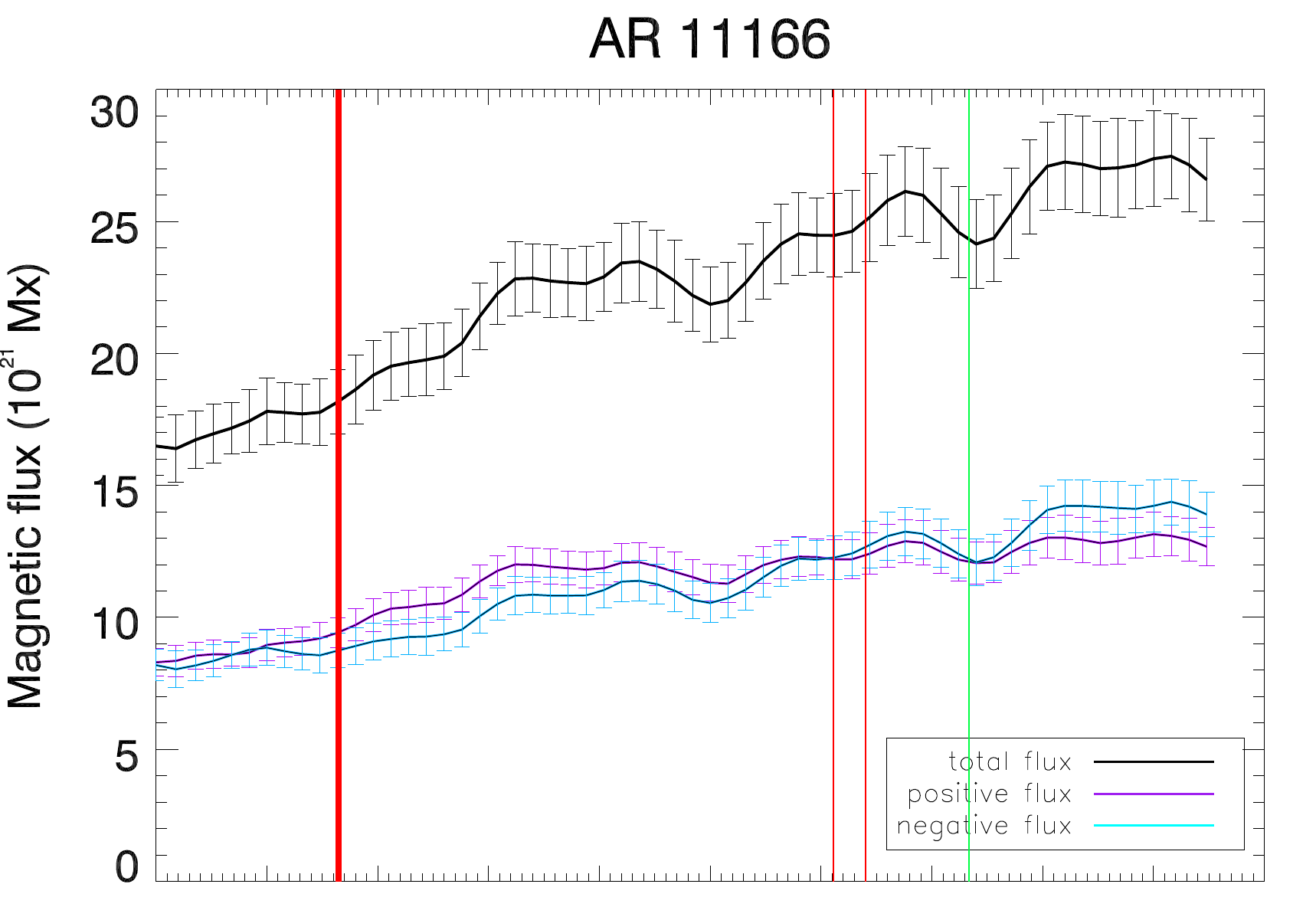}\includegraphics[scale=0.125,trim={80 285 10 215},clip,width=5.15cm]{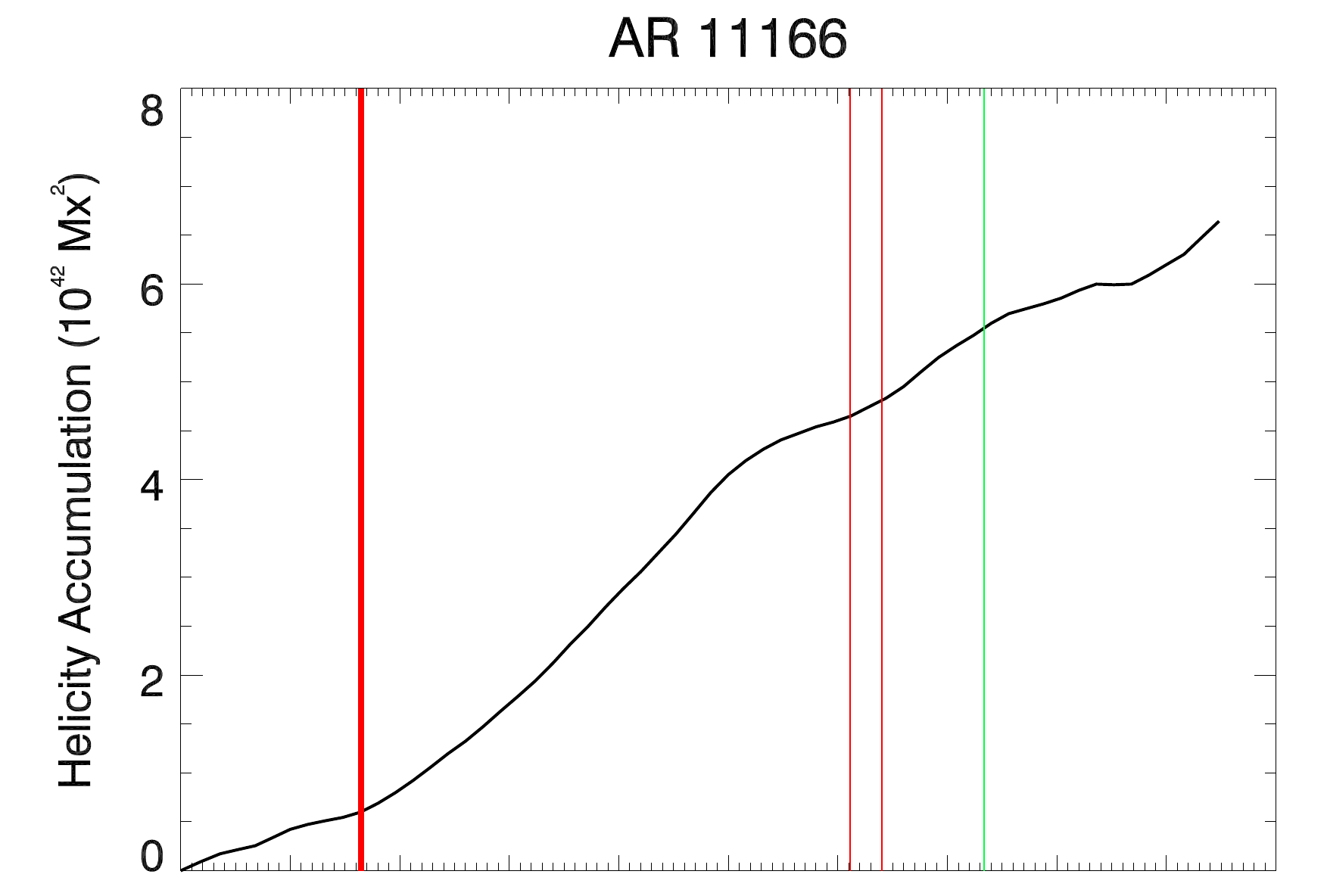}\includegraphics[scale=0.125,trim={90 285 10 215},clip,width=5.15cm]{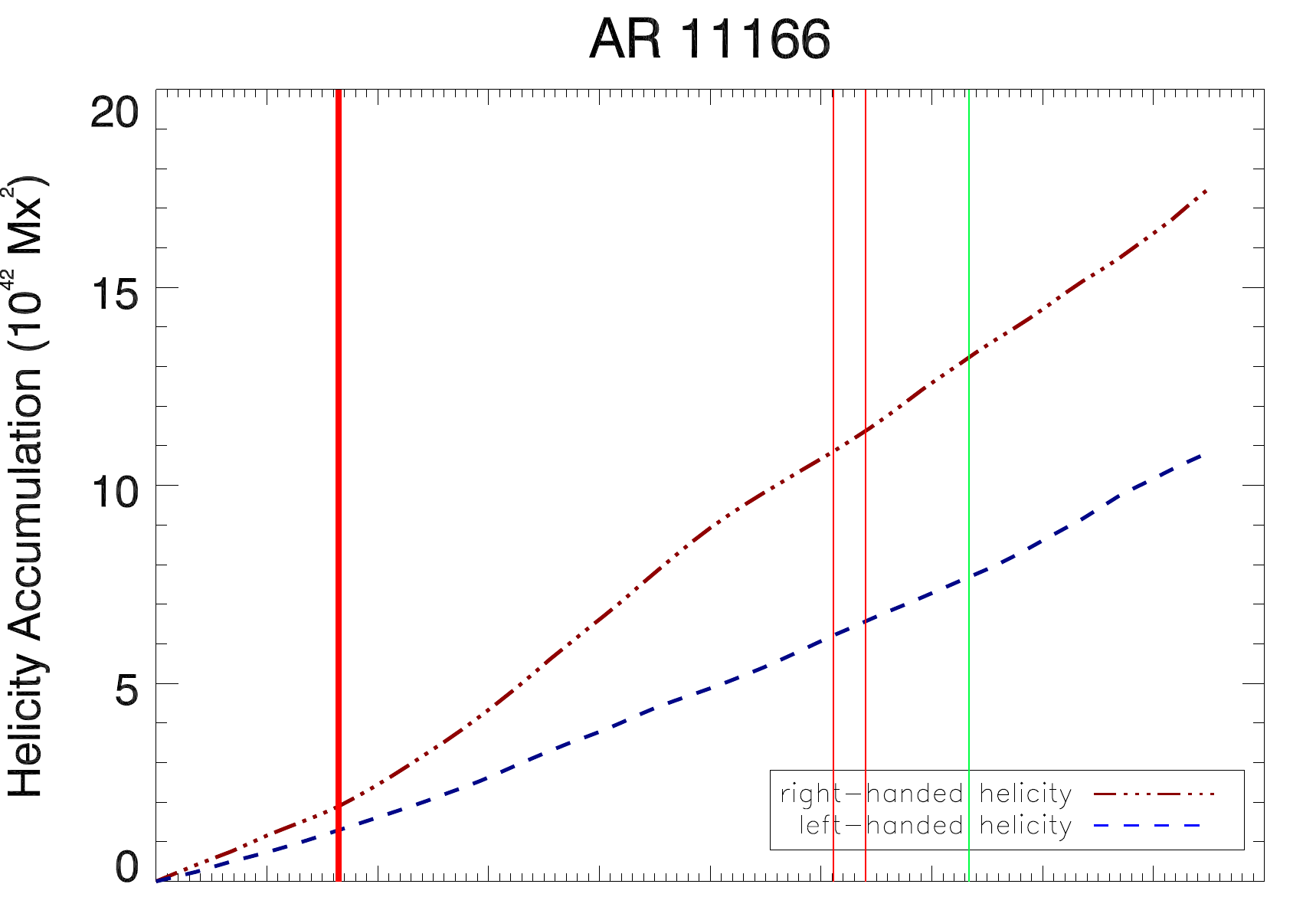}}
	\centerline{\includegraphics[scale=0.125,trim={90 285 10 215},clip,width=5.15cm]{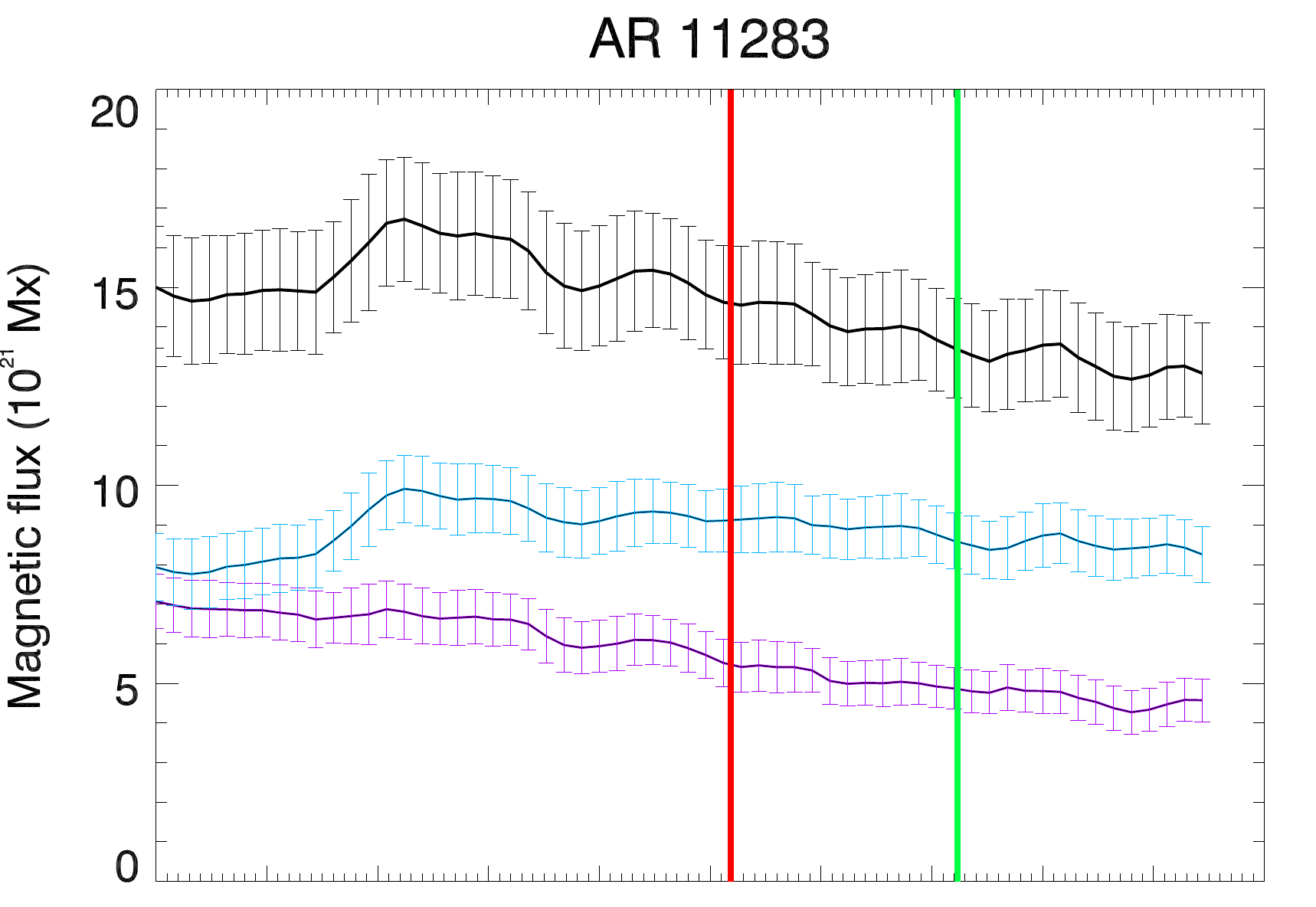}\includegraphics[scale=0.125,trim={80 285 10 215},clip,width=5.15cm]{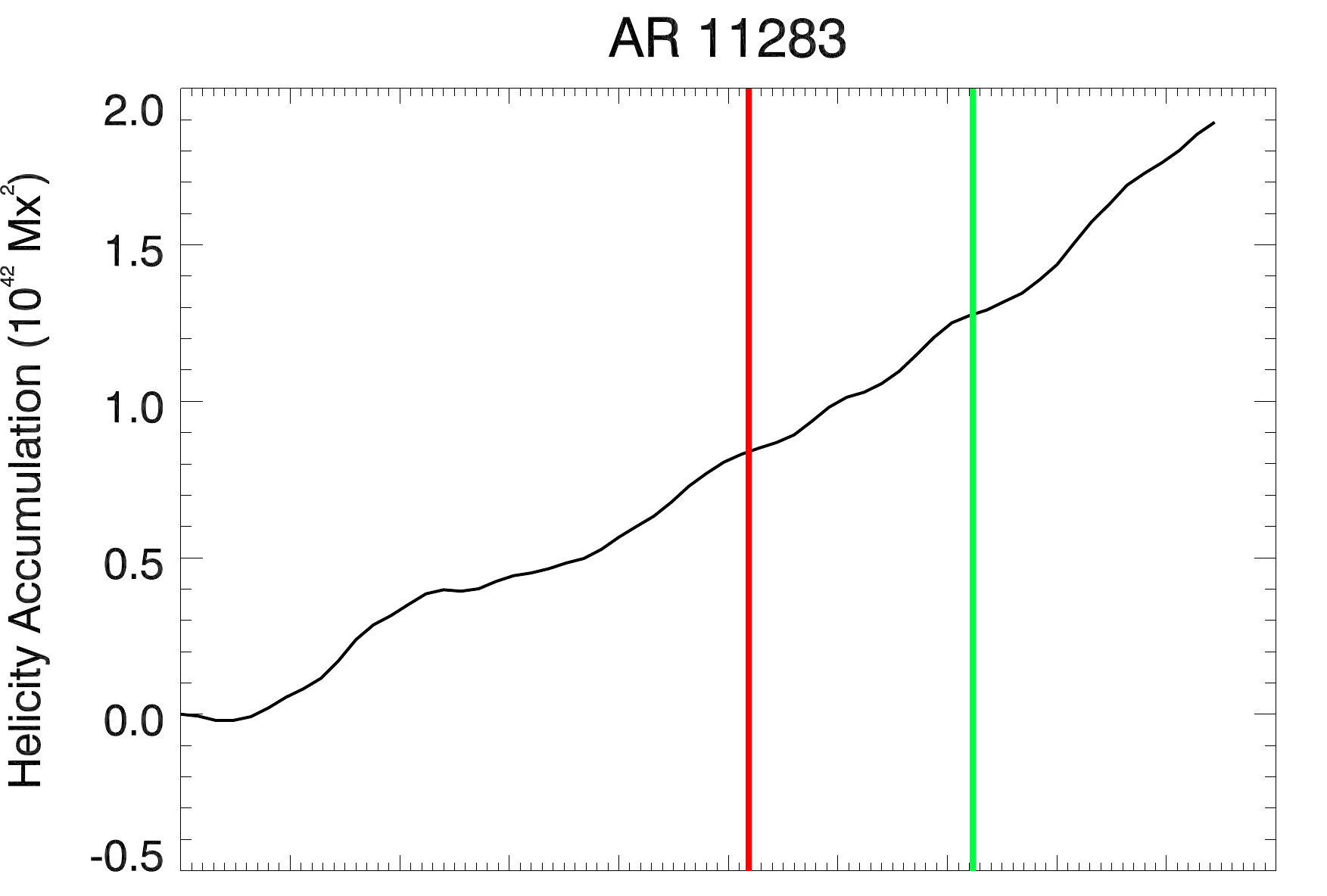}\includegraphics[scale=0.125,trim={90 285 10 215},clip,width=5.15cm]{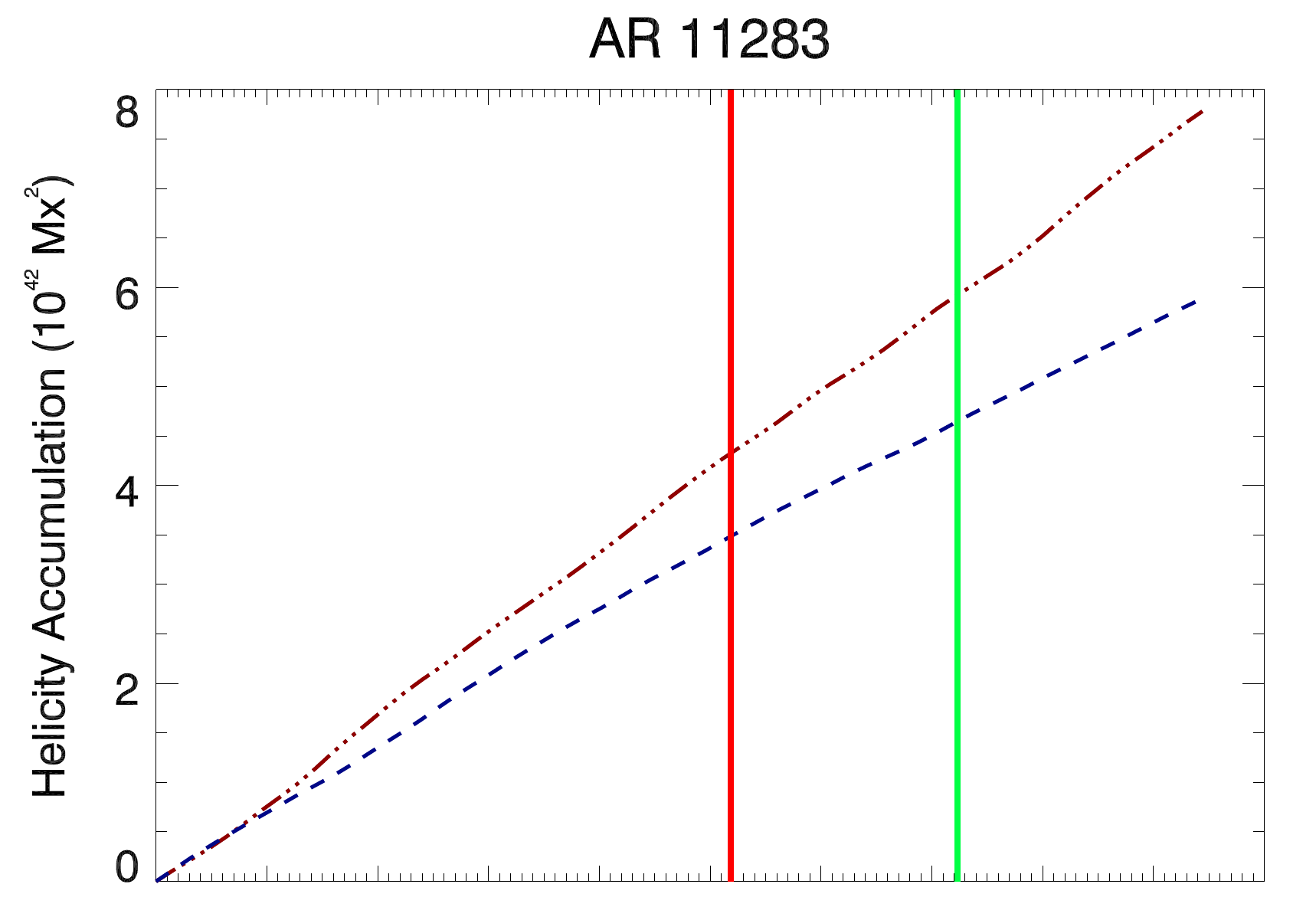}}
	\centerline{\includegraphics[scale=0.125,trim={90 285 10 215},clip,width=5.15cm]{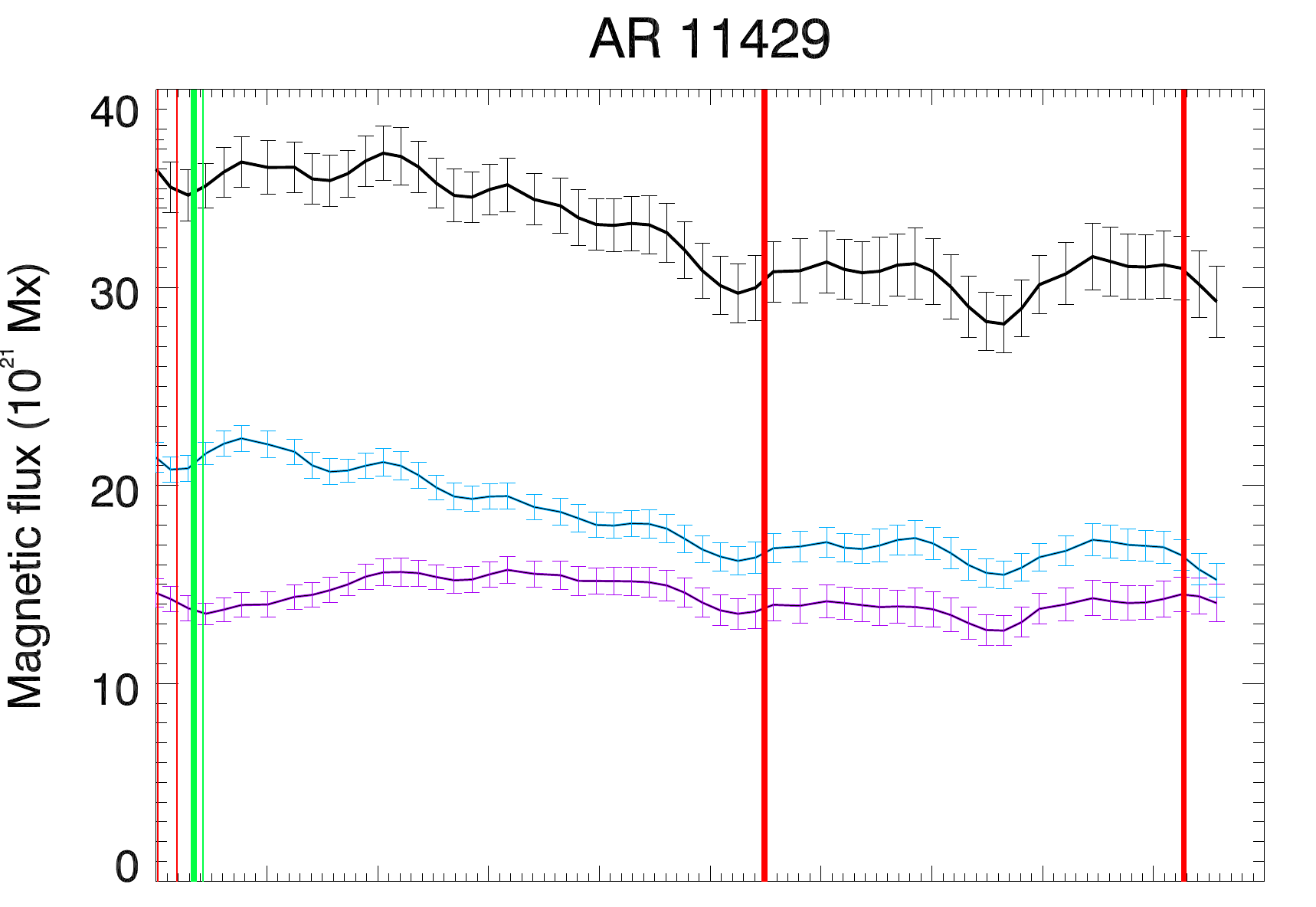}\includegraphics[scale=0.125,trim={90 285 10 215},clip,width=5.15cm]{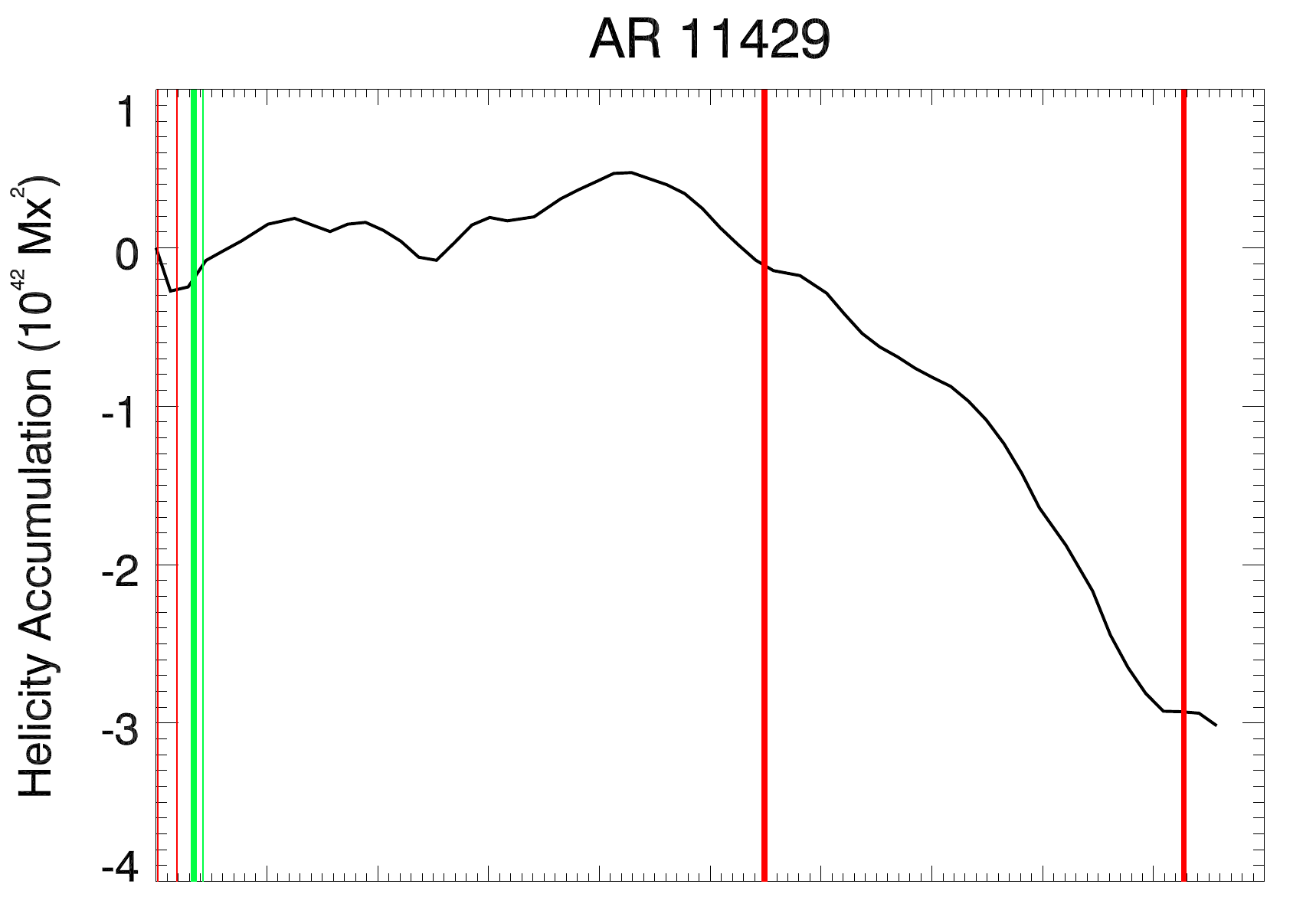}\includegraphics[scale=0.125,trim={90 285 10 215},clip,width=5.15cm]{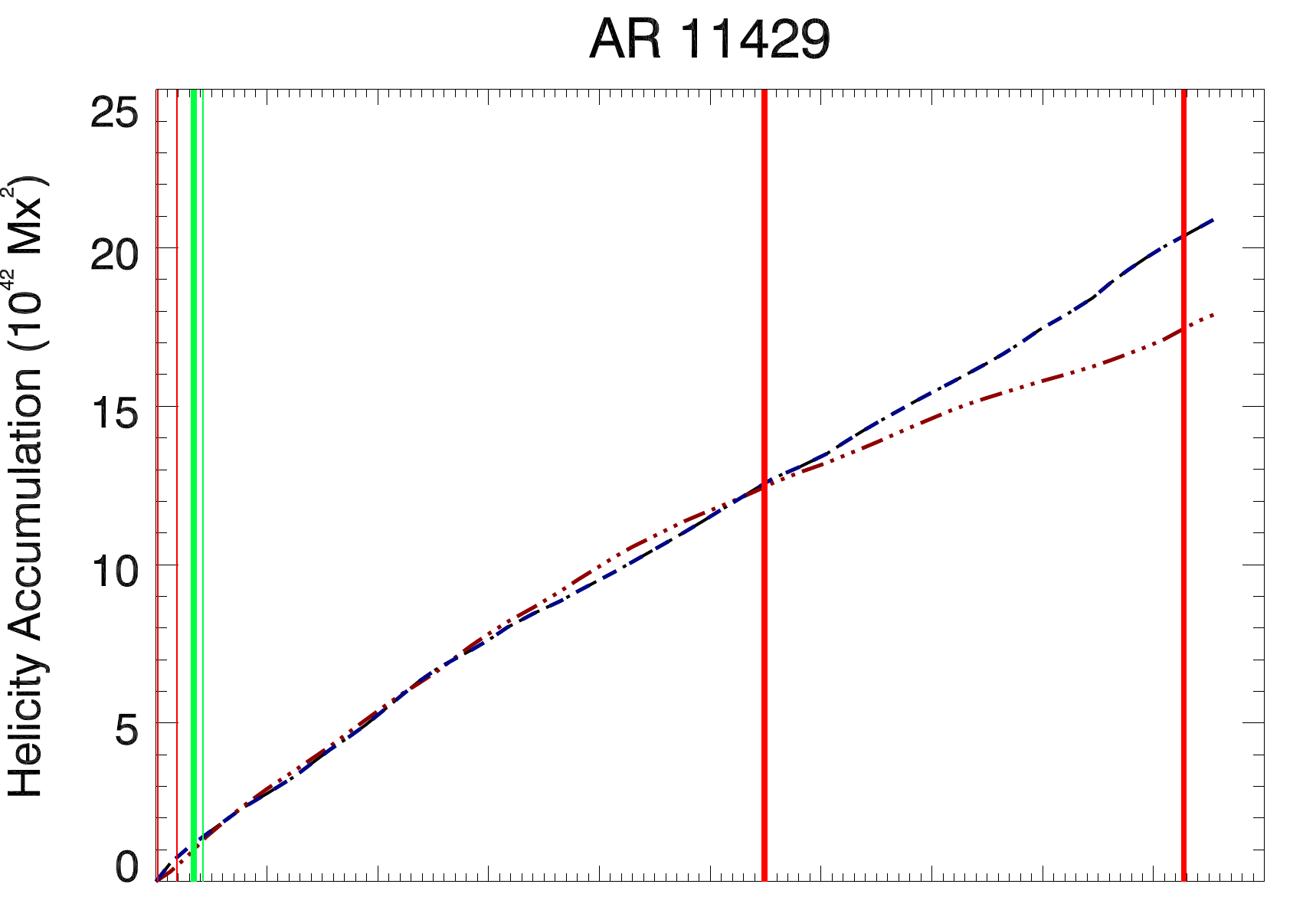}}
	\centerline{\includegraphics[scale=0.125,trim={90 285 10 215},clip,width=5.15cm]{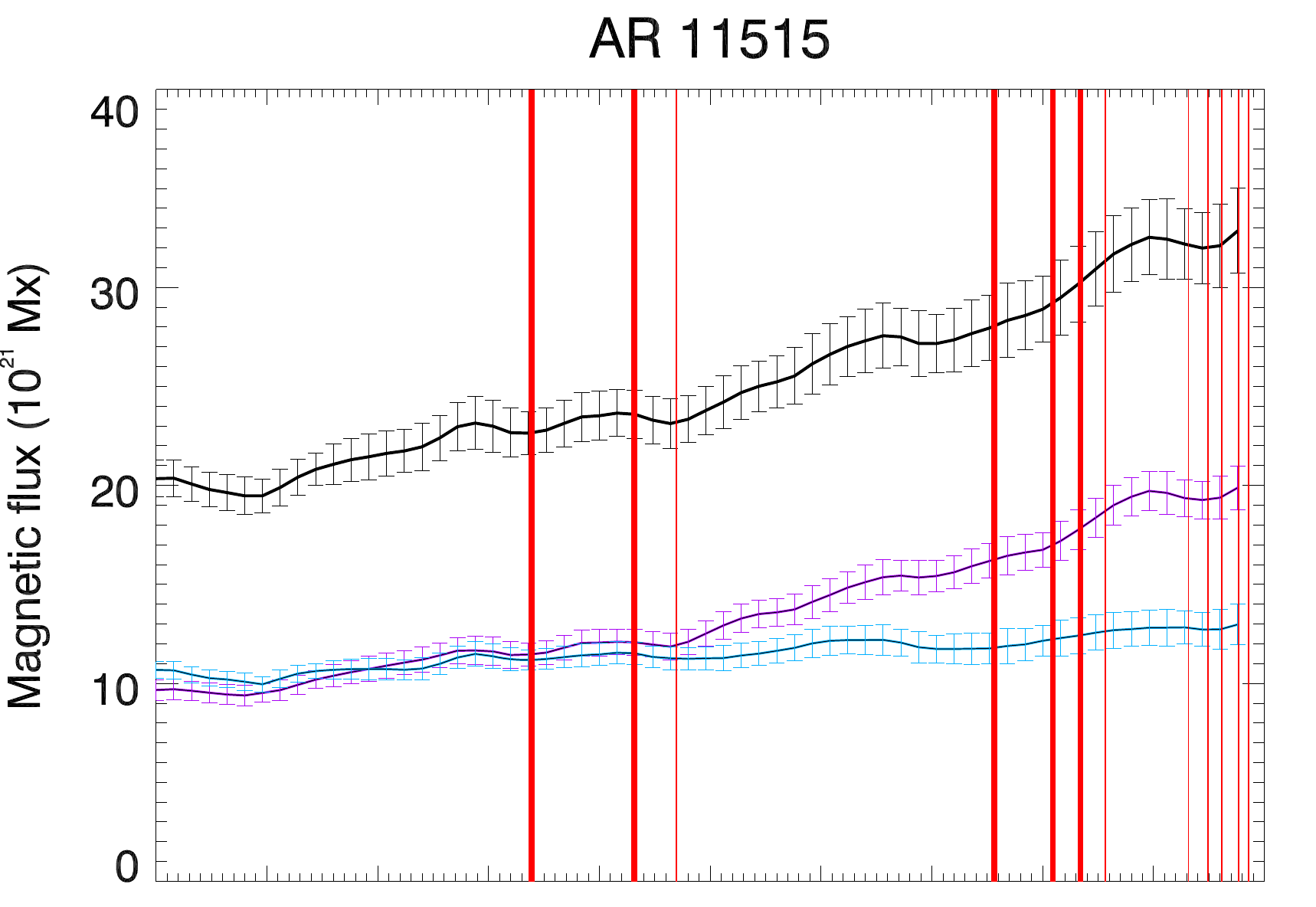}\includegraphics[scale=0.125,trim={90 285 10 215},clip,width=5.15cm]{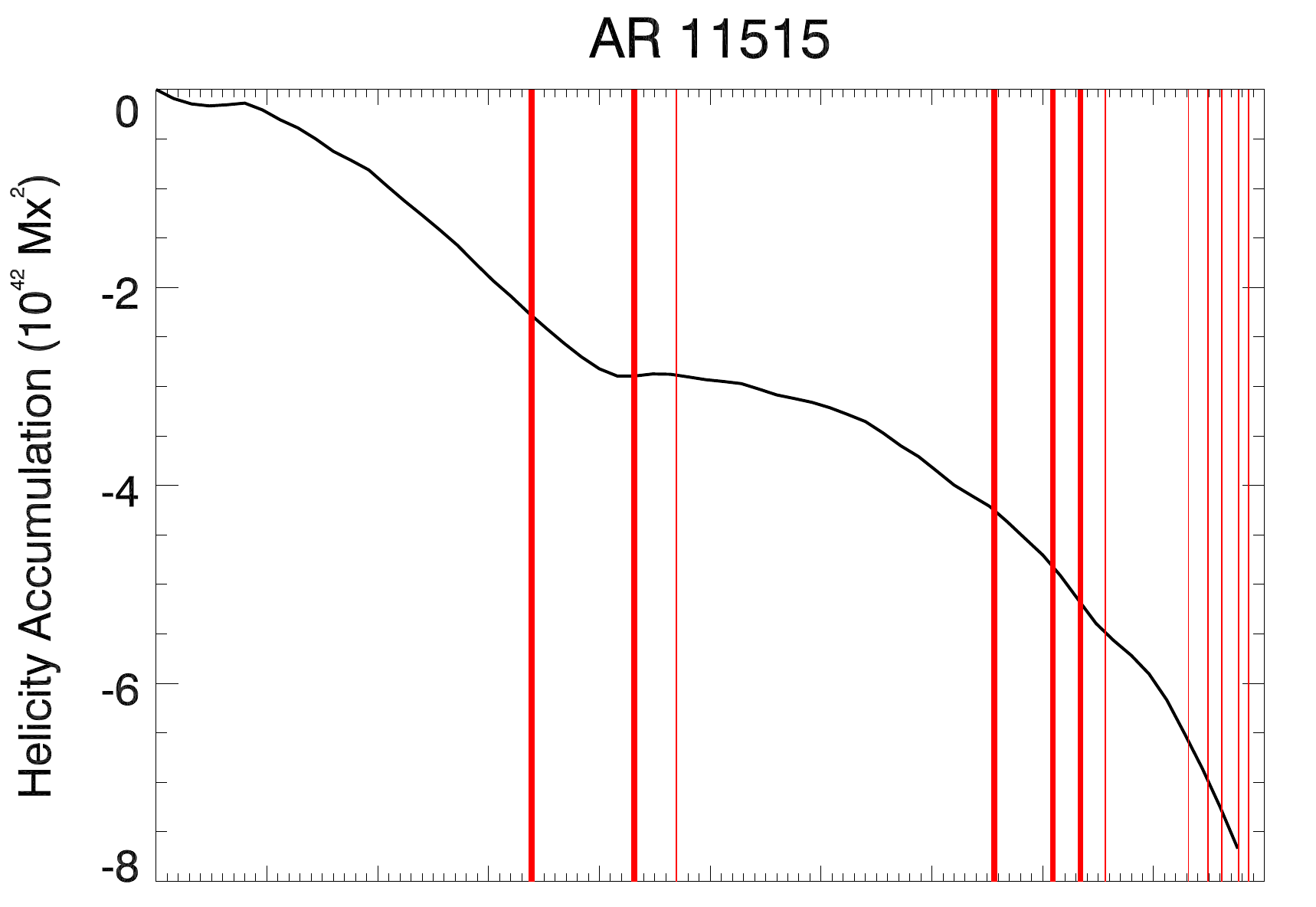}\includegraphics[scale=0.125,trim={90 285 10 215},clip,width=5.15cm]{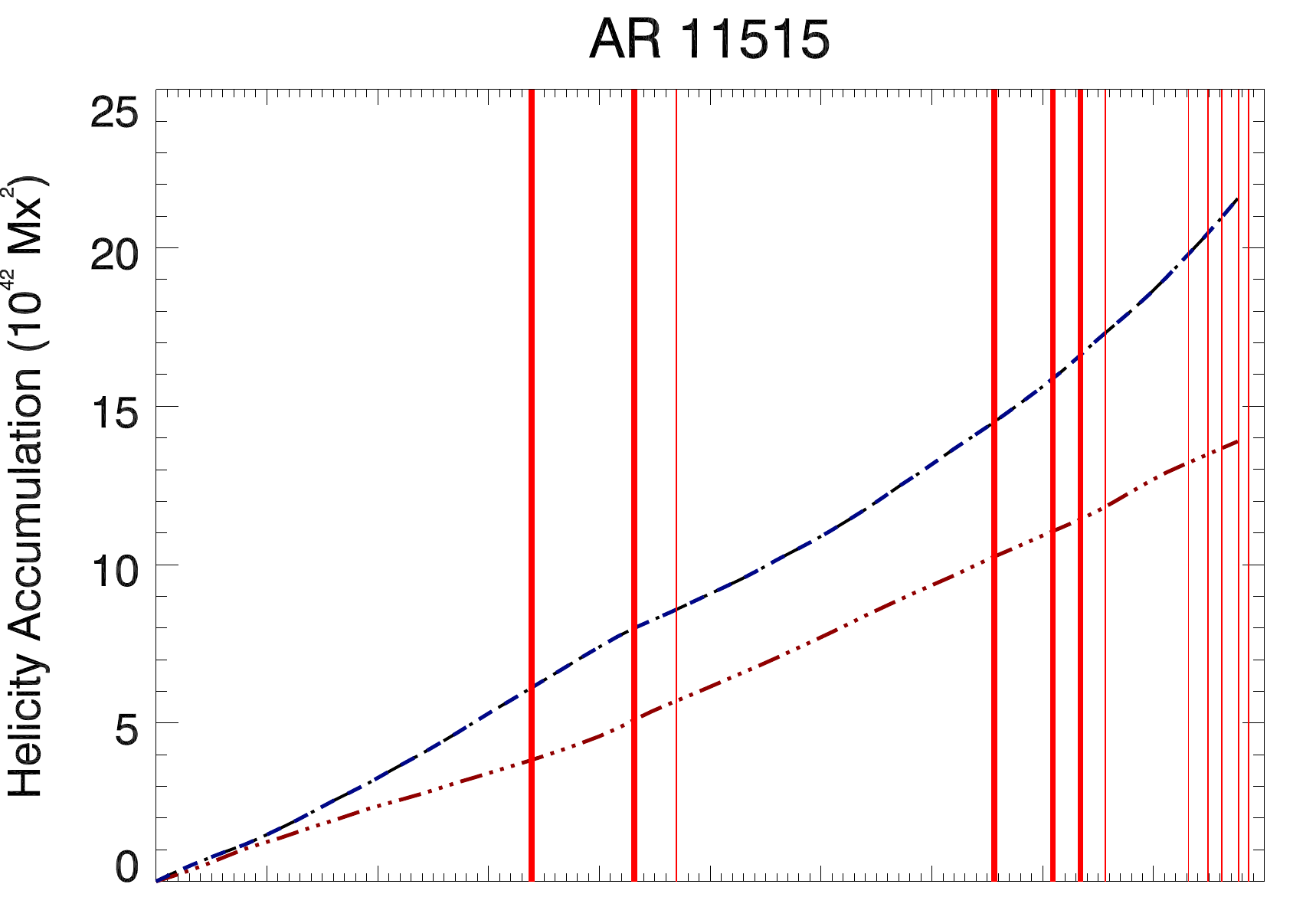}}
	\centerline{\includegraphics[scale=0.125,trim={90 220 10 215},clip,width=5.15cm]{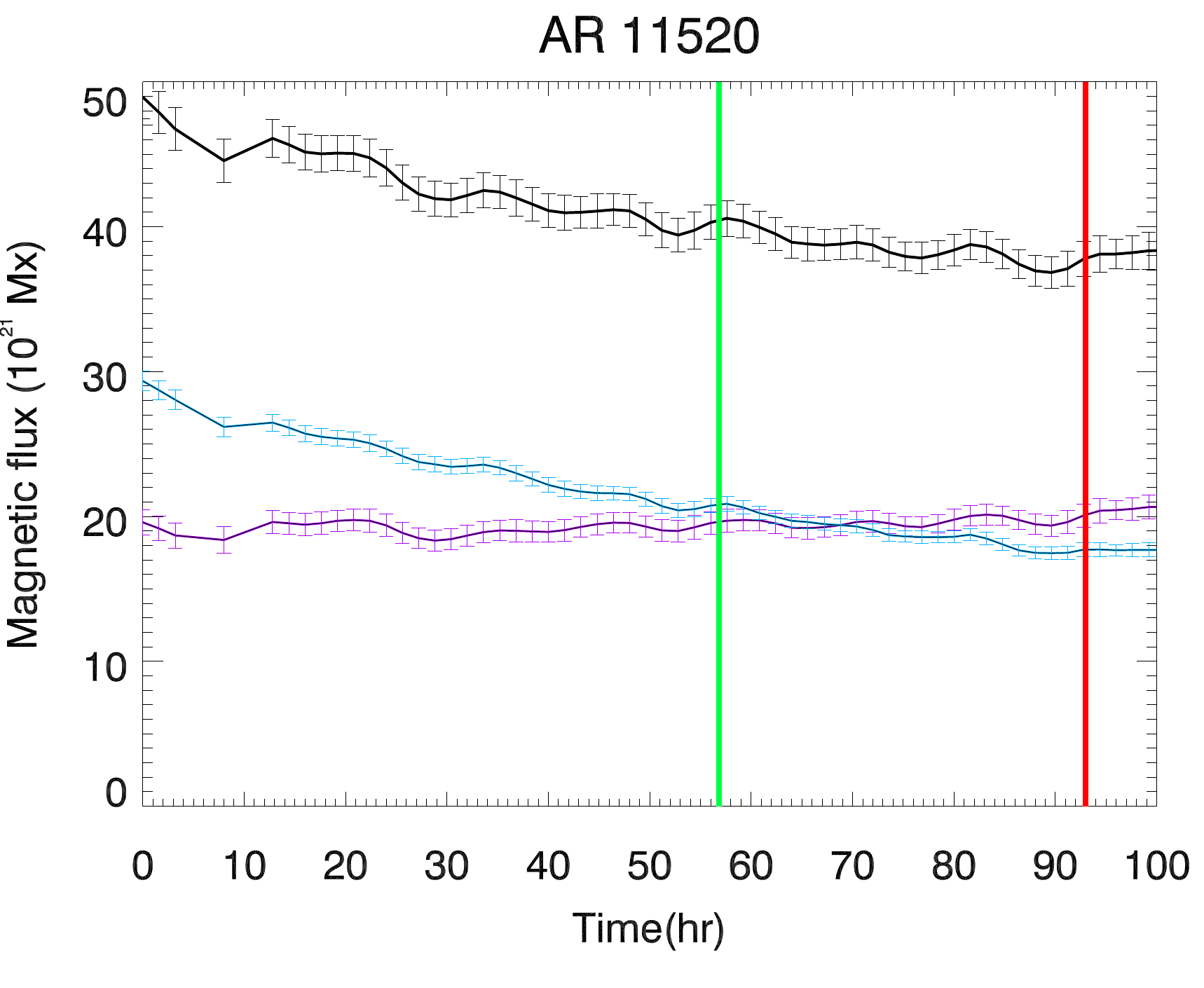}\includegraphics[scale=0.125,trim={90 220 10 215},clip,width=5.15cm]{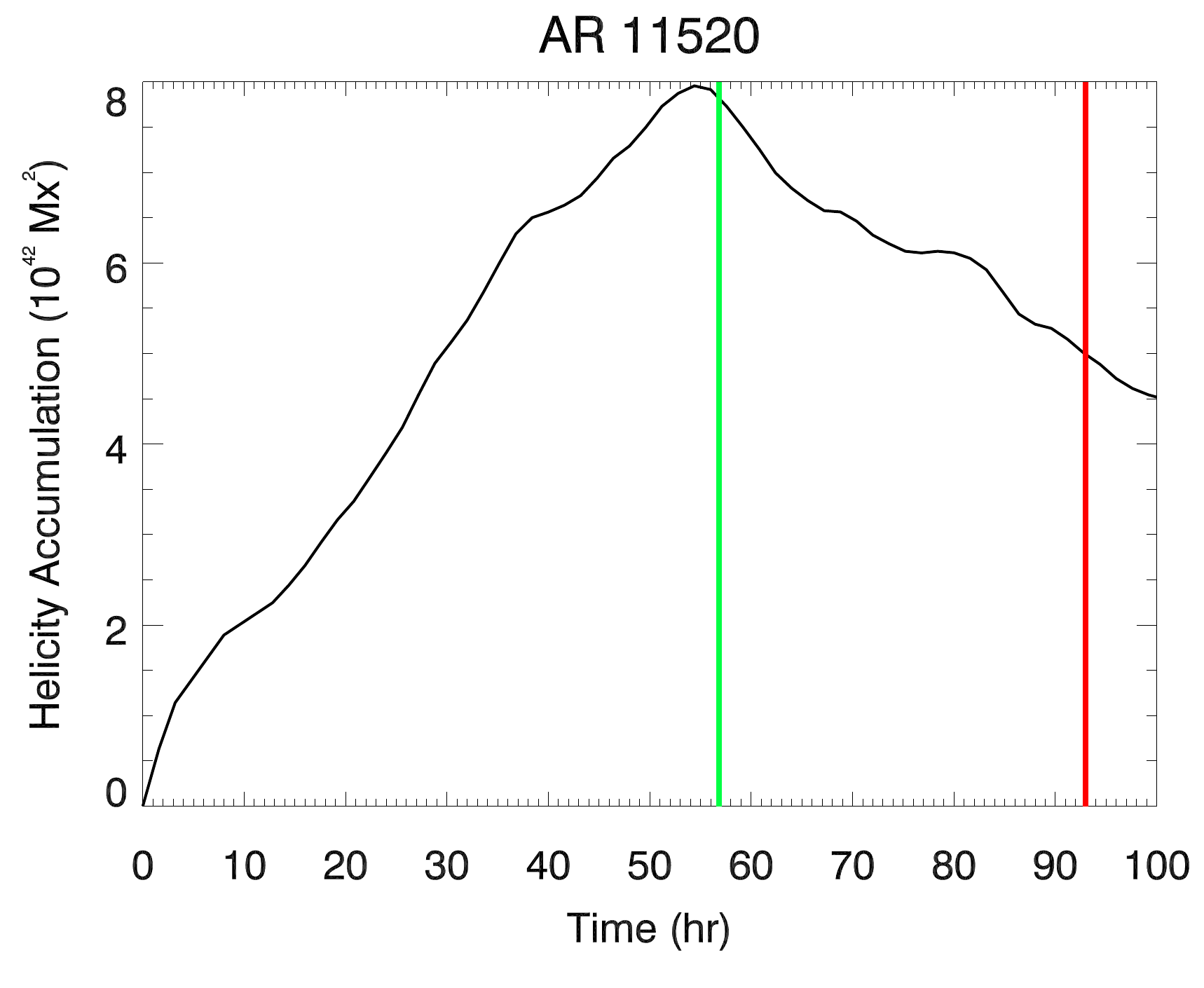}\includegraphics[scale=0.125,trim={90 220 10 215},clip,width=5.15cm]{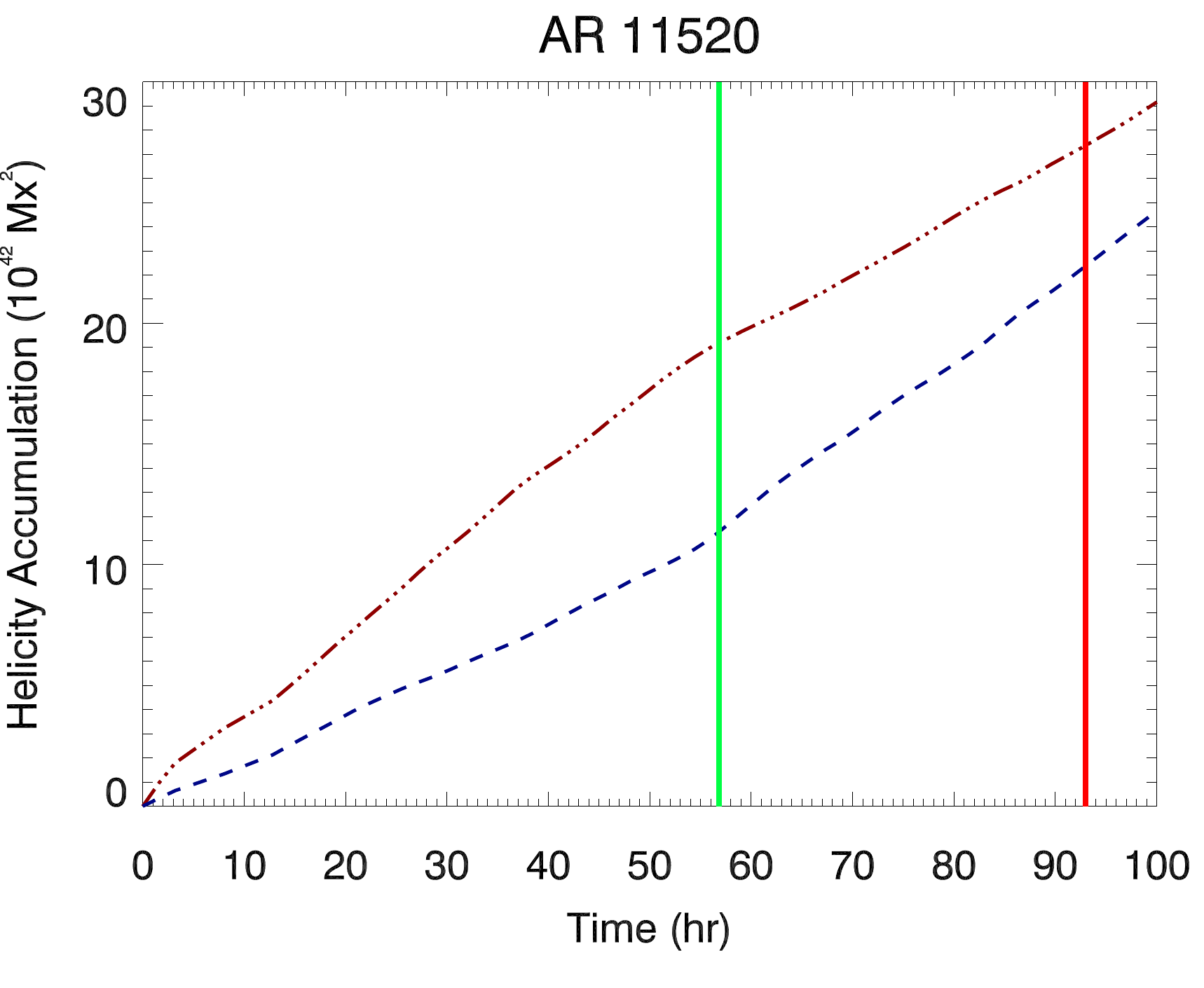}}
	\caption{\footnotesize {Trend of the magnetic flux (left-hand column), of the helicity accumulation (center column) and of the right-handed $H^+$ (dark red), and left-handed $H^-$ (in absolute value, blue) magnetic helicity accumulation (right-hand column) for the  flare-productive ARs in our sample. From top to bottom results for ARs 11166, 11283, 11429, 11515, 11520. Error bars in the magnetic flux plots indicate the standard deviation of measured values. The red (green) vertical lines indicate the time of occurrence of M-class (X-class) flares. Flares associated with CMEs are marked by thick lines. Time 0 corresponds to the start time shown in Table \ref{table1} for each analyzed AR.}}
	\label{fig1}
\end{figure*}
\clearpage}

\afterpage{
\begin{figure*}
	\centerline{\includegraphics[scale=0.125,trim={90 285 10 215},clip,width=5.15cm]{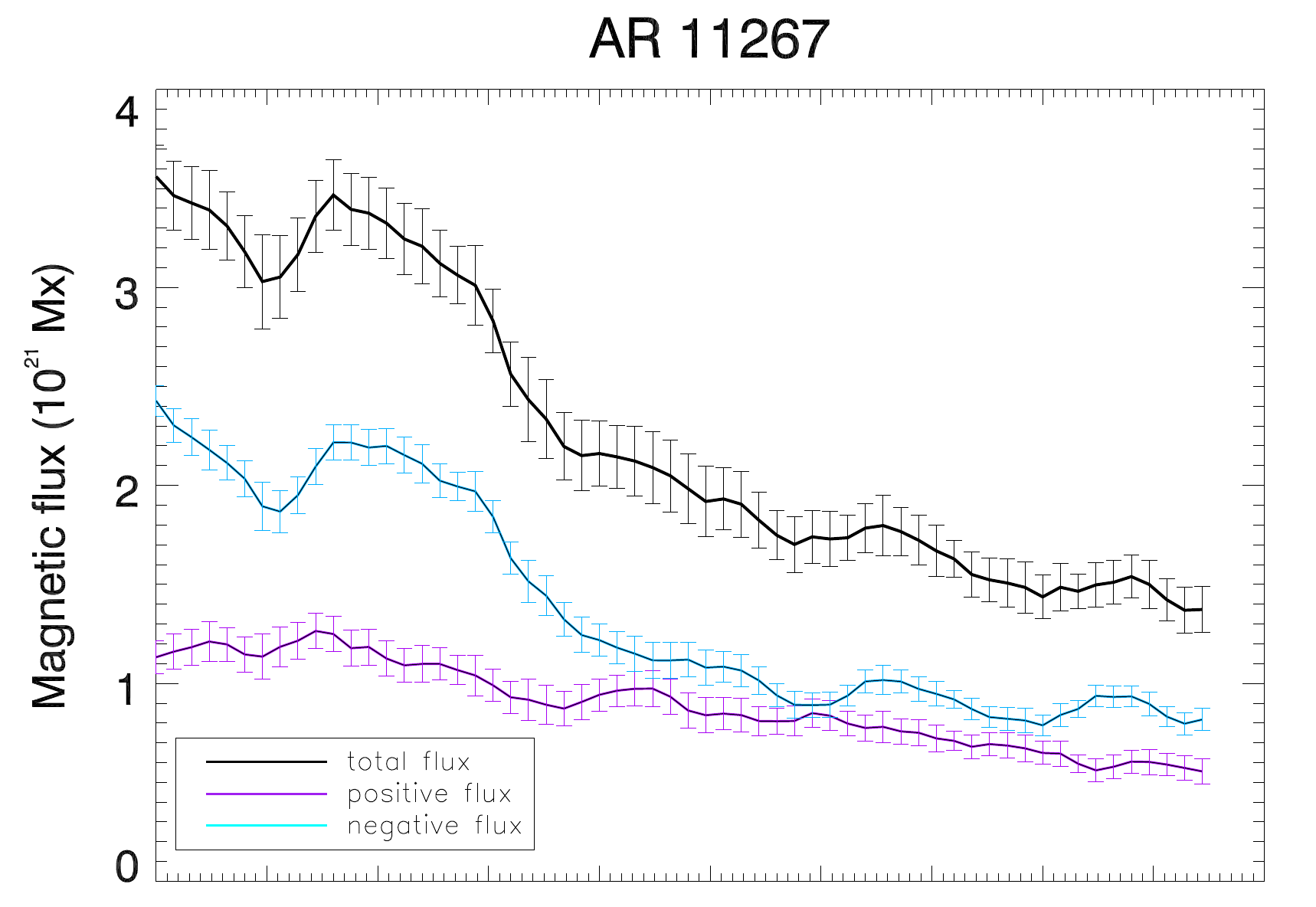}\includegraphics[scale=0.125,trim={70 285 10 215},clip,width=5.15cm]{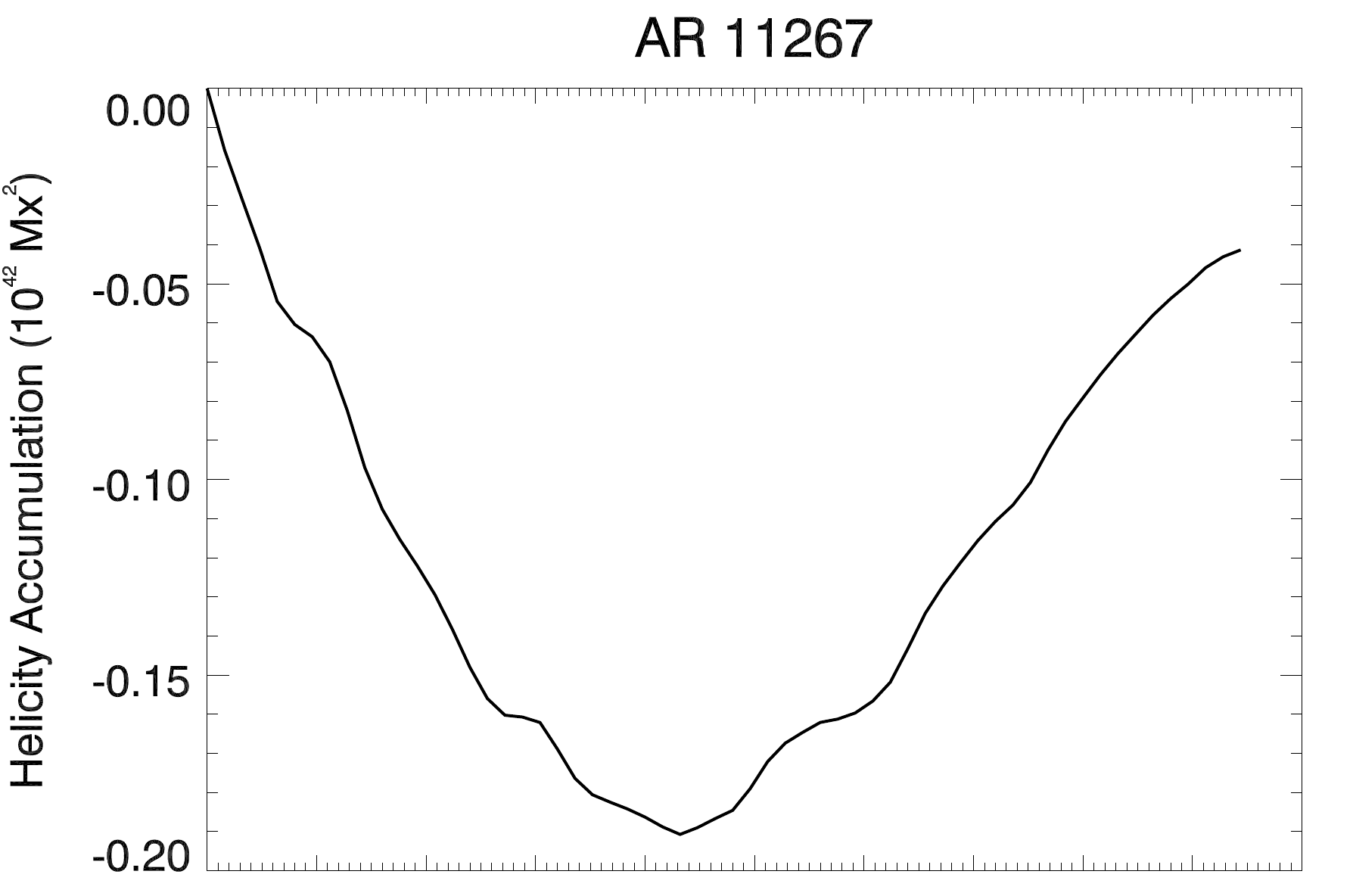}\includegraphics[scale=0.125,trim={90 285 10 215},clip,width=5.15cm]{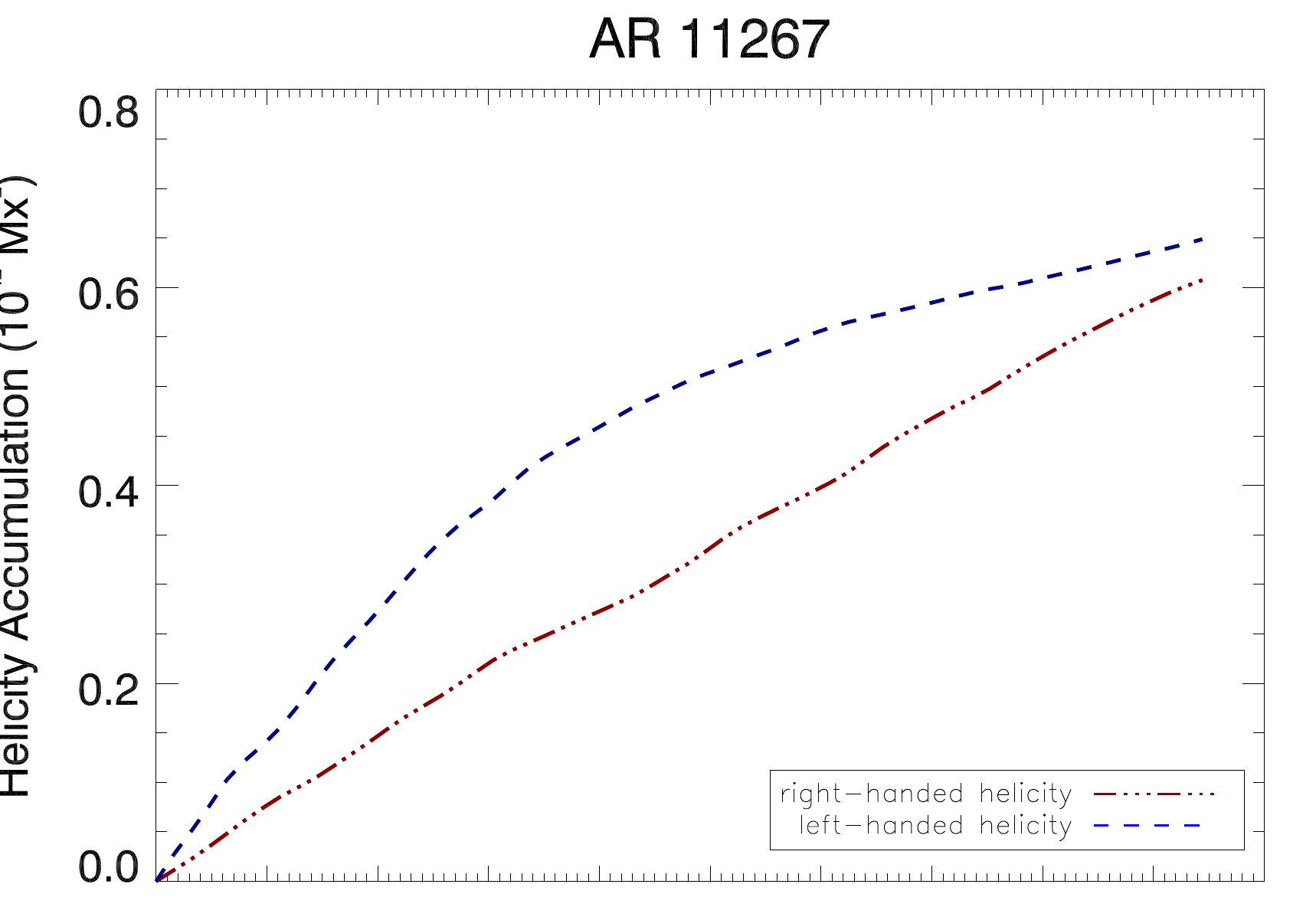}}
	\centerline{\includegraphics[scale=0.125,trim={90 285 10 215},clip,width=5.15cm]{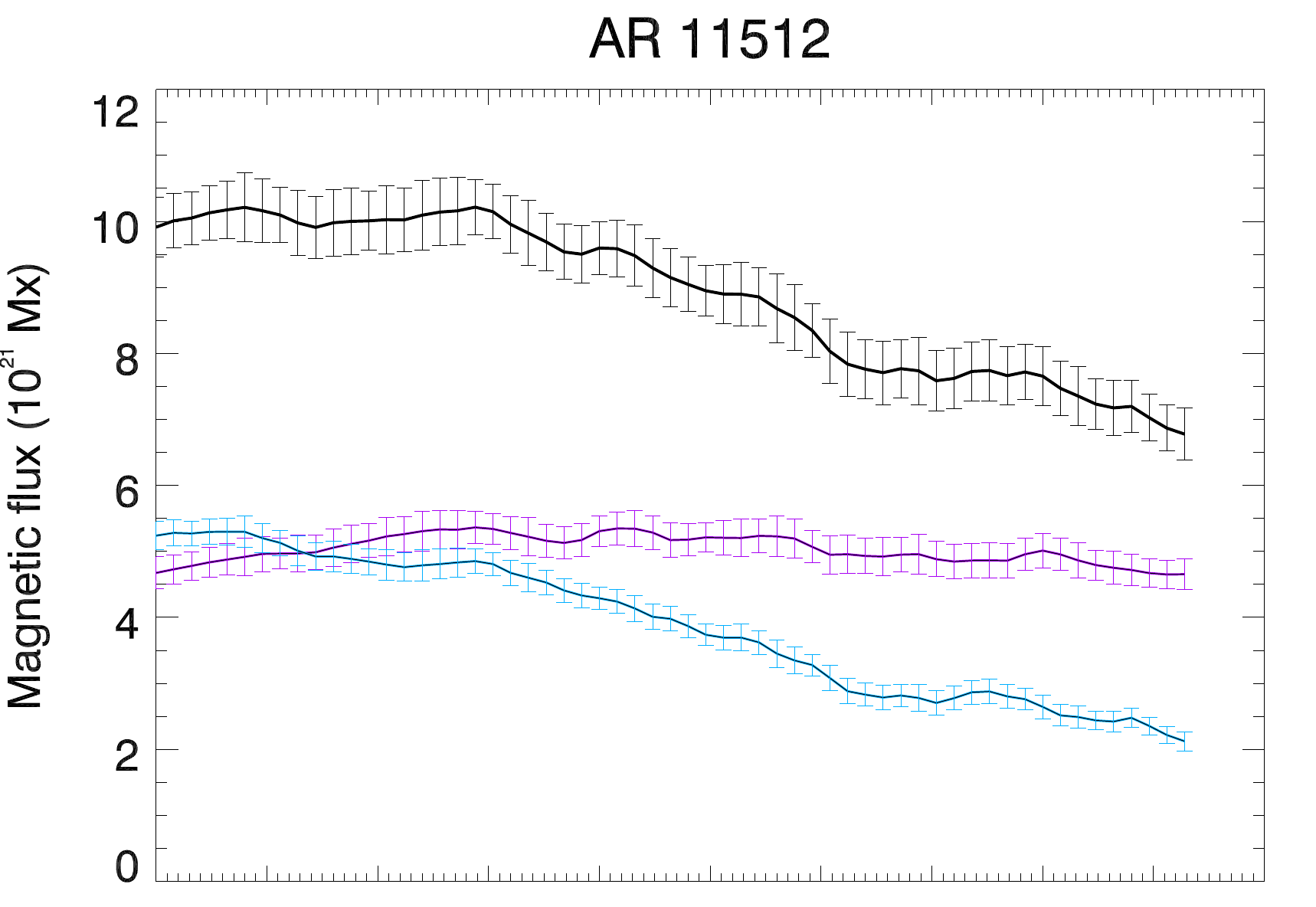}\includegraphics[scale=0.125,trim={80 285 10 215},clip,width=5.15cm]{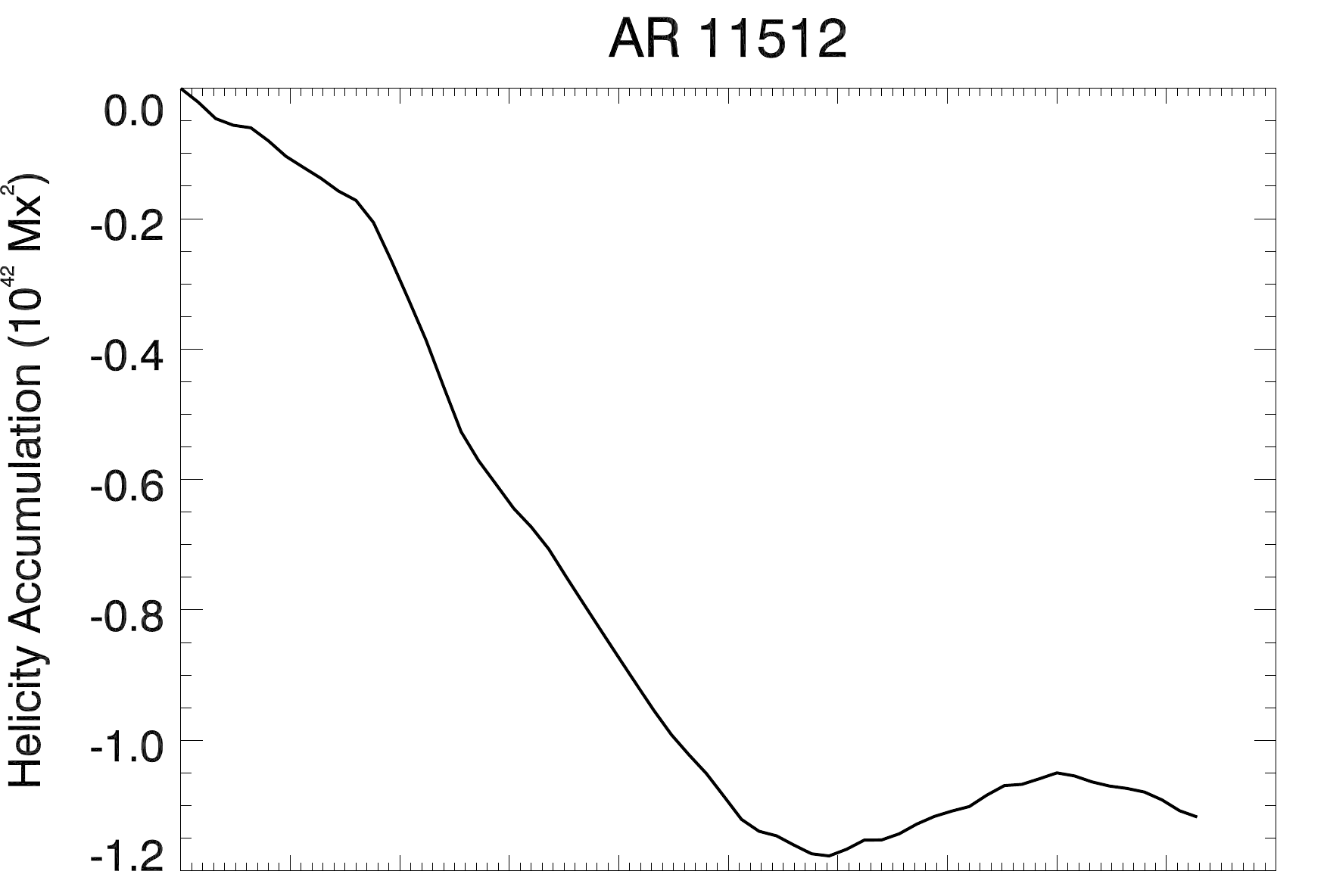}\includegraphics[scale=0.125,trim={90 285 10 215},clip,width=5.15cm]{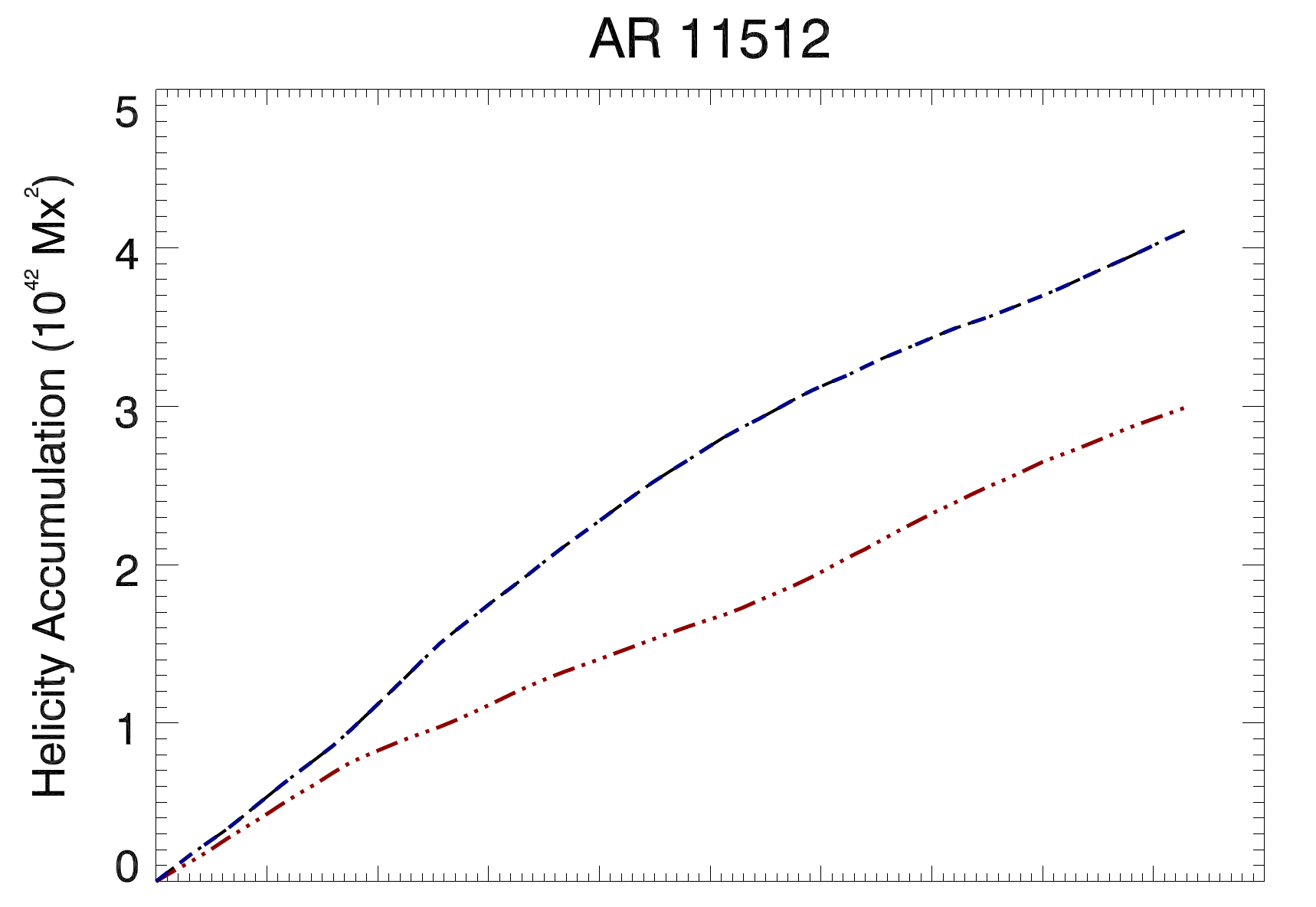}}
	\centerline{\includegraphics[scale=0.125,trim={90 285 10 215},clip,width=5.15cm]{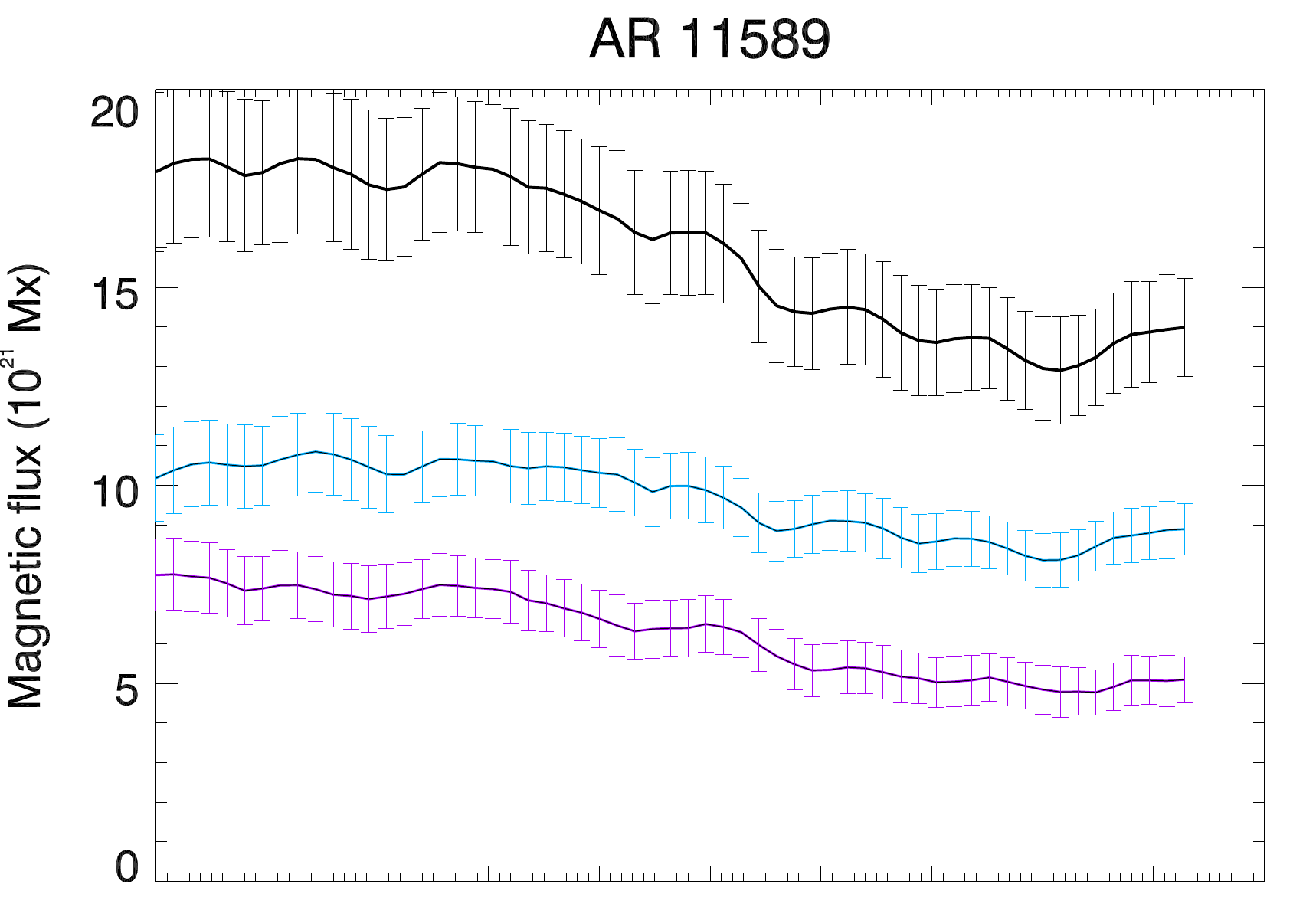}\includegraphics[scale=0.125,trim={70 285 10 215},clip,width=5.15cm]{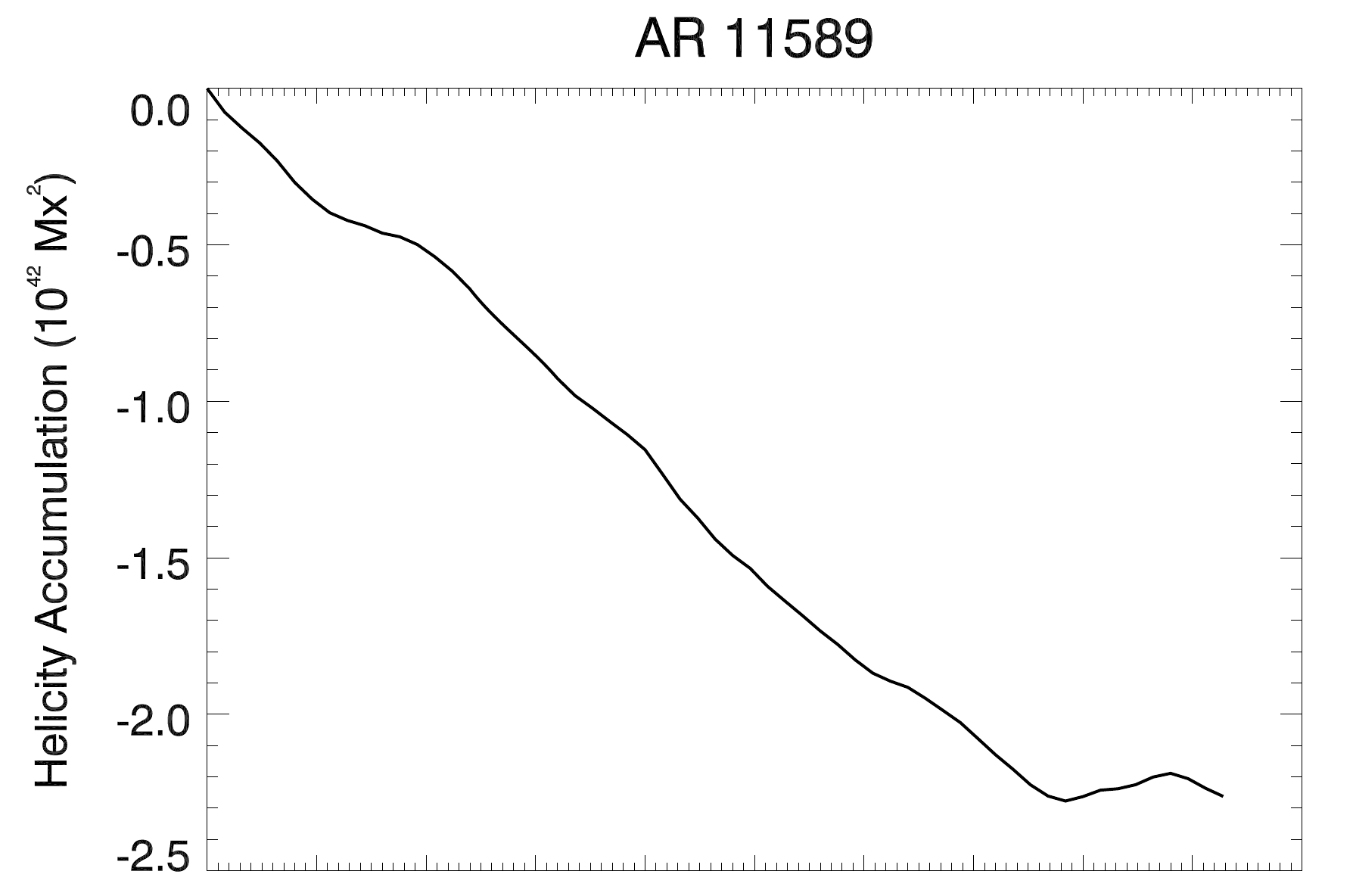}\includegraphics[scale=0.125,trim={90 285 10 215},clip,width=5.15cm]{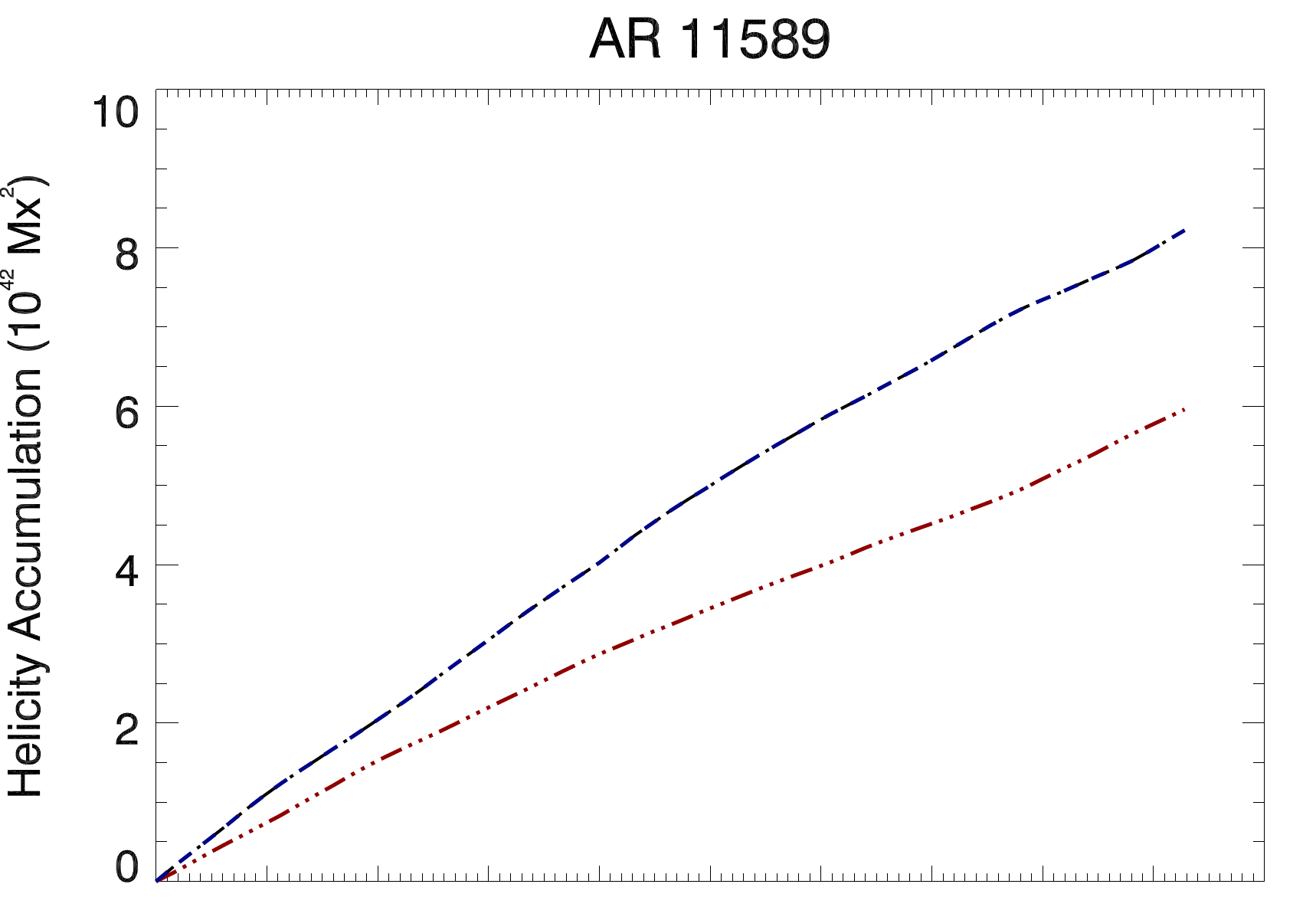}}
	\centerline{\includegraphics[scale=0.125,trim={90 285 10 215},clip,width=5.15cm]{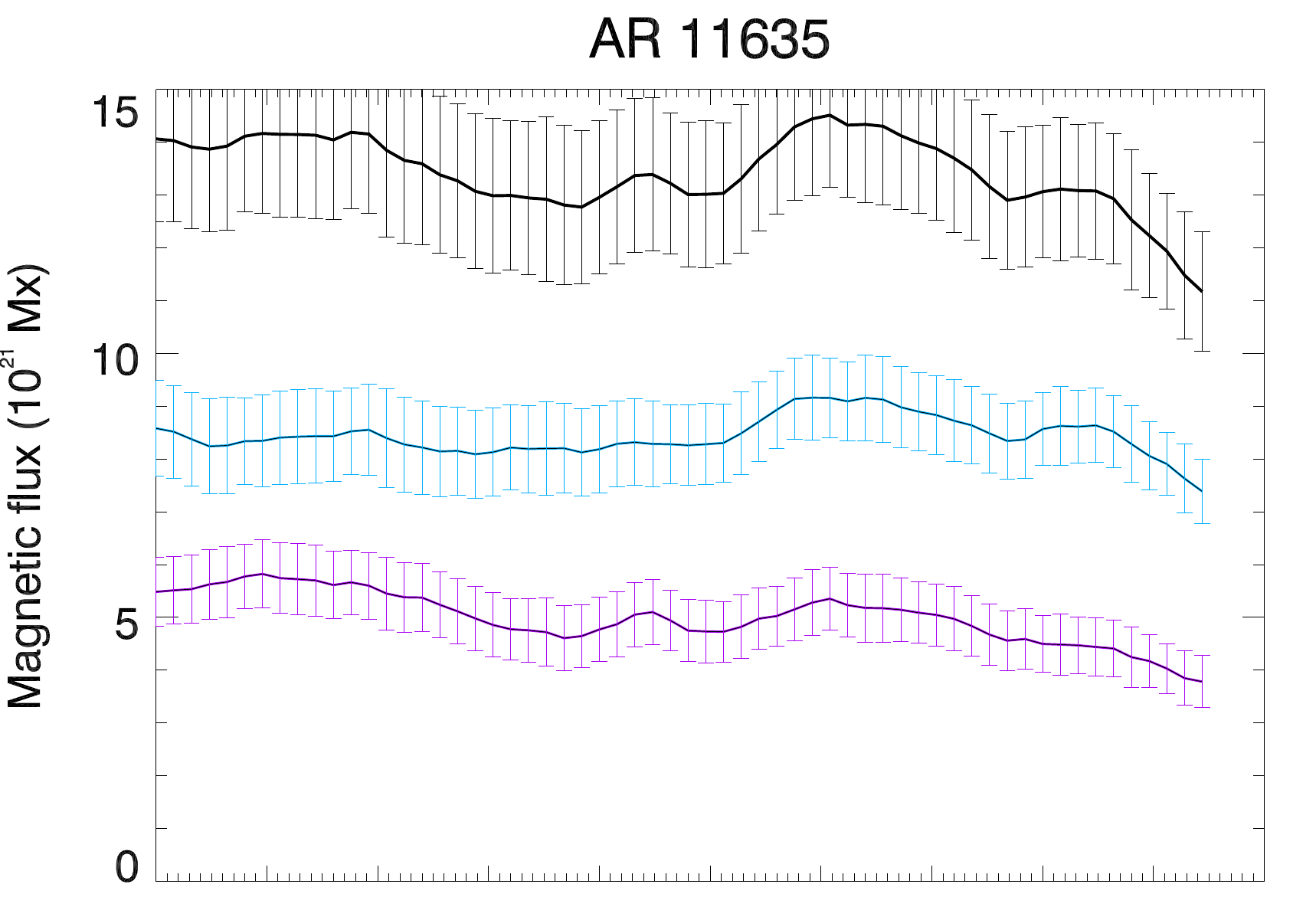}\includegraphics[scale=0.125,trim={90 285 10 215},clip,width=5.15cm]{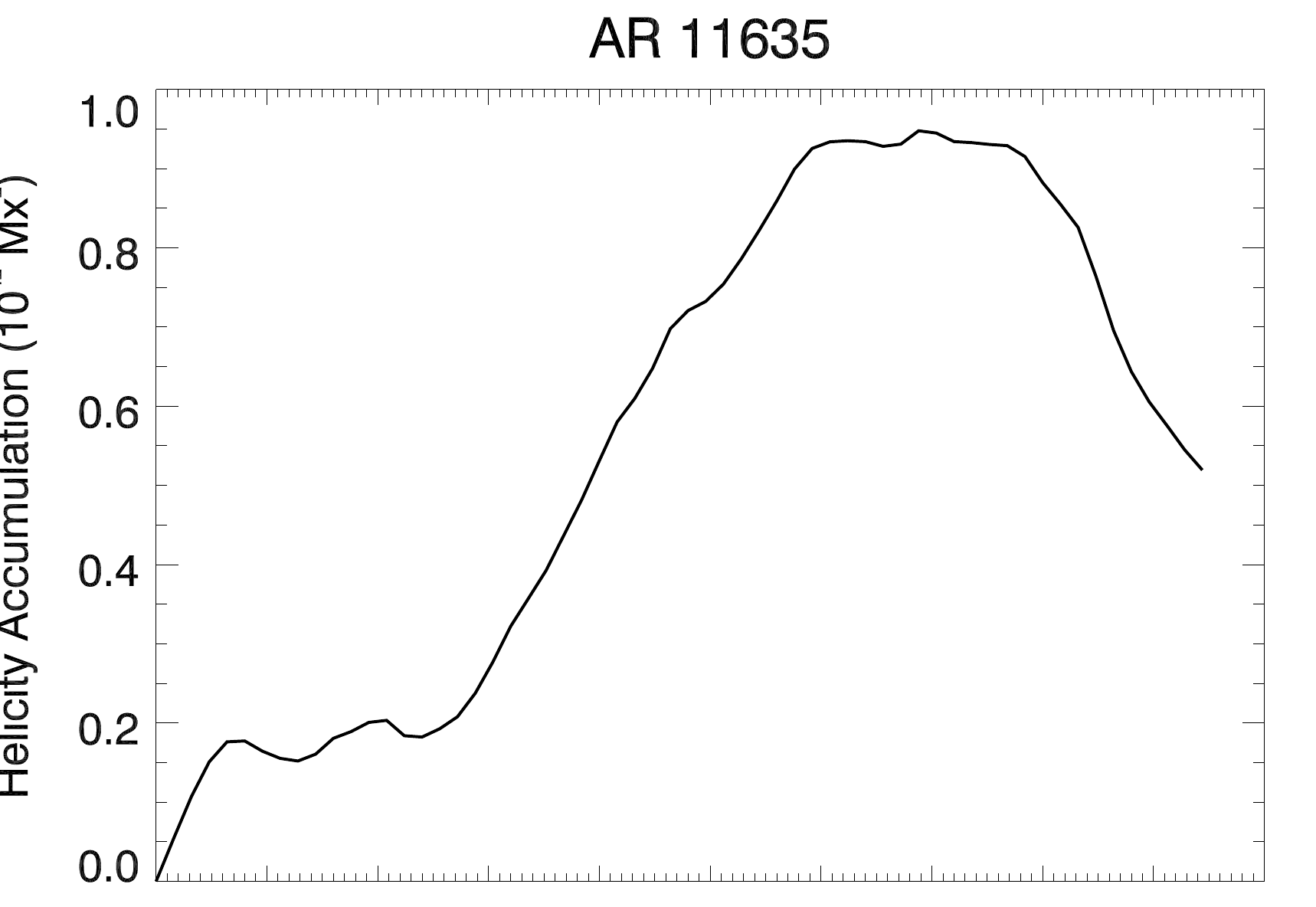}\includegraphics[scale=0.125,trim={90 285 10 215},clip,width=5.15cm]{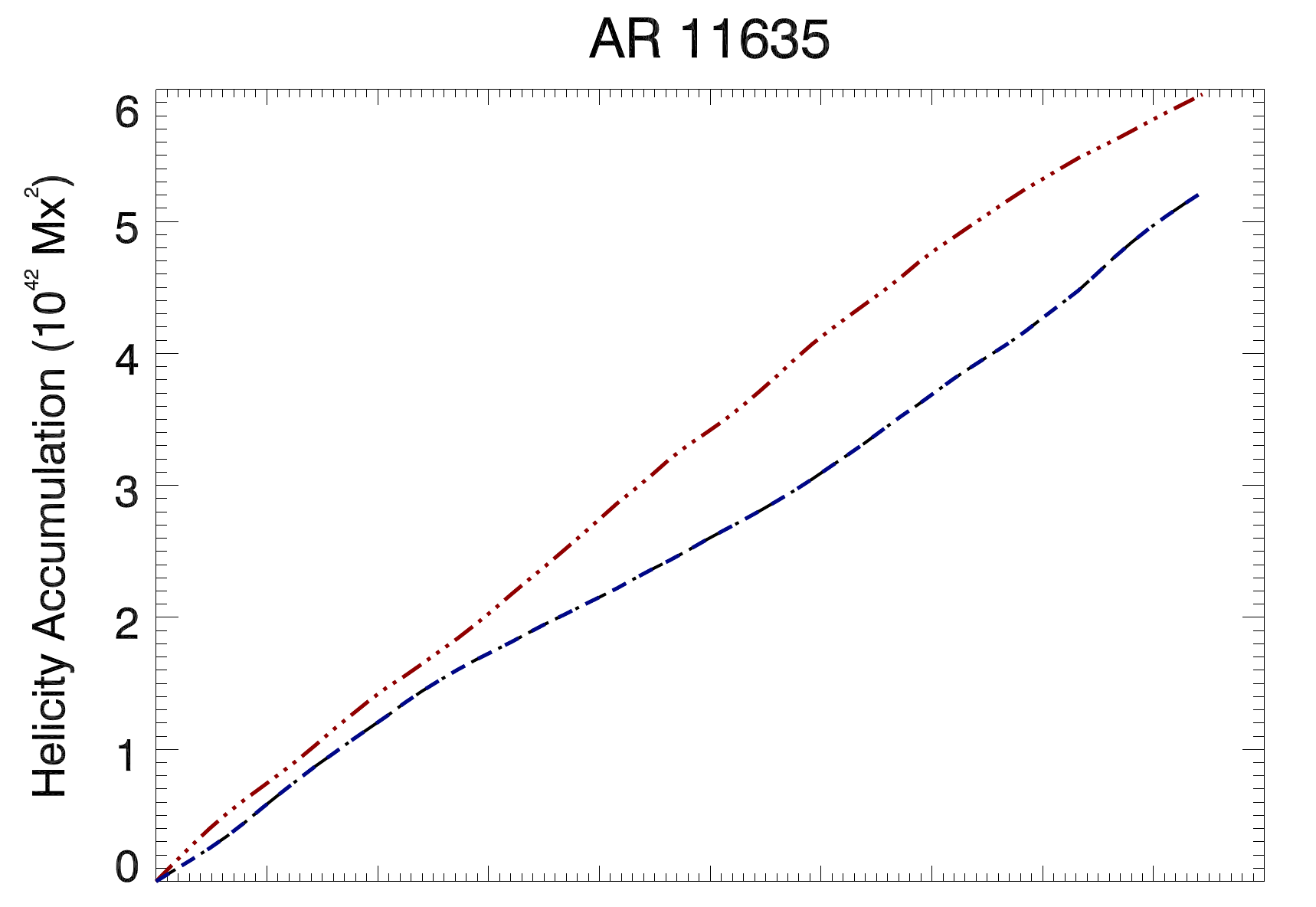}}
	\centerline{\includegraphics[scale=0.125,trim={90 220 10 215},clip,width=5.15cm]{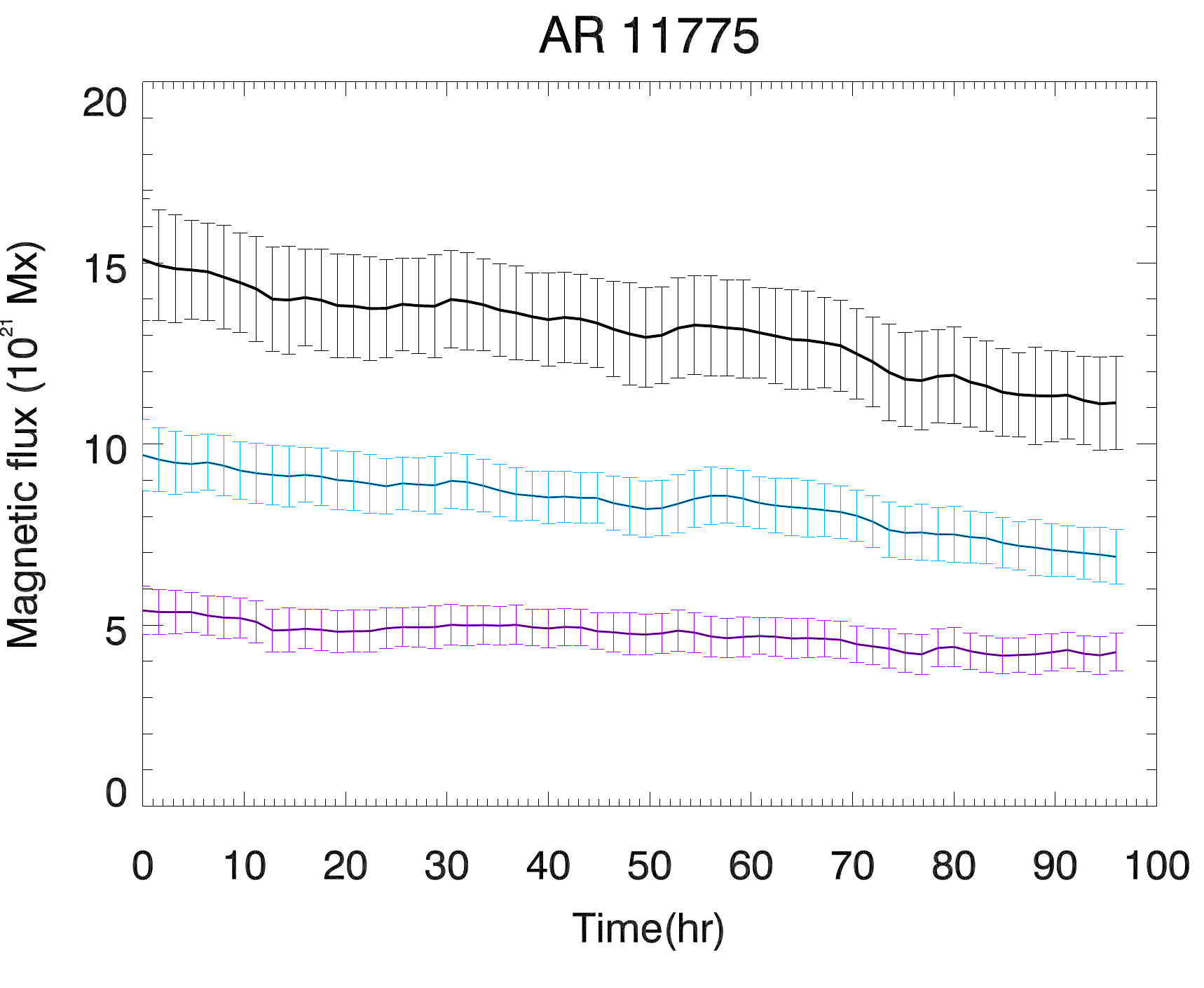}\includegraphics[scale=0.125,trim={80 220 10 215},clip,width=5.15cm]{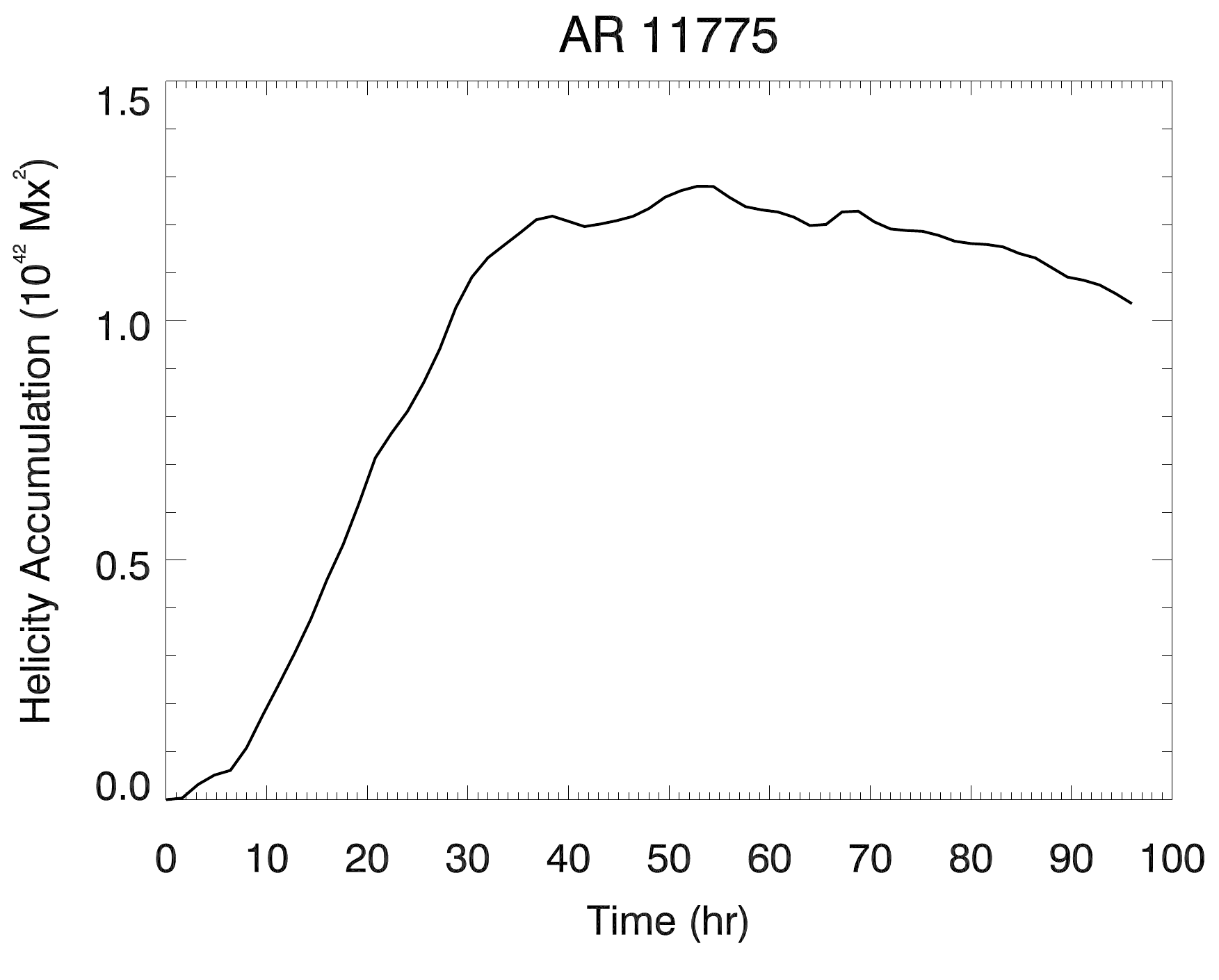}\includegraphics[scale=0.125,trim={90 220 10 215},clip,width=5.15cm]{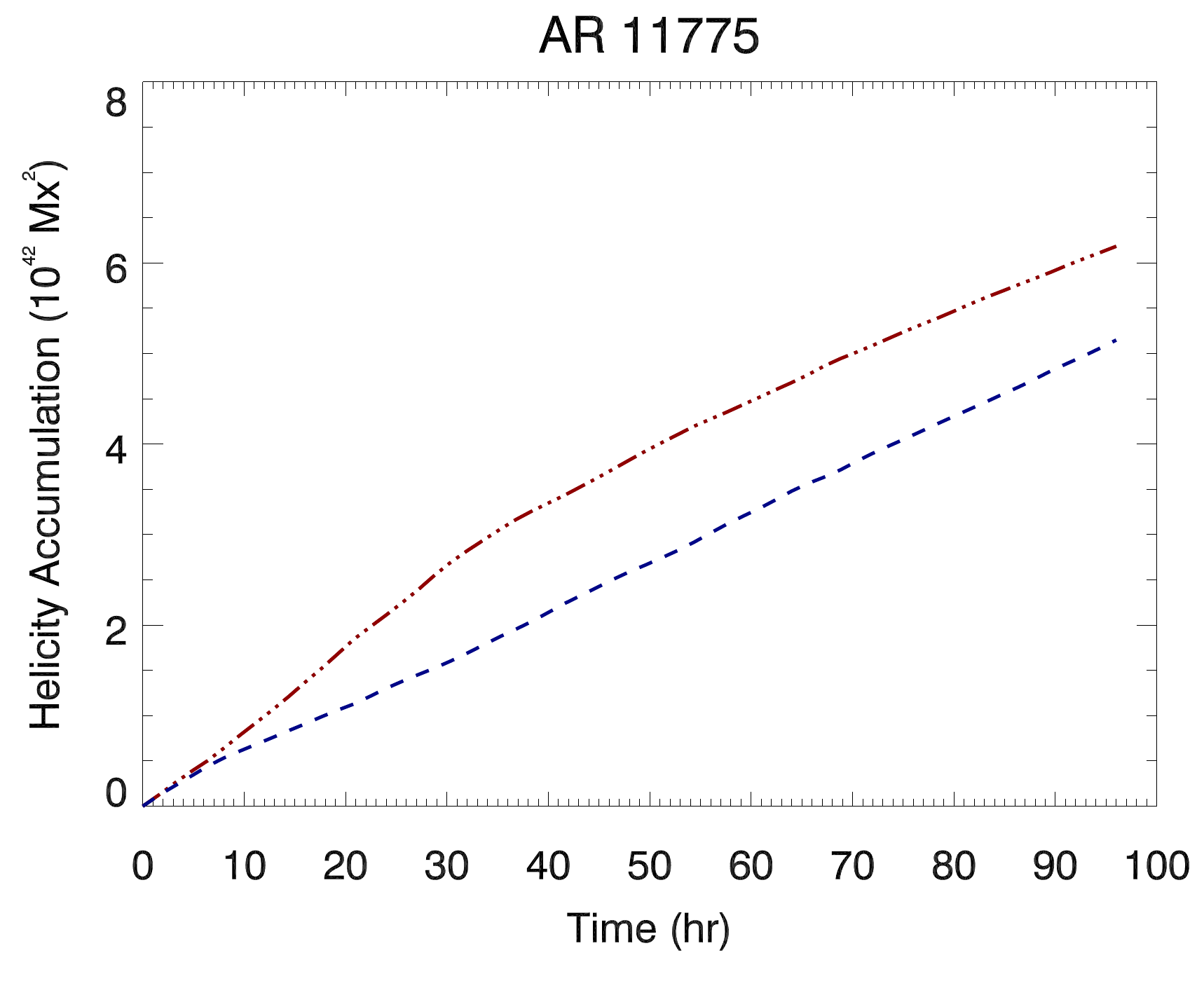}}
	\caption{\footnotesize {Same parameters as in Fig.~\ref{fig1} for flare-quiet ARs 11267, 11512, 11589, 11635, 11775.}}
	\label{fig1_bis}
\end{figure*}
\clearpage}

\newpage

\subsection{Fractal and multi-fractal properties}

Figures~{\ref{ff1} to~\ref{ff2b}} show the time series of the fractal and multifractal measurements on  analyzed data. The error associated with the measured values is equal to the 2-sigma uncertainty for the parameters (see  \citet{ermolli2014} for more details). For the sake of clarity, this uncertainty is shown 
only for the results derived from unsigned flux data of the analyzed ARs. Vertical lines in each plot indicate the time of occurrence of  M- (red), and X- (green) class flares; when the flare was associated with a CME, the thickness of the vertical line is enhanced.

Figures~\ref{ff1} to~\ref{ff2b} show small differences among the values of the fractal and multifractal parameters measured on the various ARs, as well as among the evolution of the measured values.

This is confirmed by the results summarized in Table \ref{tblff}, which lists the values of the parameters measured on each AR when considering unsigned flux data. Table \ref{tblff} also reports the flaring level of the analyzed ARs. We estimated this quantity by using the flare index (FI,  \citealp[see][]{Li2004}), which accounts for the flare history of the region during its disk transit as depicted from the NOAA’s GOES X-ray archive. Table~\ref{tblff} also lists the FI value of the most intense event hosted by each region (hereafter referred to as Max FI), and the average and standard deviation of the values derived from the two classes of studied ARs for each measured parameter.

 In Table \ref{tblff}, the values confirm previous published results on a significant fractality of the morphology of the magnetic flux concentration in both flare-productive and flare-quiet ARs. Indeed, the fractal parameter $D_0$ derived from all the ARs ranges between $\approx 1.64$ and $\approx 1.90$.
The values in Table~\ref{tblff} show that the fractal and multifractal parameters of the flare-productive and flare-quiet regions can overlap, as already reported by \cite{giorgi2014}. It is worth nothing that, although a thresholded warning method based on measurements of fractal/multifractal parameters may result in some misclassification of the flare and flare-quiet classes, using these measures imply an automatic and robust analysis of observations that is suitable for a quick initial taxonomy of new solar regions. 

Figures \ref{ff1} to~\ref{ff2b}
also display the evolution of the various analysed parameters  when taking into account the  trailing and leading flux  data of the analysed AR. The various trends confirm previous findings that there is a systematic larger variance of the values derived from  trailing flux data in the flaring ARs than obtained from both unsigned and leading flux data. 
From analyzing the series of the measured parameters, we notice that several M- and X-class flares occur during a decreasing phase of the $D_{\mathrm{div}}$ and an increasing phase of the $D_8$ values estimated by considering unsigned and signed flux data of the leading polarity of the AR hemisphere. However, these features of the parameter trends seem not to represent a consistent pre-flare signature in the whole  sample of analyzed flaring ARs and events. 

Finally, we notice a clear resemblance between the trends of the fractal and multifractal parameters shown in Figures~\ref{ff1} to~\ref{ff2b}  and those of the  magnetic flux displayed in Figures~\ref{fig1} to~\ref{fig1_bis}.

\afterpage{
\begin{figure*}   
    \centerline{\includegraphics[trim={1.cm 2.cm  1.2cm 0.2cm},clip,width=6.cm]{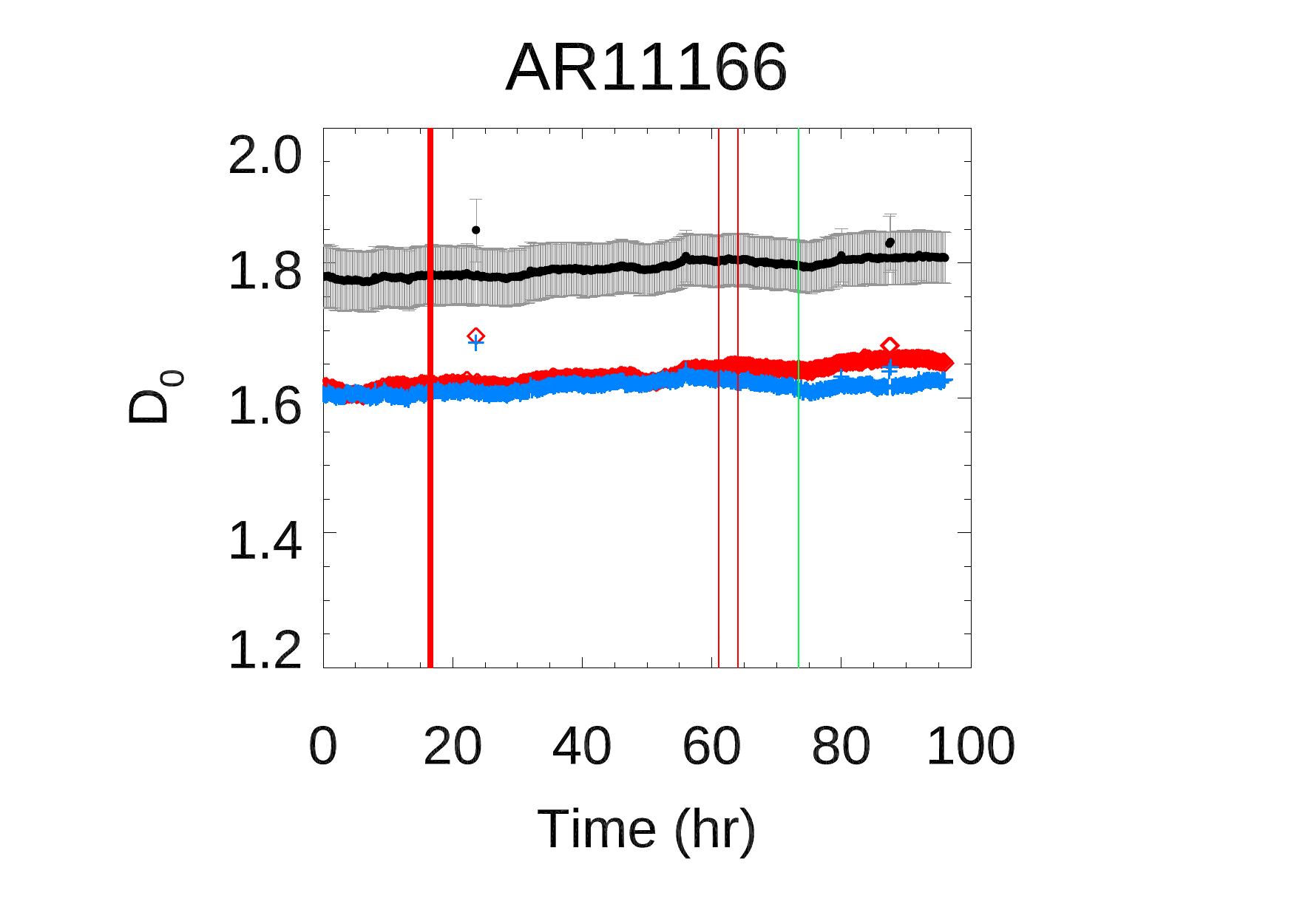}\includegraphics[trim={1.cm 2.cm  1.2cm 0.2cm},clip,width=6.cm]{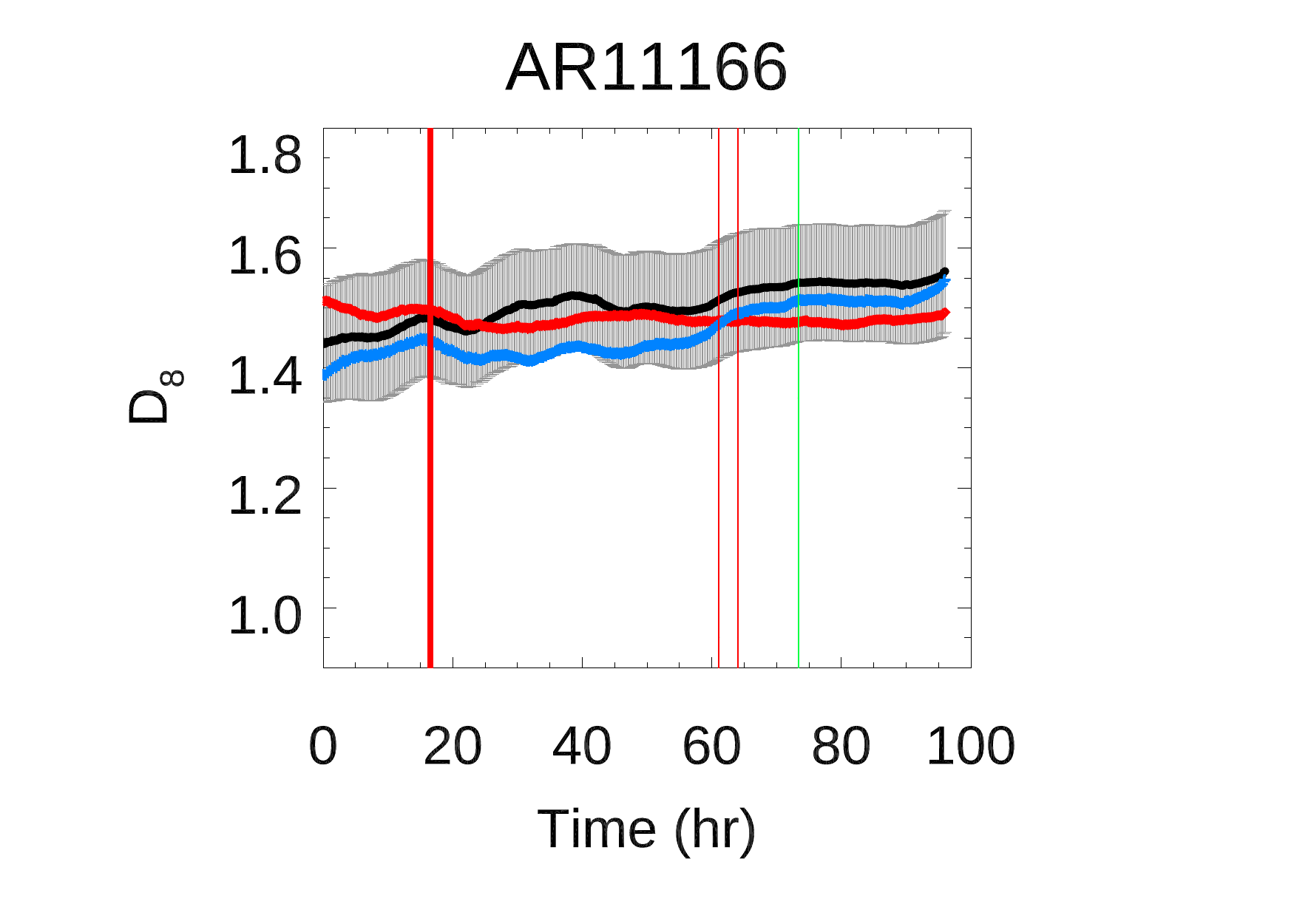}}
    \centerline{\includegraphics[trim={1.cm 2.cm  1.2cm  0.2cm},clip,width=6.cm]{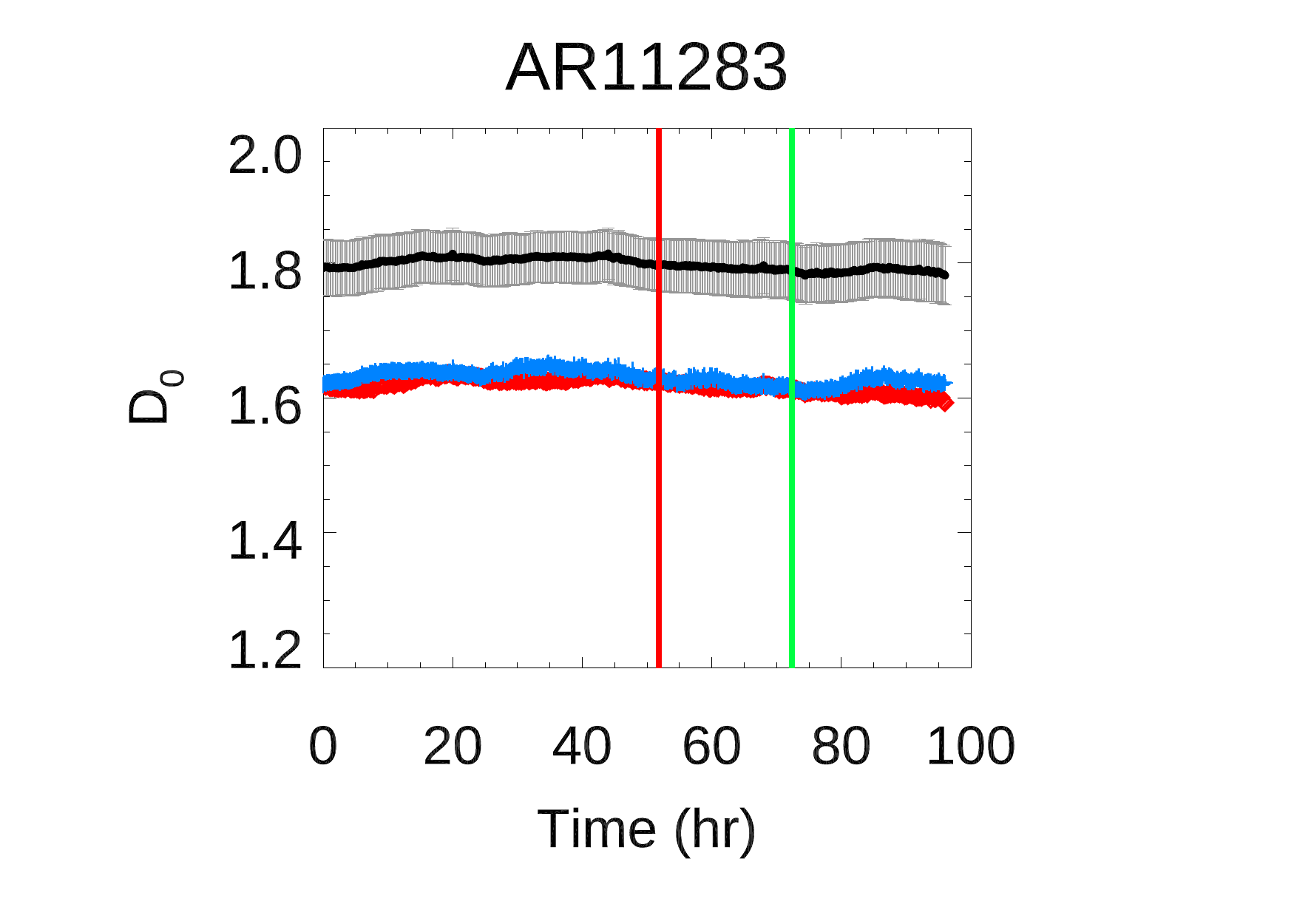}\includegraphics[trim={1.cm 2.cm  1.2cm  0.2cm},clip,width=6.cm]{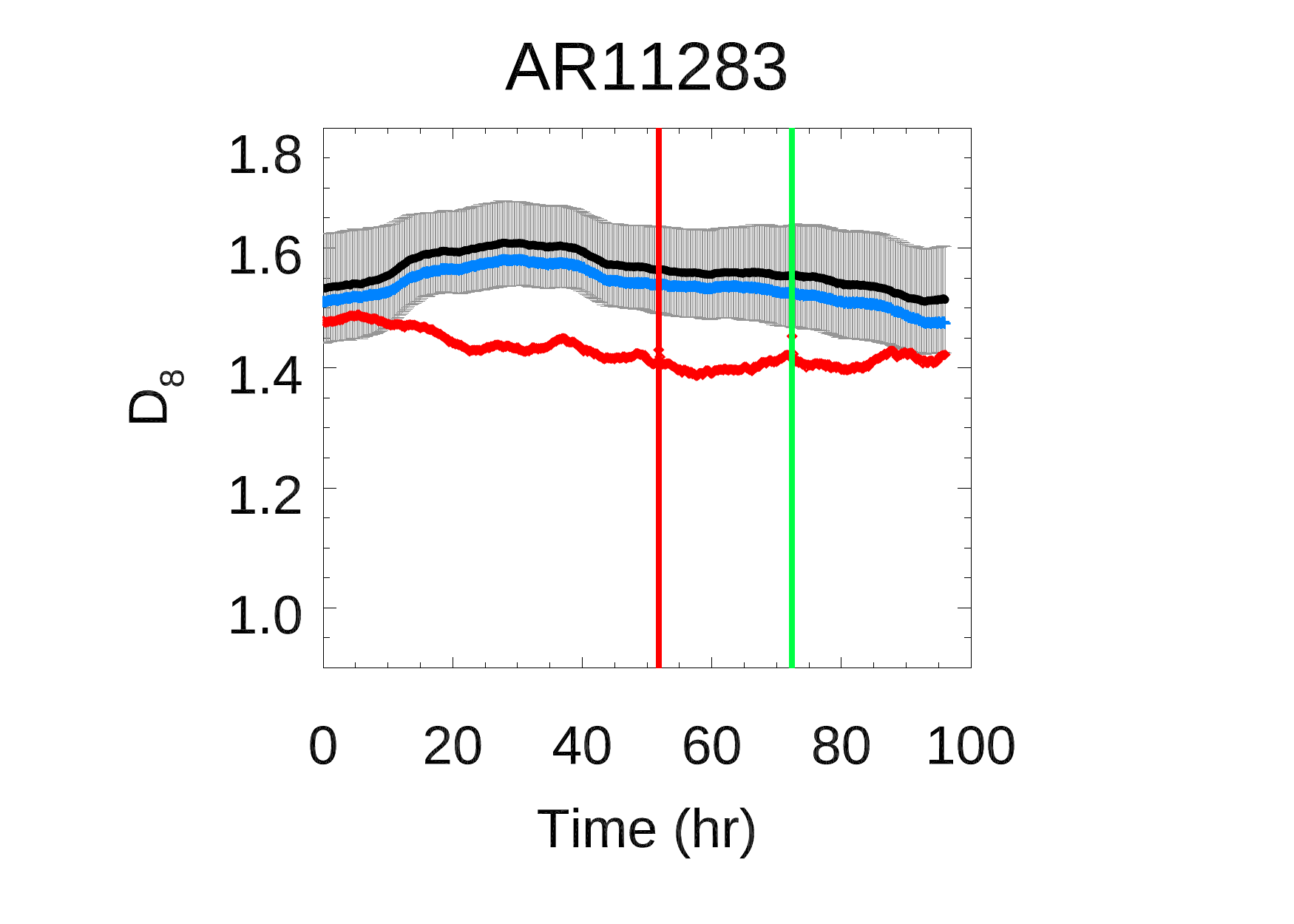}}
    \centerline{\includegraphics[trim={1.cm 2.cm  1.2cm  0.2cm},clip,width=6.cm]{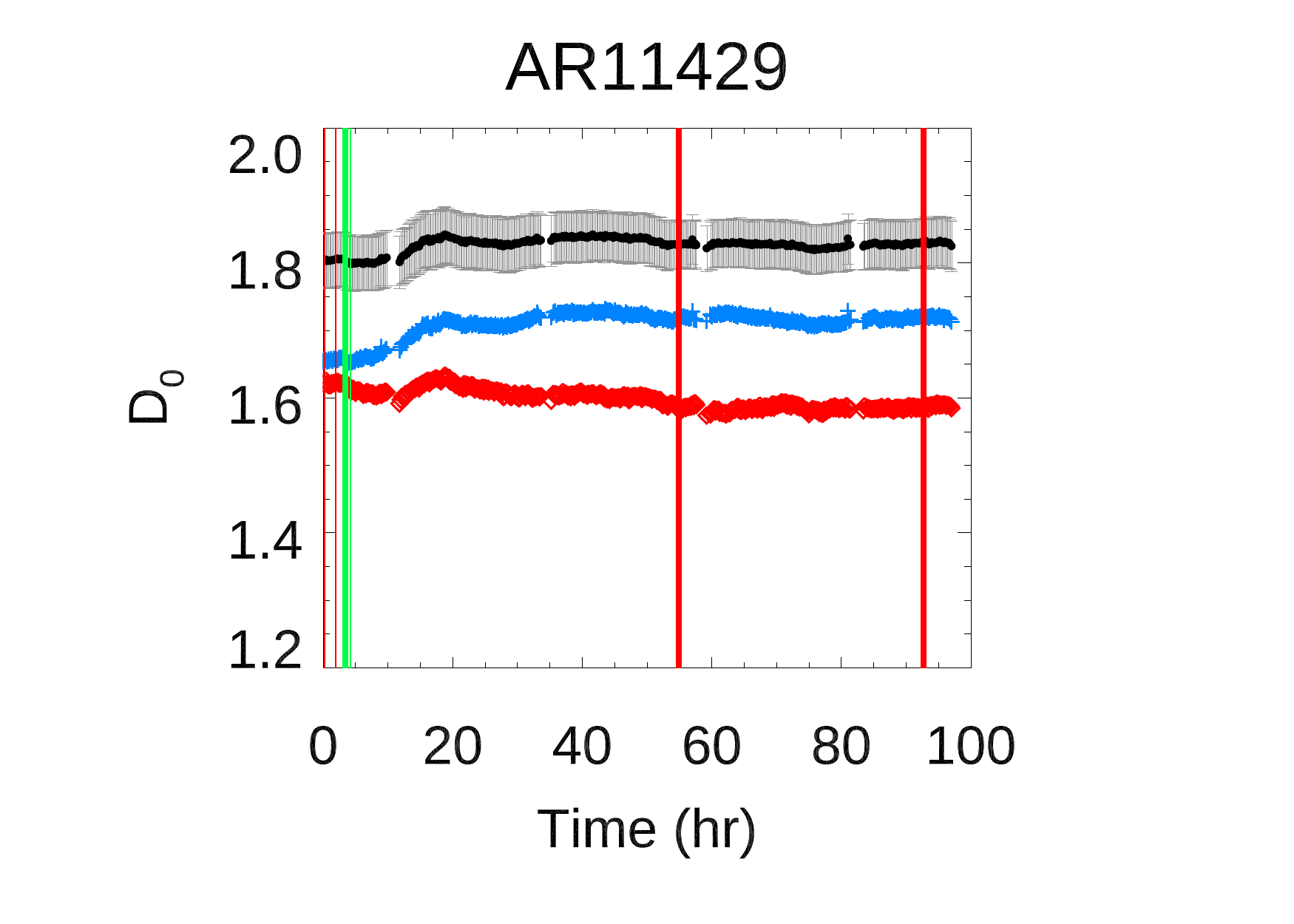}\includegraphics[trim={1.cm 2.cm  1.2cm  0.2cm},clip,width=6.cm]{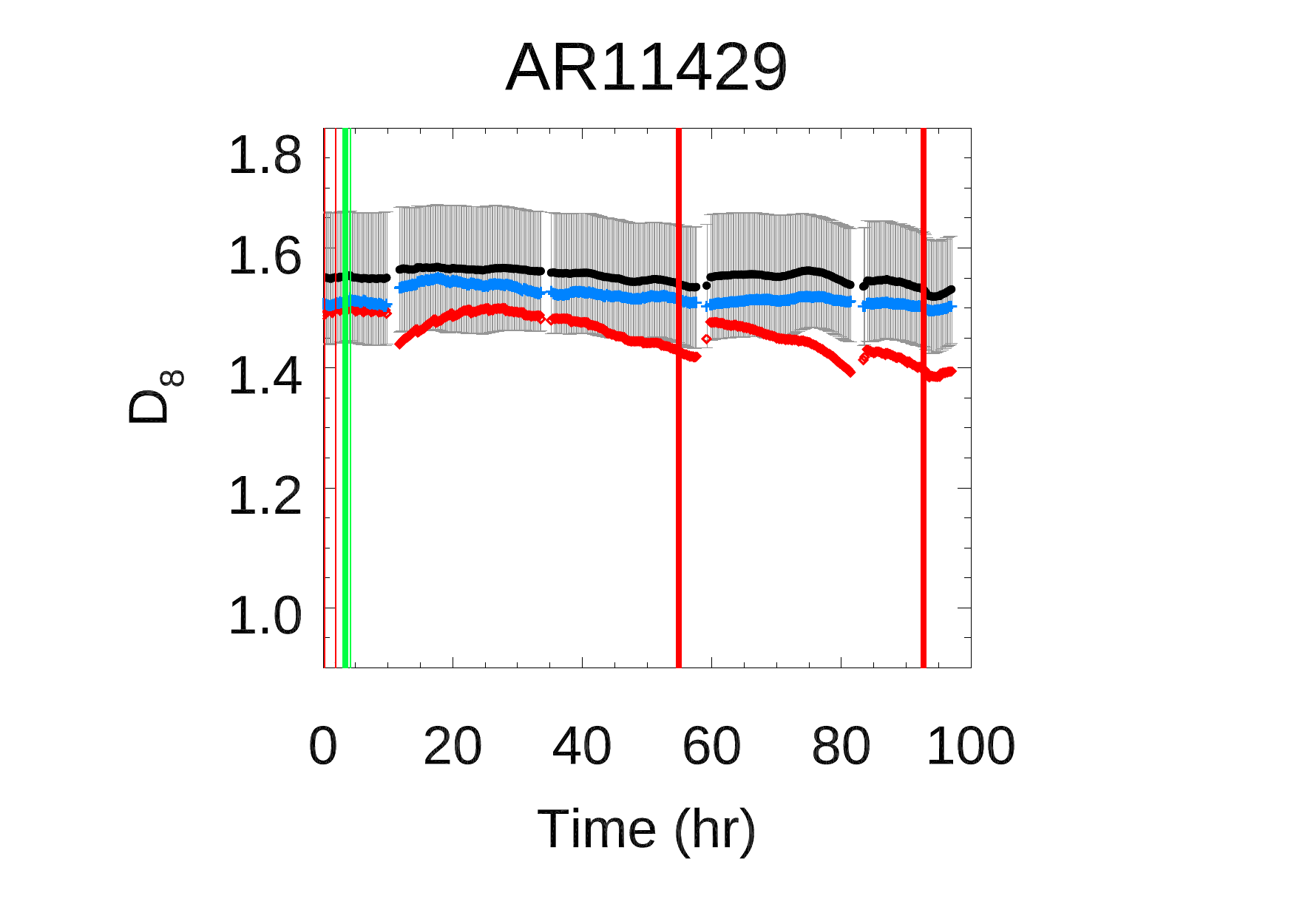}}
    \centerline{\includegraphics[trim={1.cm 2.cm  1.2cm  0.2cm},clip,width=6.cm]{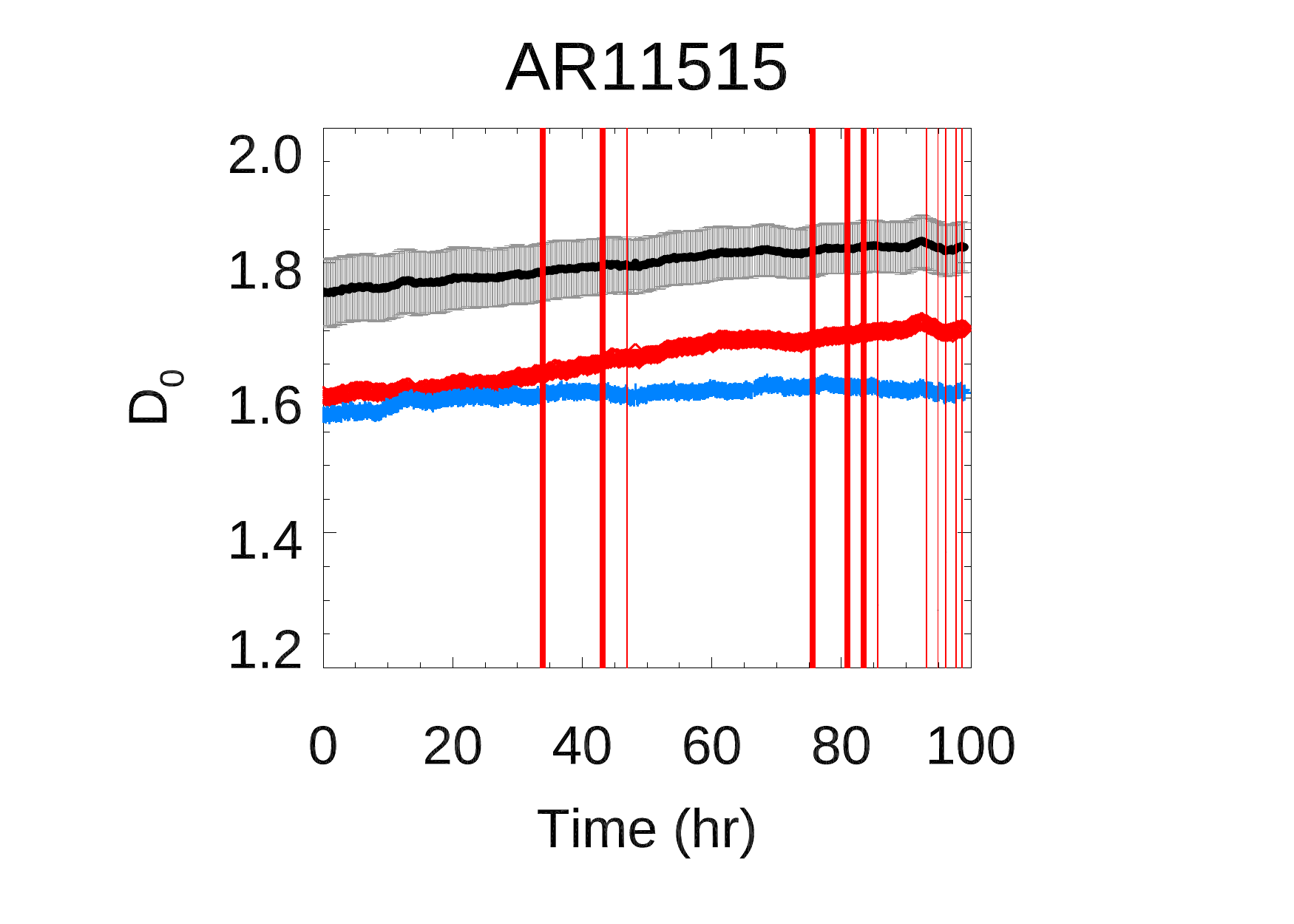}\includegraphics[trim={1.cm 2.cm  1.2cm  0.2cm},clip,width=6.cm]{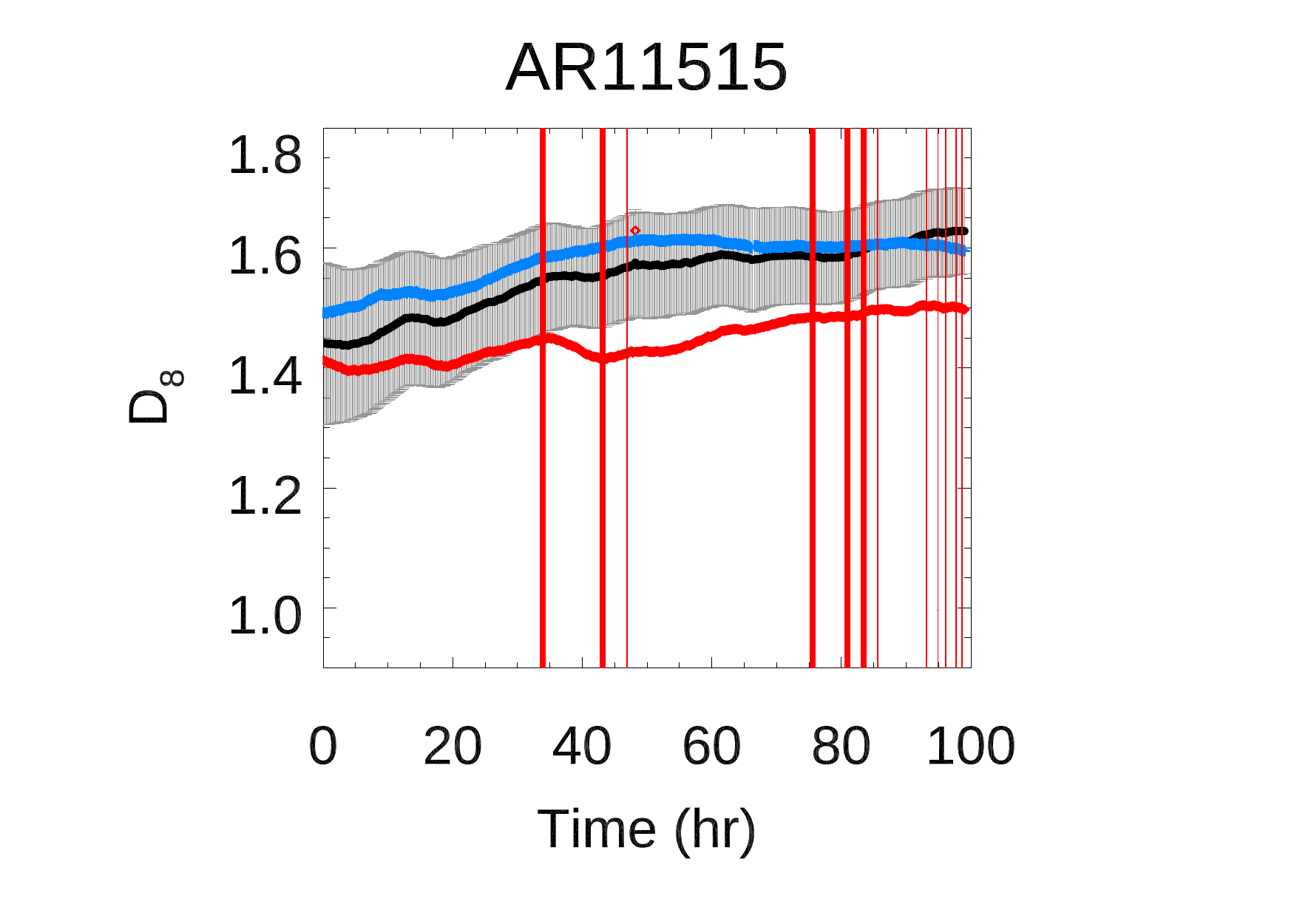}}
    \centerline{\includegraphics[trim={1.cm 0.5cm  1.2cm  0.2cm},clip,width=6.cm]{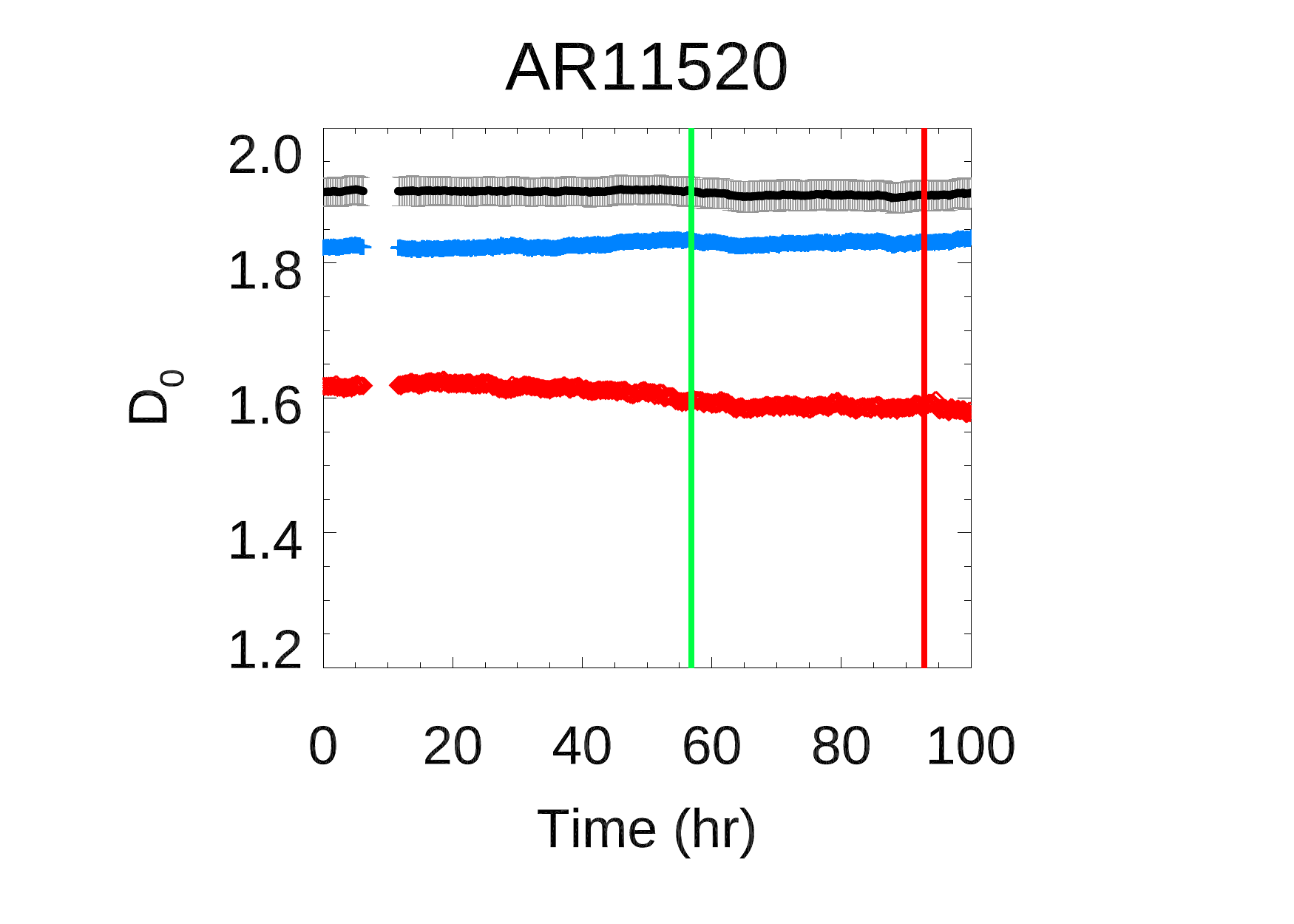}\includegraphics[trim={1.cm 0.5cm  1.2cm  0.2cm},clip,width=6.cm]{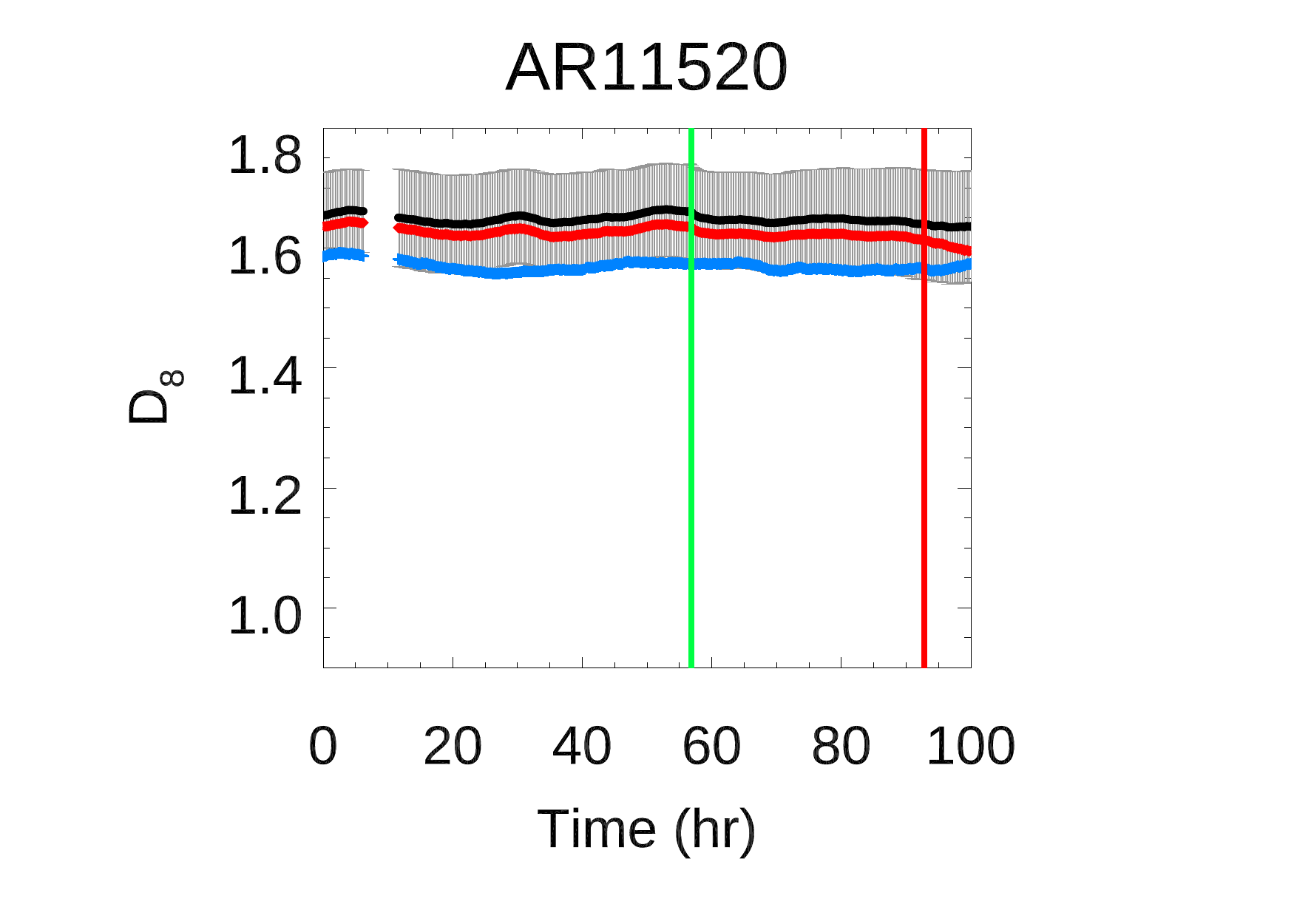}} 
              \caption{\linespread{0.7} \footnotesize {Time series of the fractal parameters $D_0$ (left-hand column) and $D_8$ (right-hand column)  measured on the five selected productive ARs, by considering both unsigned (black symbols) and signed (positive and negative, red and blue symbols, respectively) flux data in the analyzed regions. From top to bottom results for ARs 11166, 11283, 11429, 11515, 11520. Vertical bars indicate the flare activity of the AR as specified in the caption of Fig.~\ref{fig1}. Error bars show the uncertainty associated with the measured values, details are given in the text. For clarity, the error bars are only shown for the results from unsigned flux data. The gaps in the time series are due to the lack of SDO/HMI observations. 
              }}
   \label{ff1}
   \end{figure*}
\clearpage}

\afterpage{
\begin{figure*}
    \centerline{\includegraphics[trim={1.cm 2.cm  1.2cm 0.2cm},clip,width=6.cm]{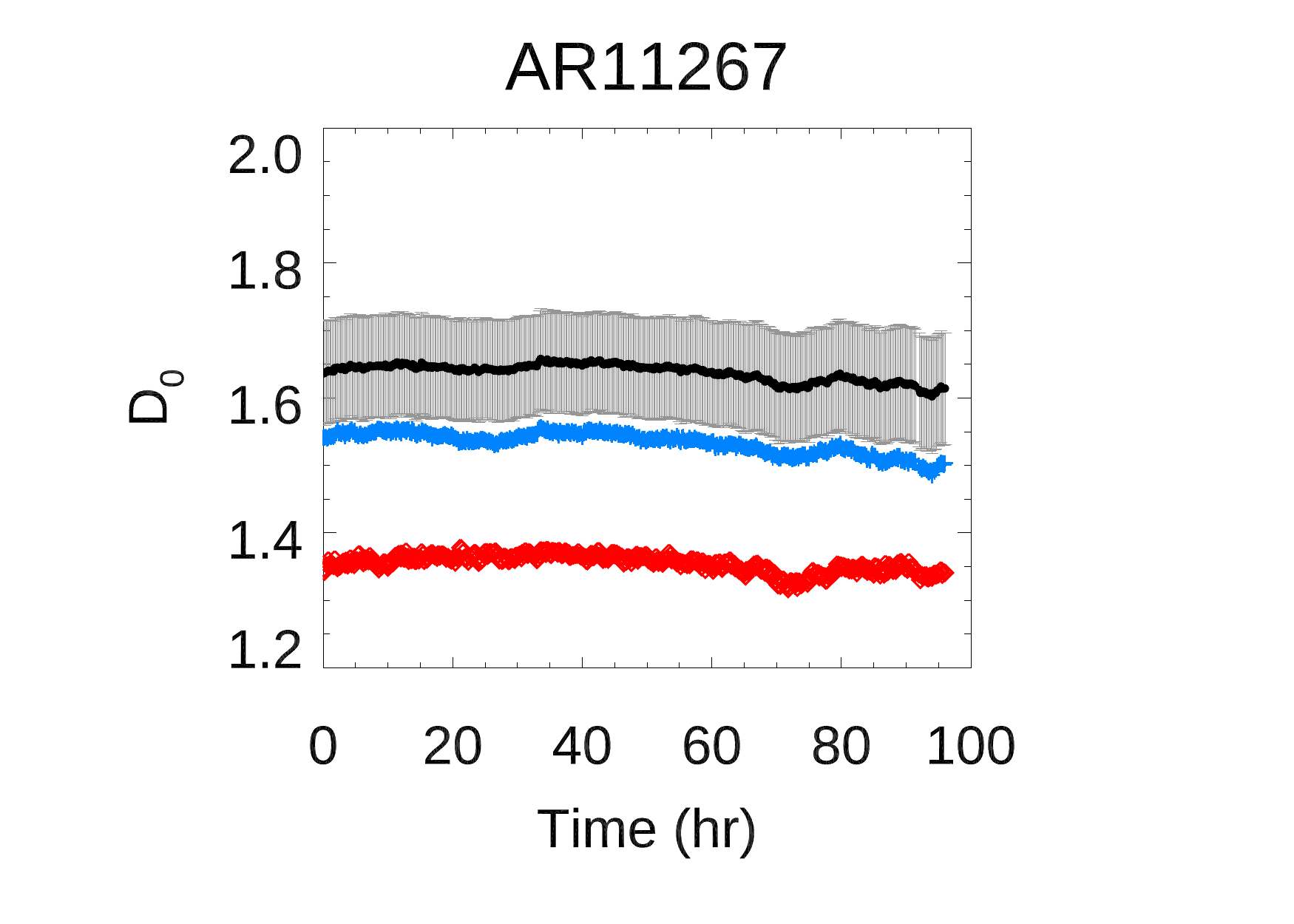}\includegraphics[trim={1.cm 2.cm  1.2cm 0.2cm},clip,width=6.cm]{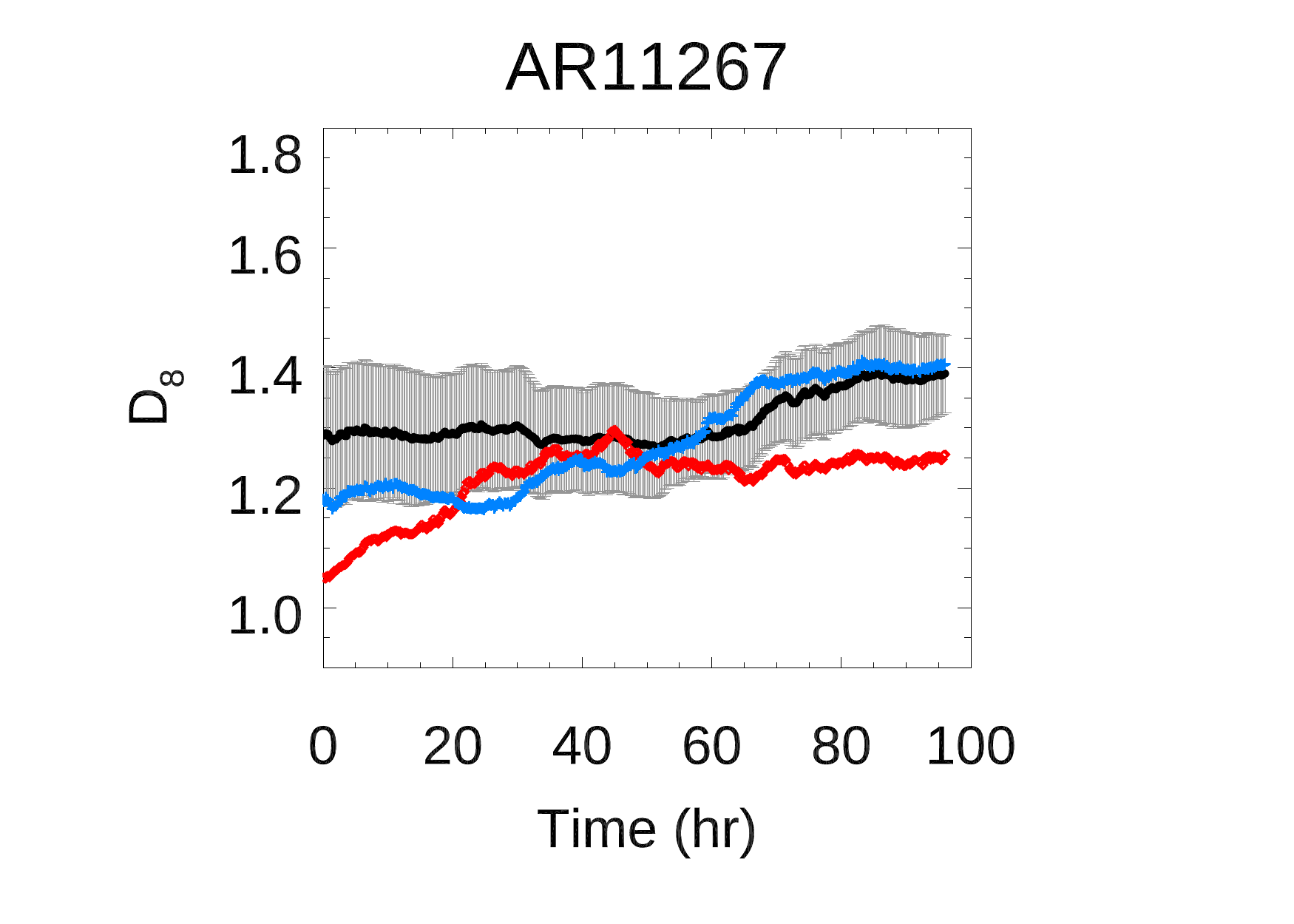}}
    \centerline{\includegraphics[trim={1.cm 2.cm  1.2cm  0.2cm},clip,width=6.cm]{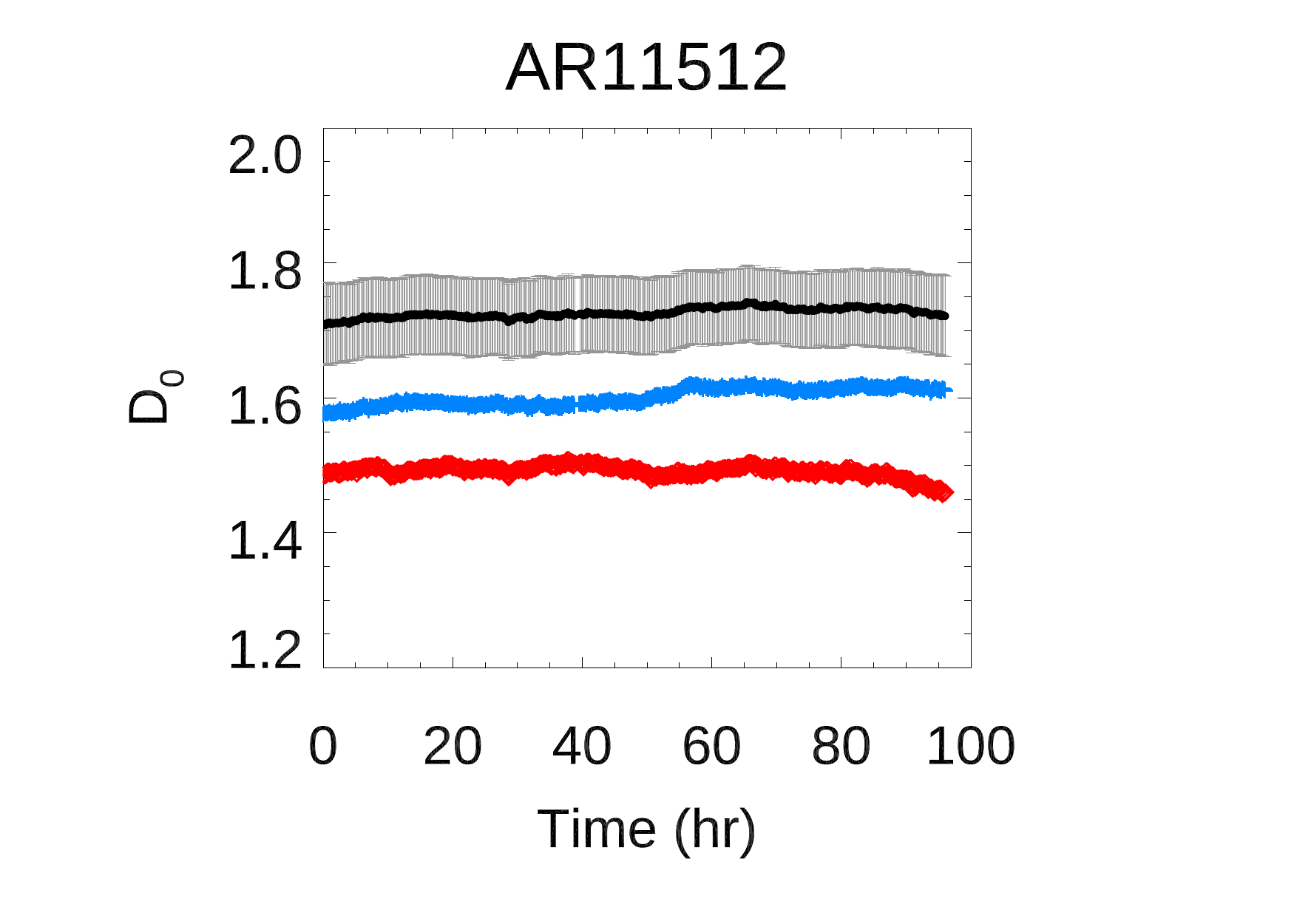}\includegraphics[trim={1.cm 2.cm  1.2cm  0.2cm},clip,width=6.cm]{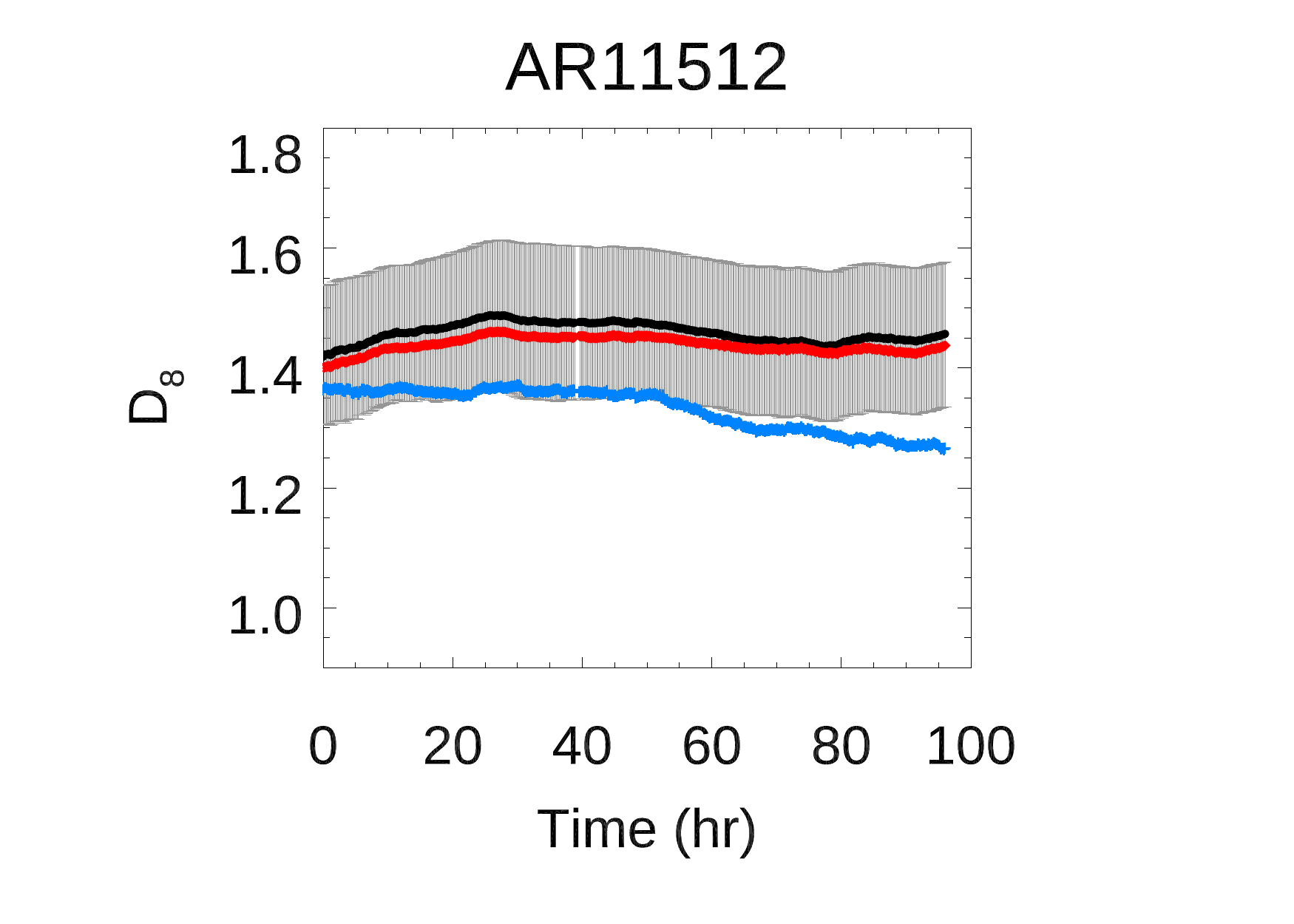}}
    \centerline{\includegraphics[trim={1.cm 2.cm  1.2cm  0.2cm},clip,width=6.cm]{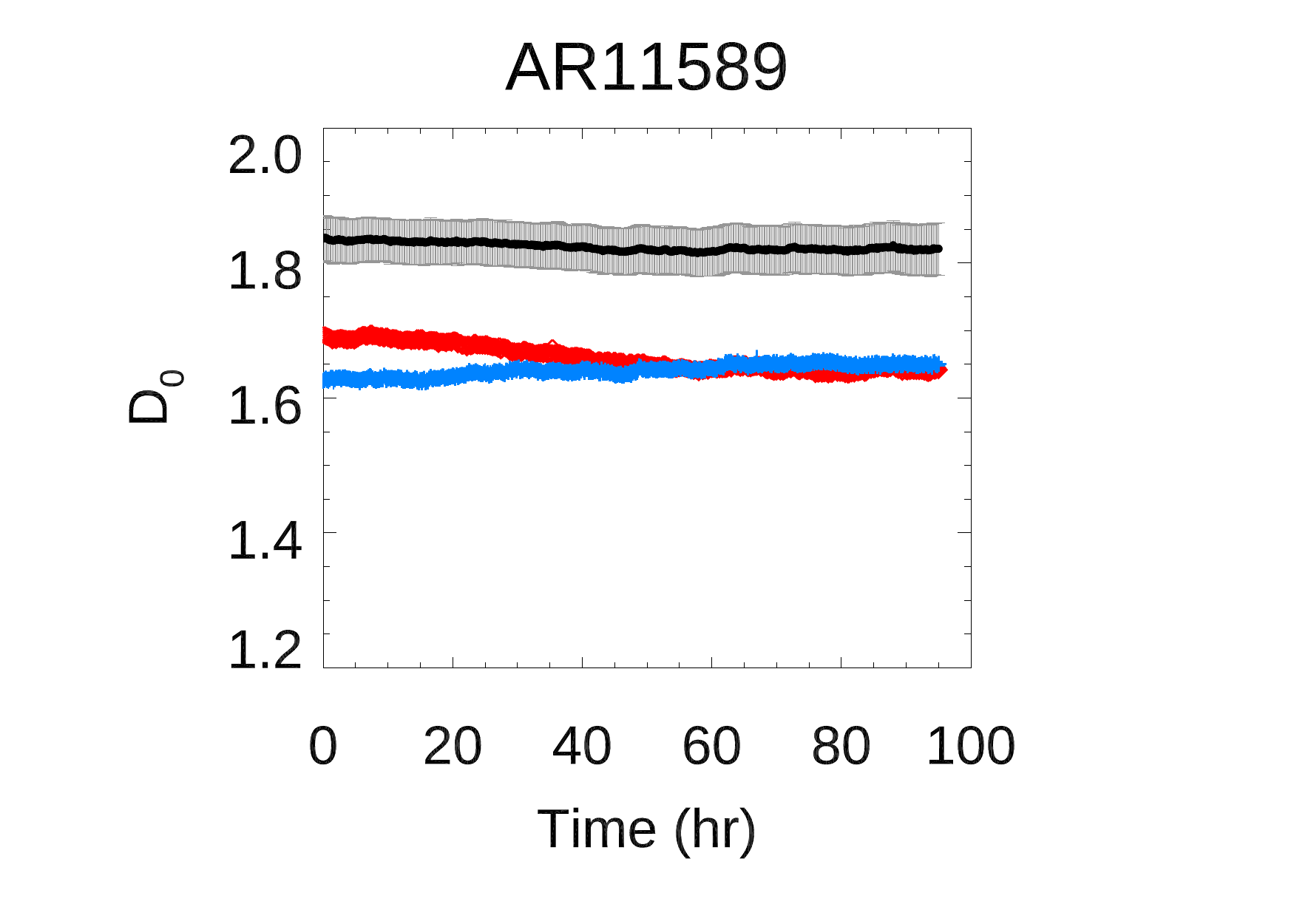}\includegraphics[trim={1.cm 2.cm  1.2cm  0.2cm},clip,width=6.cm]{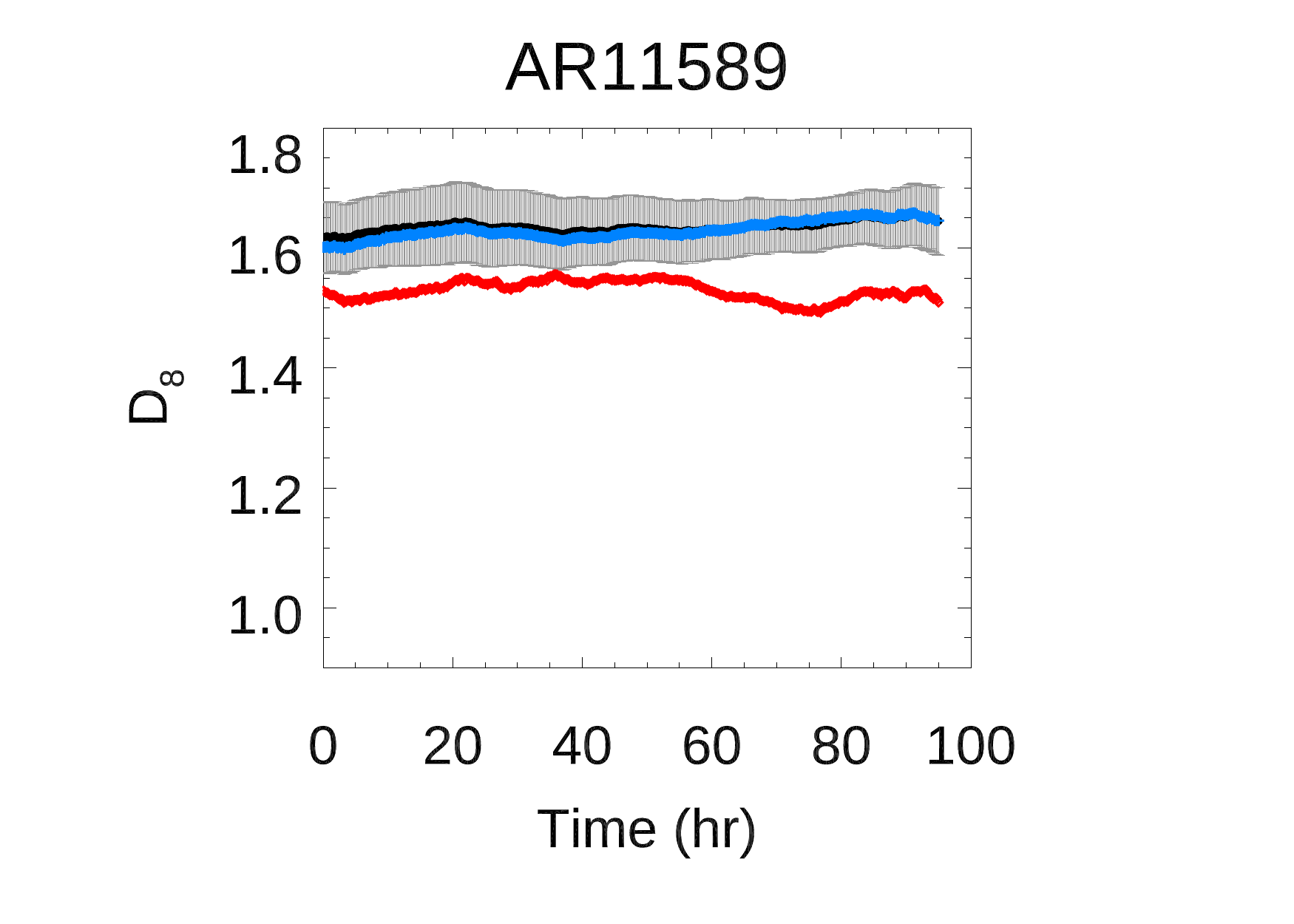}}
    \centerline{\includegraphics[trim={1.cm 2.cm  1.2cm  0.2cm},clip,width=6.cm]{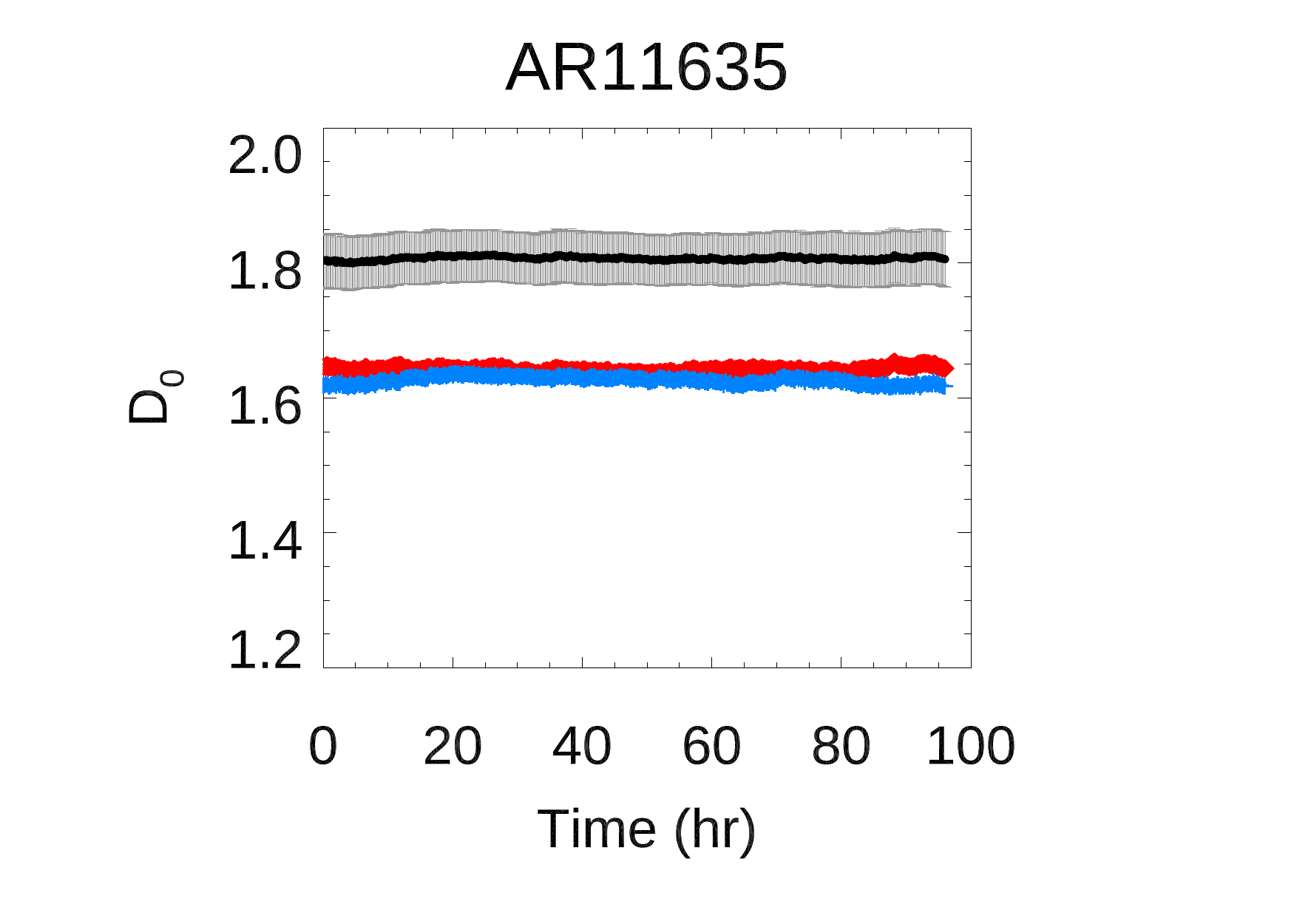}\includegraphics[trim={1.cm 2.cm  1.2cm  0.2cm},clip,width=6.cm]{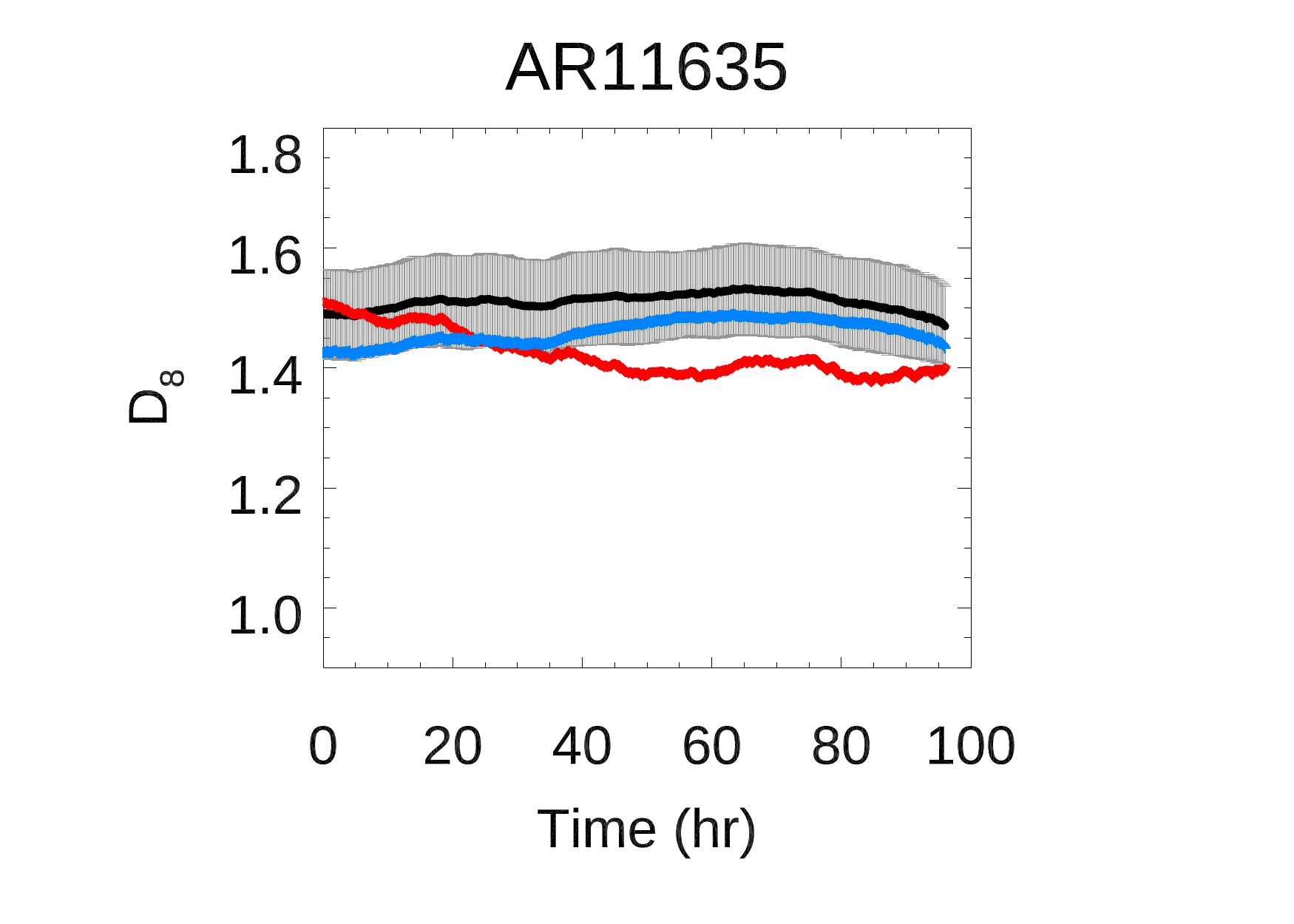}}
    \centerline{\includegraphics[trim={1.cm 0.5cm  1.2cm  0.2cm},clip,width=6.cm]{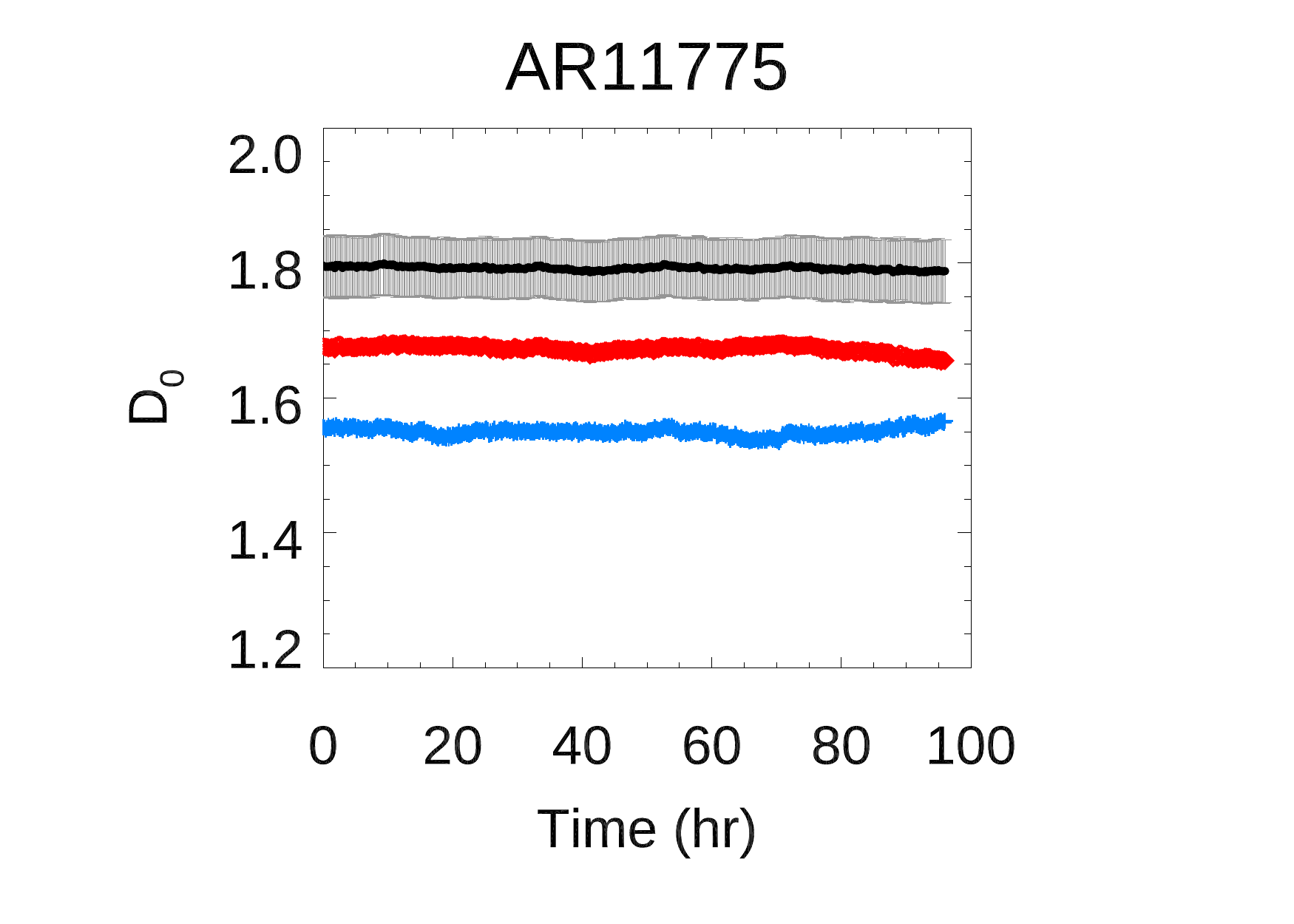}\includegraphics[trim={1.cm 0.5cm  1.2cm  0.2cm},clip,width=6.cm]{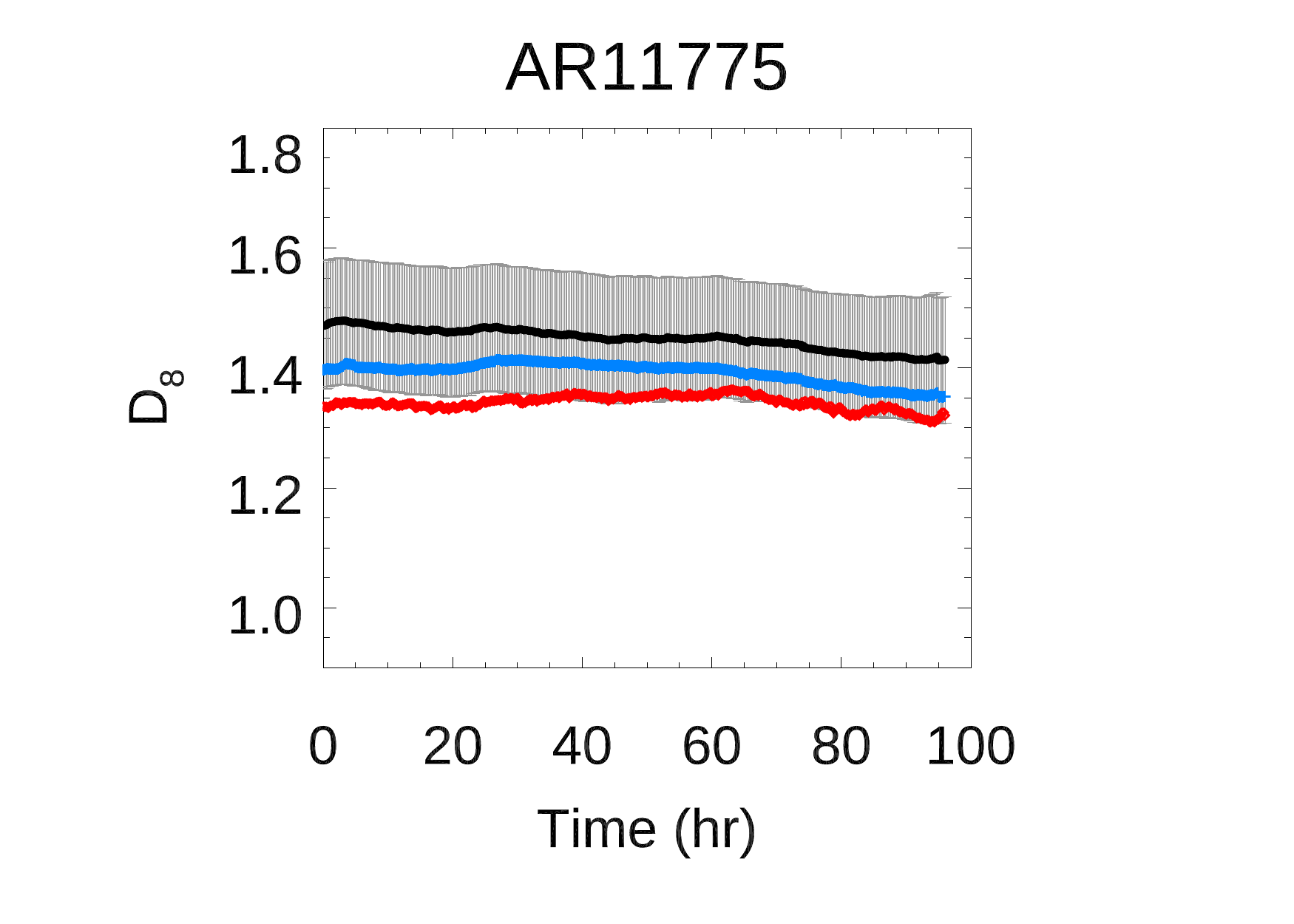}} 
              \caption{\linespread{0.7} \footnotesize {Same parameters as in Fig.~\ref{ff1} for ARs 11267, 11512, 11589, 11635, 11775 hosting only B and C flares.}
              }
   \label{ff1b}
   \end{figure*}
\clearpage
}
\afterpage{
     \begin{figure*}  
      \centerline{\includegraphics[trim={1.cm 2.cm  1.2cm  0.2cm},clip,width=6.cm]{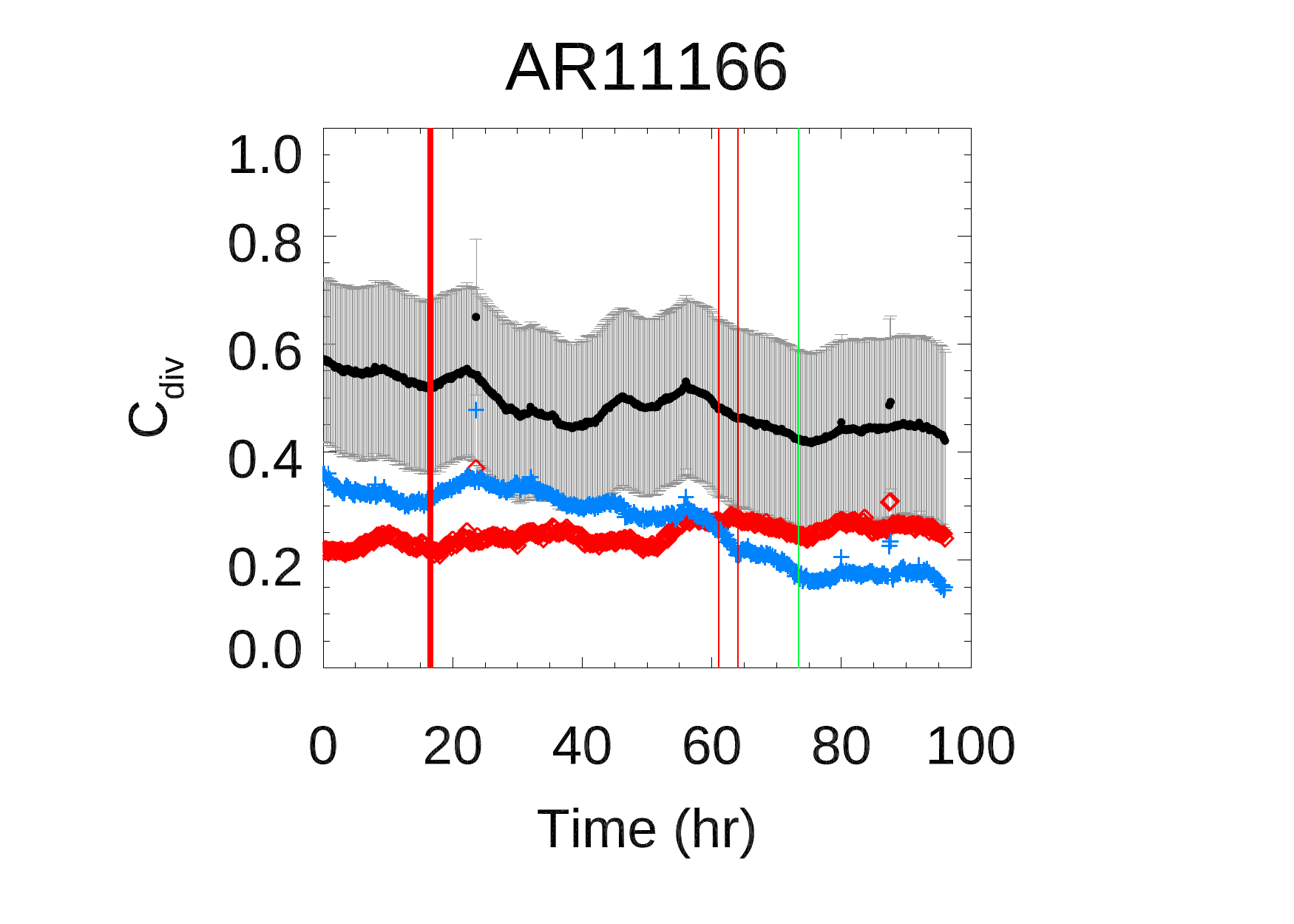}\includegraphics[trim={1.cm 2.cm  1.2cm  0.2cm},clip,width=6.cm]{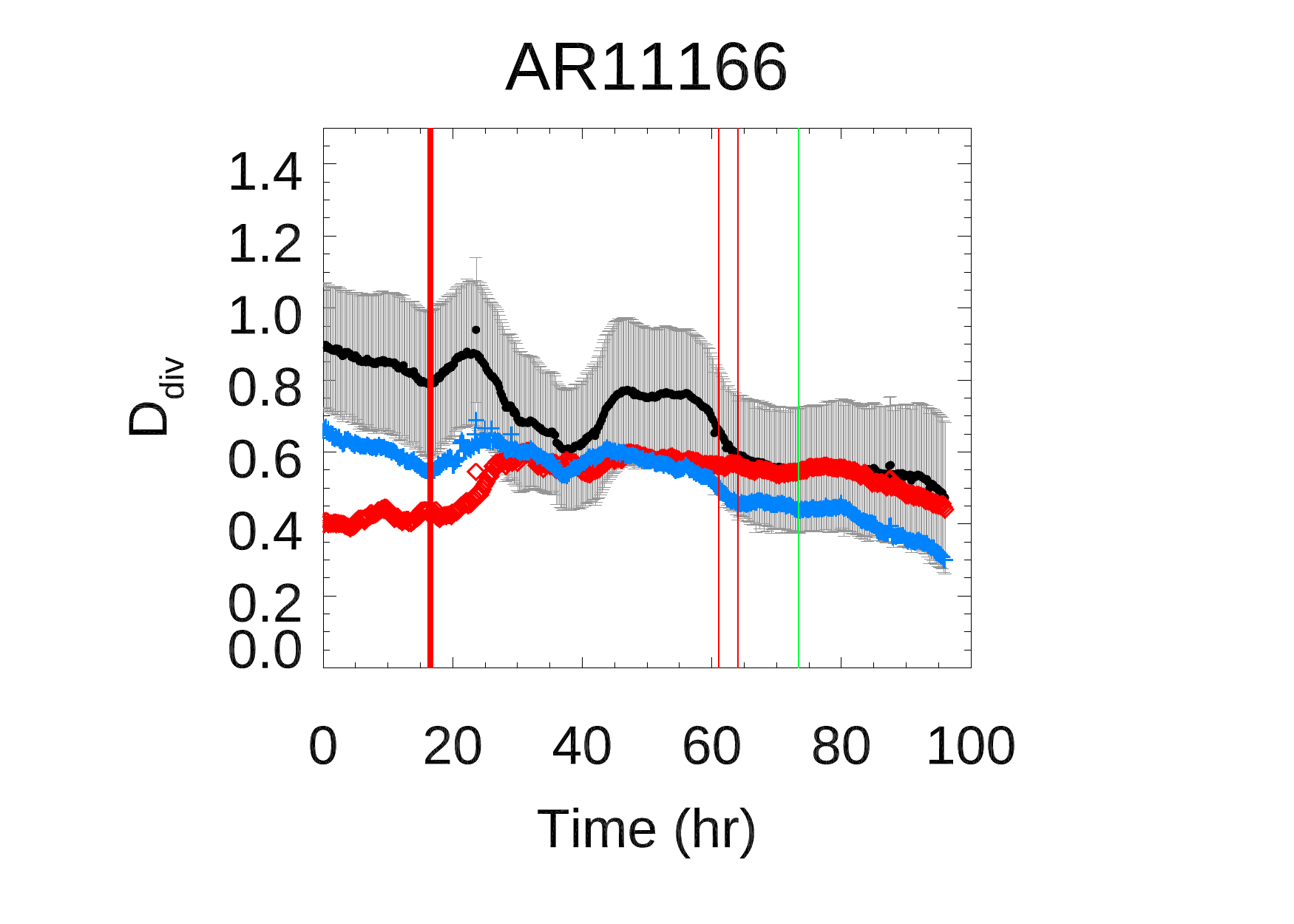}}
      \centerline{\includegraphics[trim={1.cm 2.cm  1.2cm  0.2cm},clip,width=6.cm]{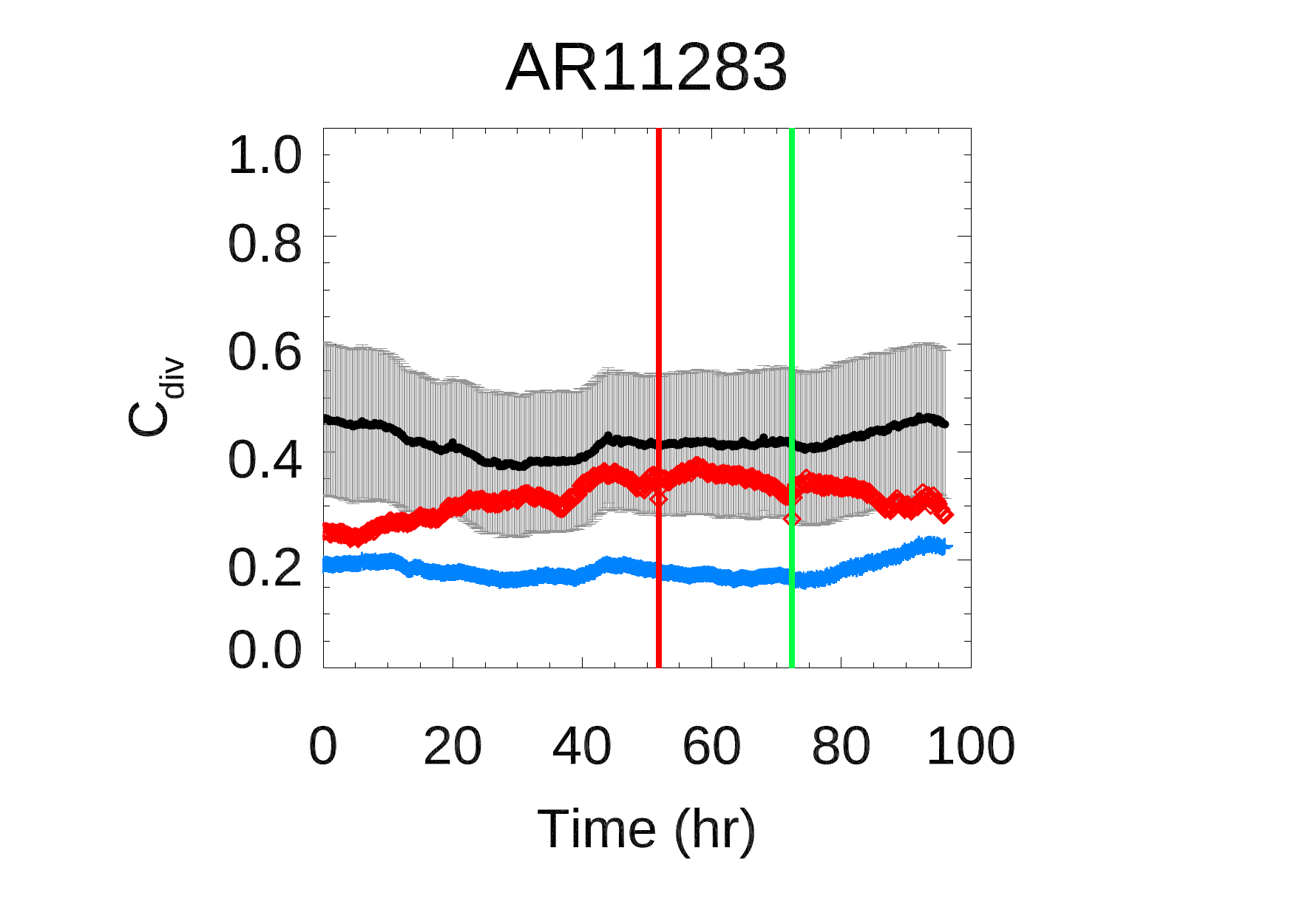}\includegraphics[trim={1.cm 2.cm  1.2cm  0.2cm},clip,width=6.cm]{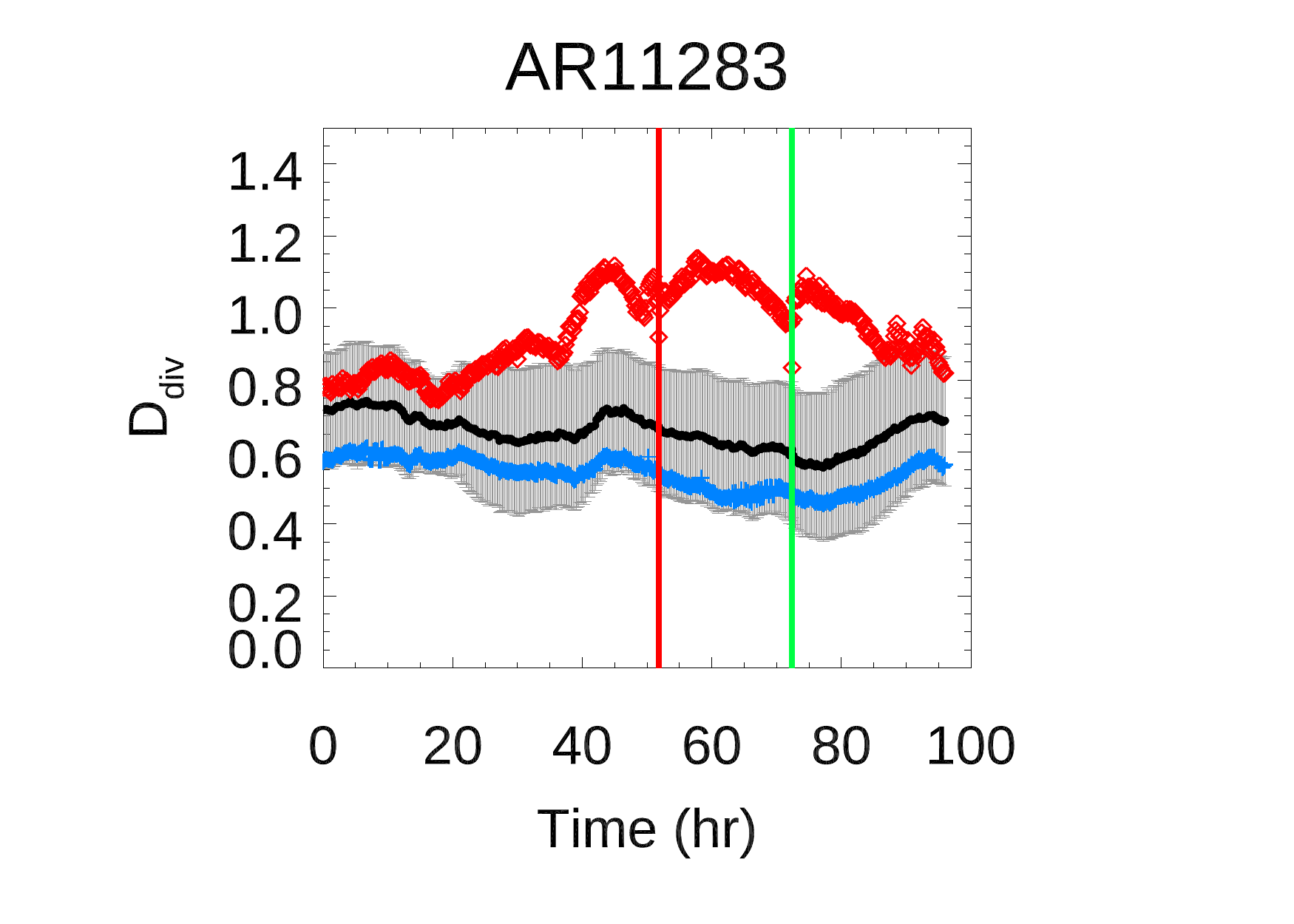}}
      \centerline{\includegraphics[trim={1.cm 2.cm  1.2cm  0.2cm},clip,width=6.cm]{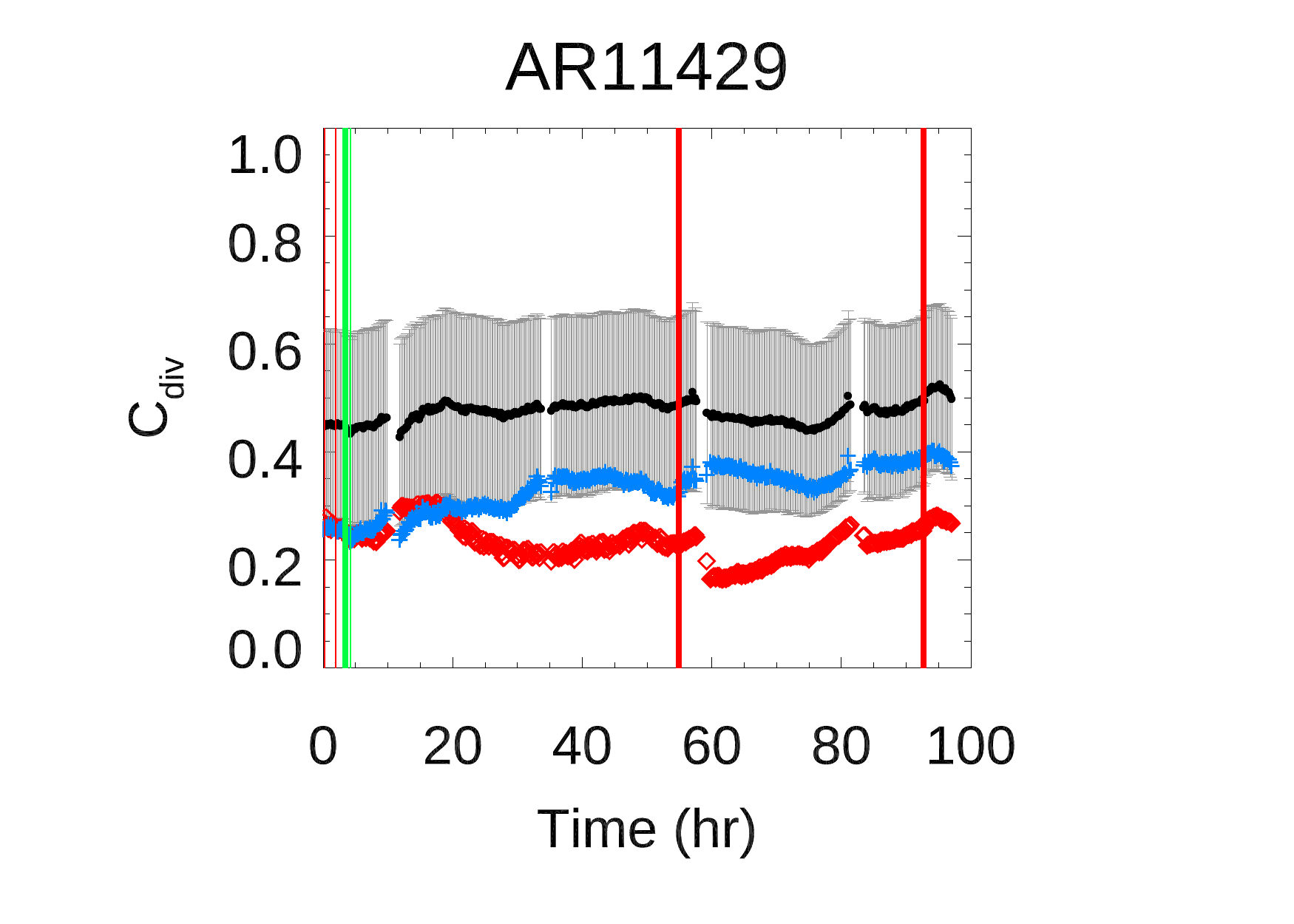}\includegraphics[trim={1.cm 2.cm  1.2cm  0.2cm},clip,width=6.cm]{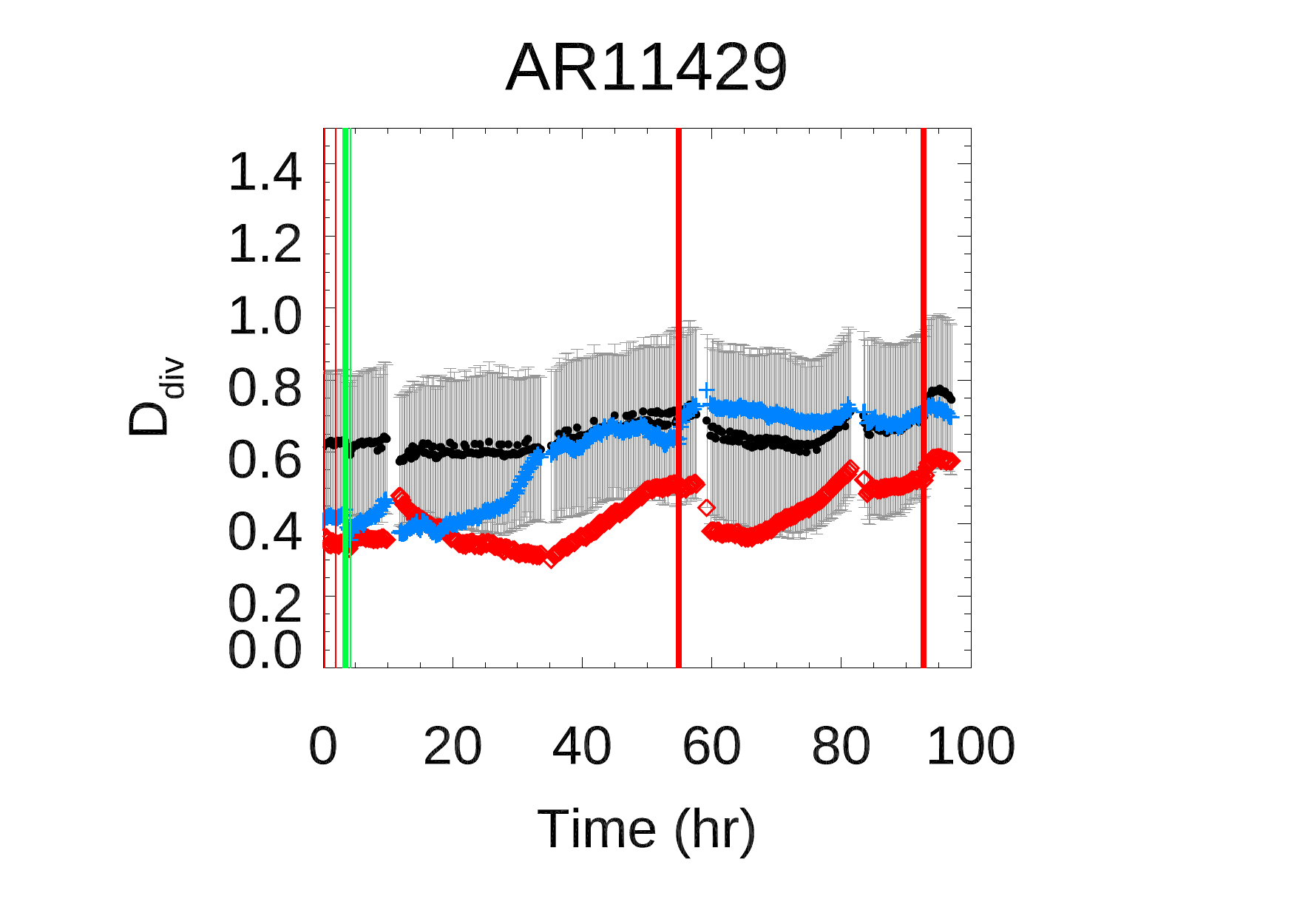}}
      \centerline{\includegraphics[trim={1.cm 2.cm  1.2cm  0.2cm},clip,width=6.cm]{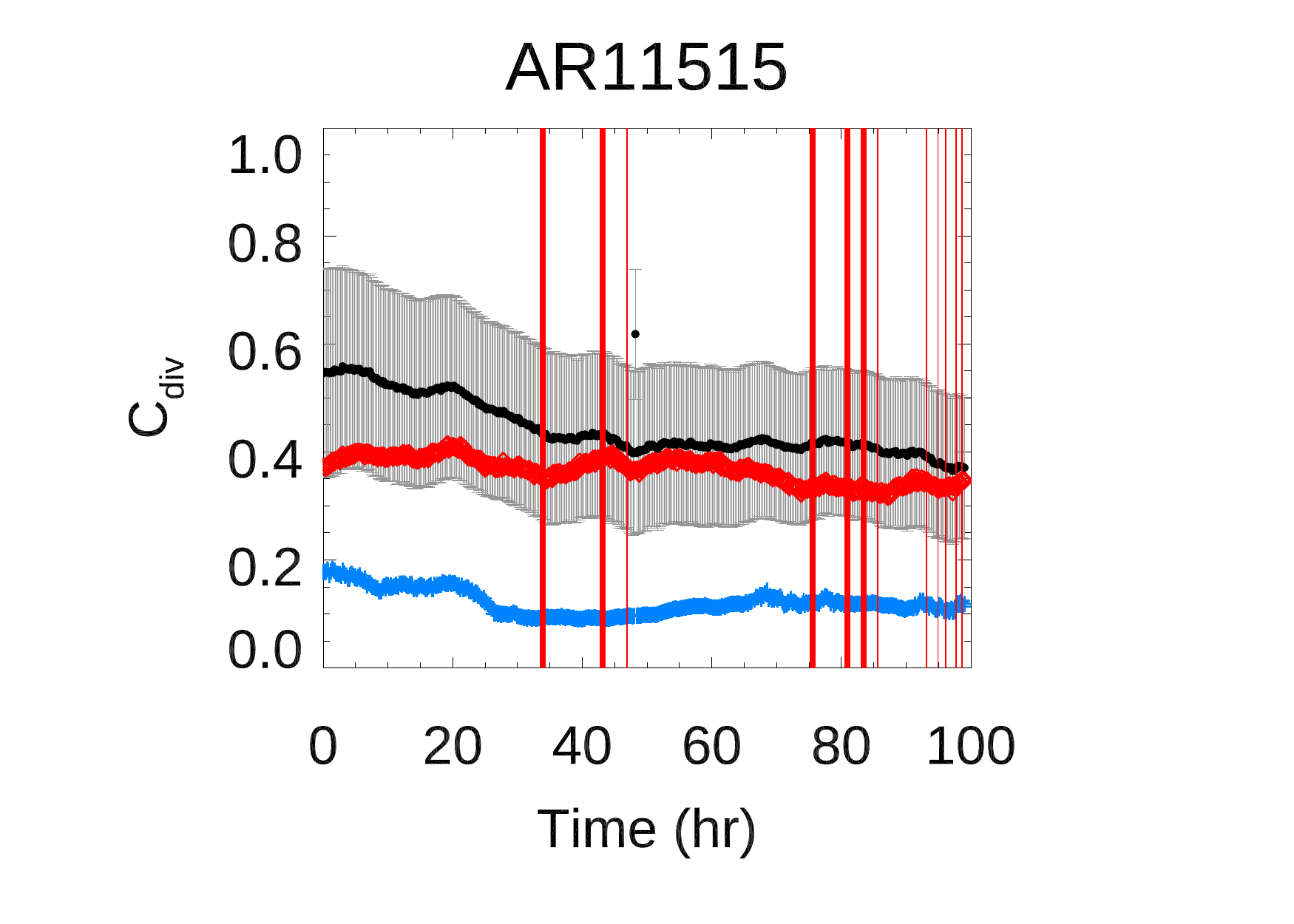}\includegraphics[trim={1.cm 2.cm  1.2cm  0.2cm},clip,width=6.cm]{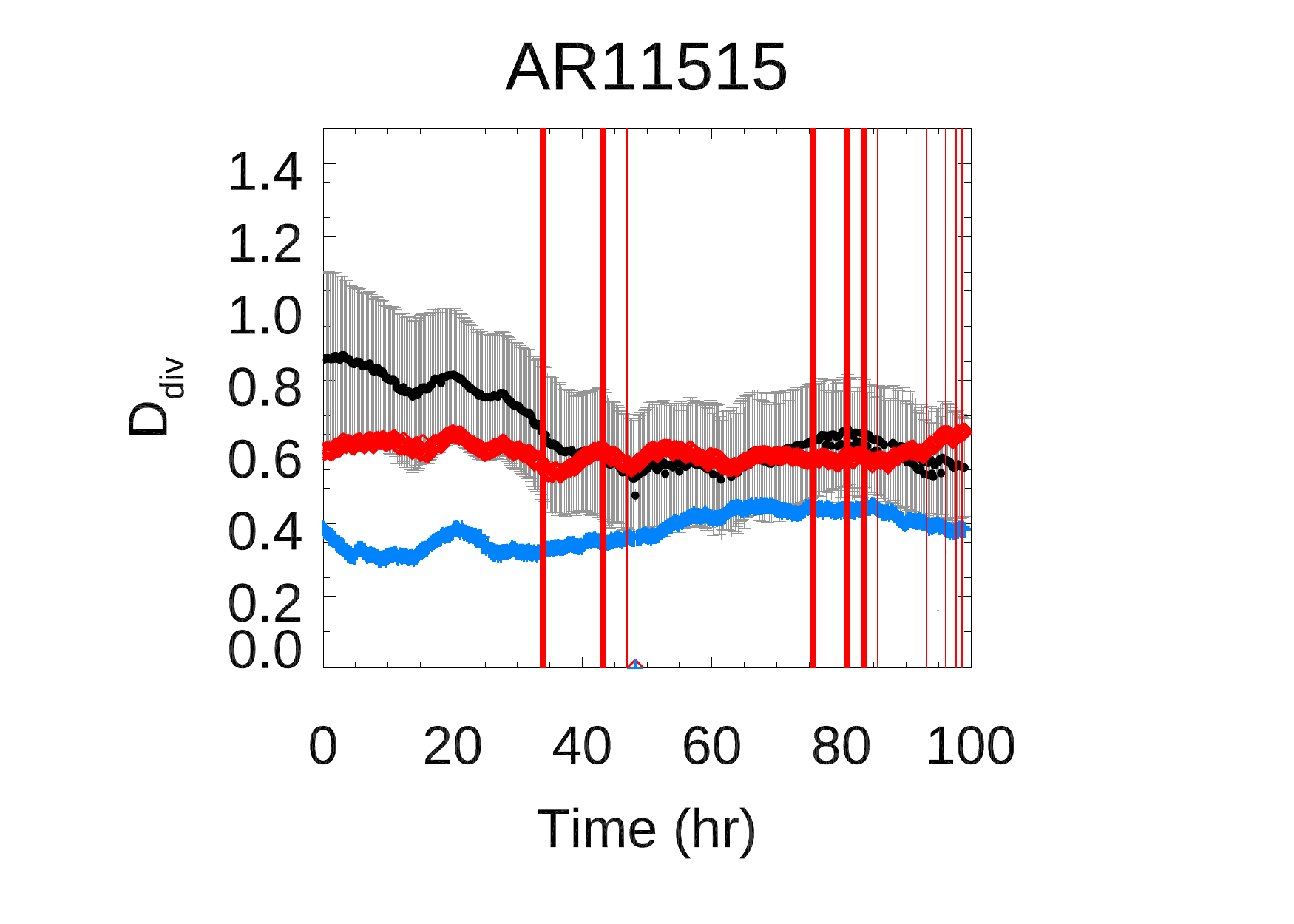}}
      \centerline{\includegraphics[trim={1.cm 0.5cm  1.2cm  0.2cm},clip,width=6.cm]{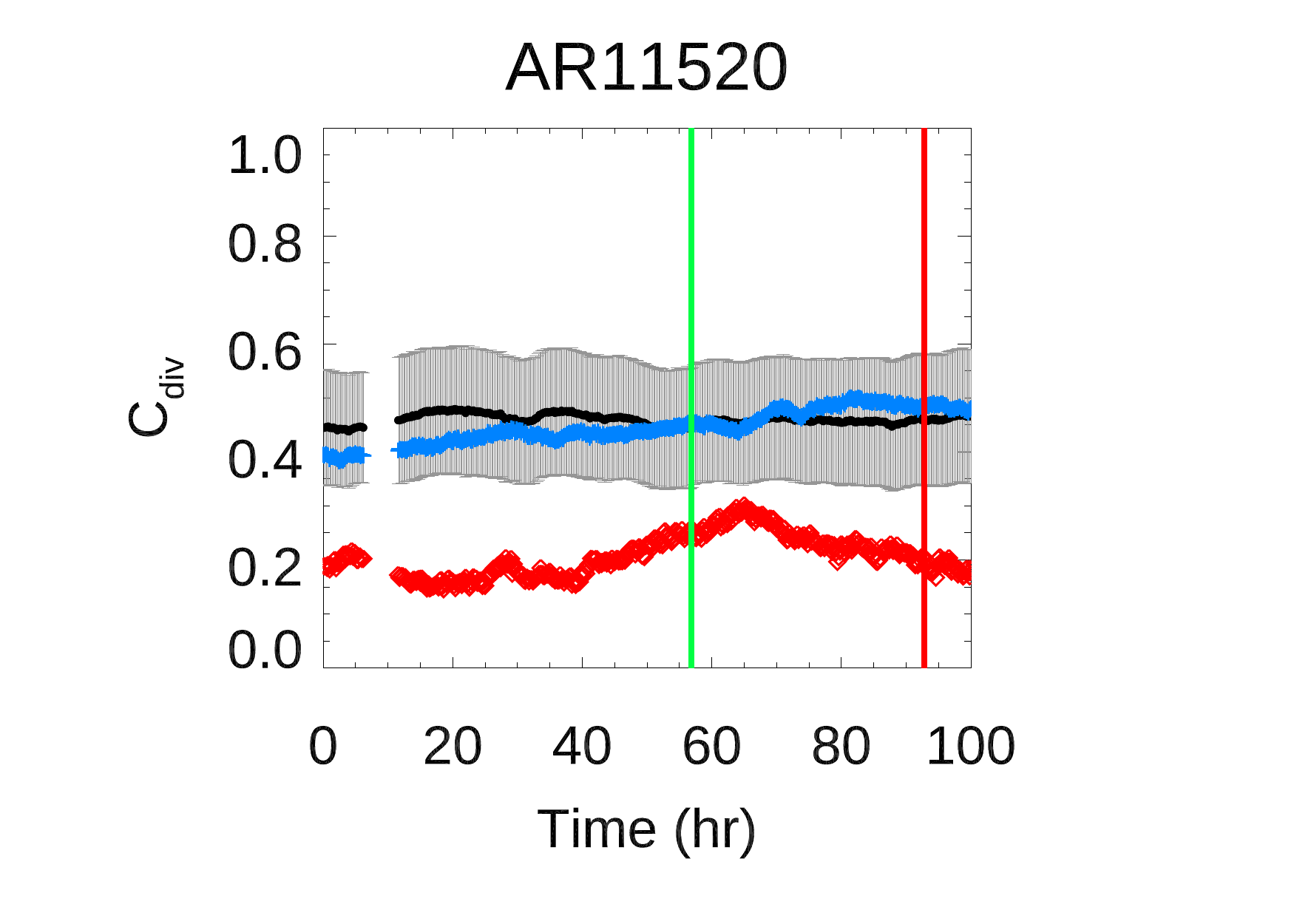}\includegraphics[trim={1.cm 0.5cm  1.2cm  0.2cm},clip,width=6.cm]{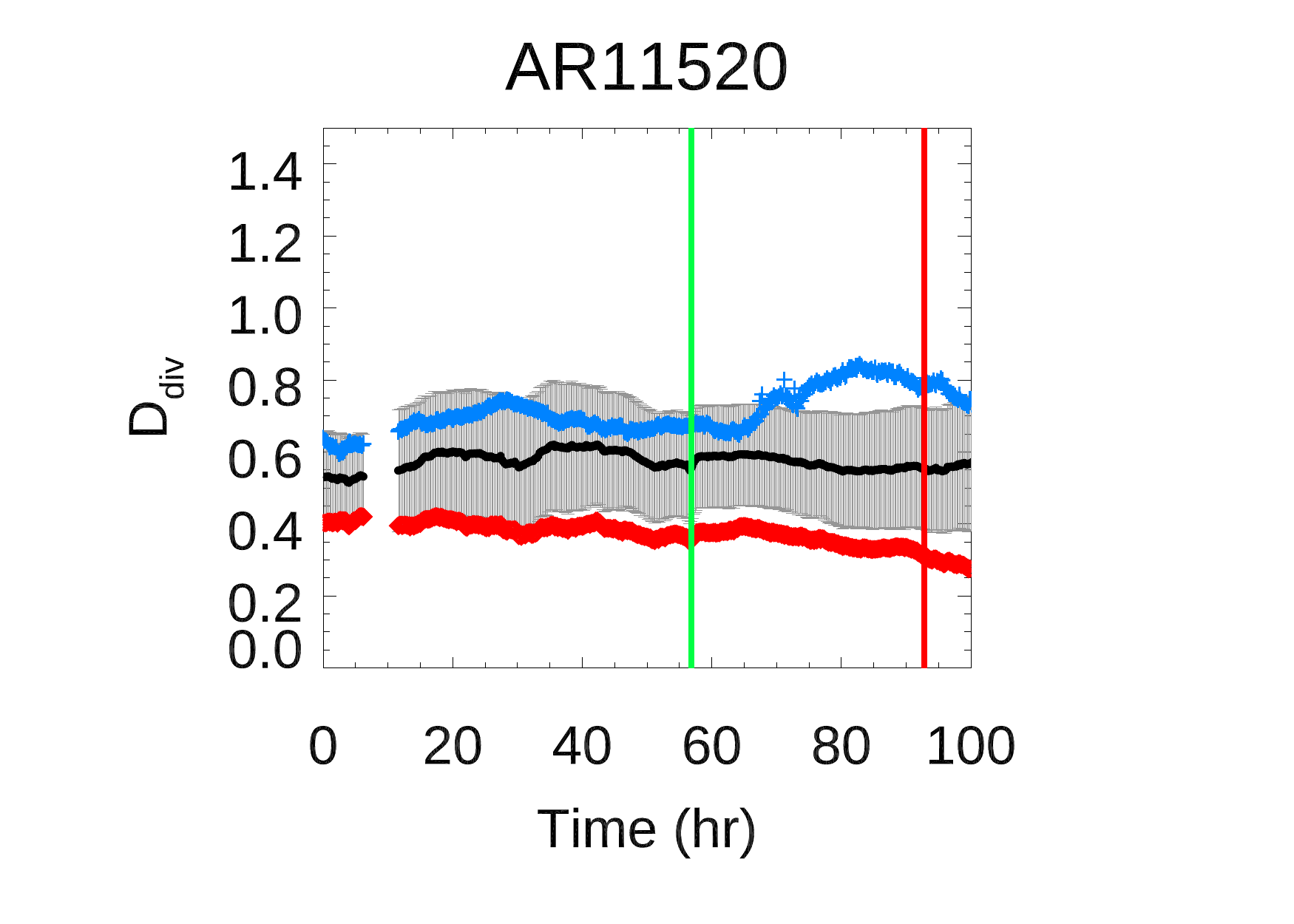}}
              \caption{ \footnotesize {Time series of the multifractal parameters $C_{\mathrm{div}}$ (left-hand column) and $D_{\mathrm{div}}$ (right-hand column) measured on the five selected productive ARs, by considering both unsigned (black symbols) and signed (positive and negative, red and blue symbols, respectively) flux data in the analyzed regions. From top to bottom results for ARs 11166, 11283, 11429, 11515, 11520.  See caption of Fig.~\ref{ff1} for more details. }}
   \label{ff2}
   \end{figure*}
\clearpage}

\afterpage{
     \begin{figure*}   
     \centerline{\includegraphics[trim={1.cm 2.cm  1.2cm 0.2cm},clip,width=6.1cm]{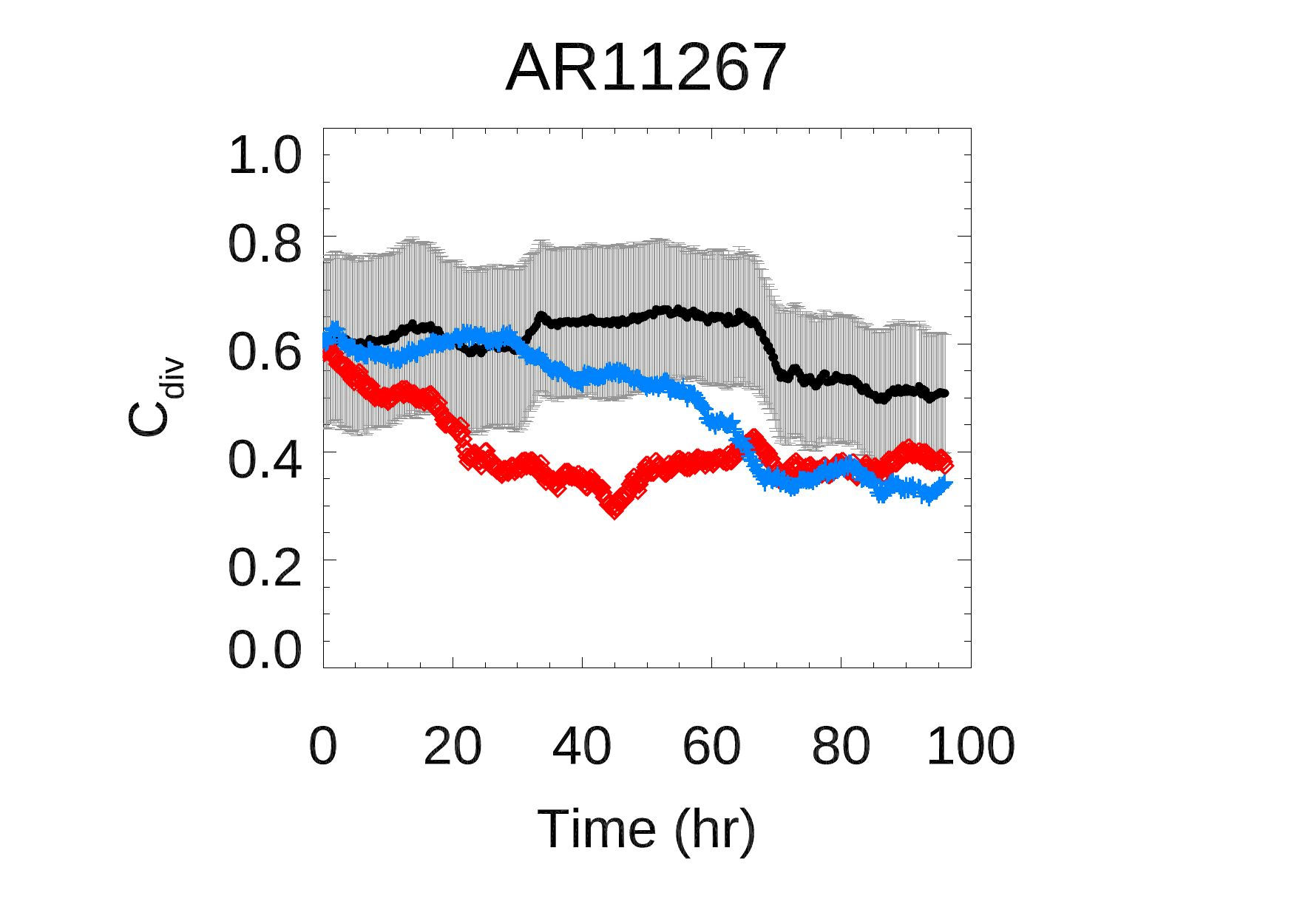}\includegraphics[trim={1.cm 2.cm  1.2cm 0.2cm},clip,width=6.1cm]{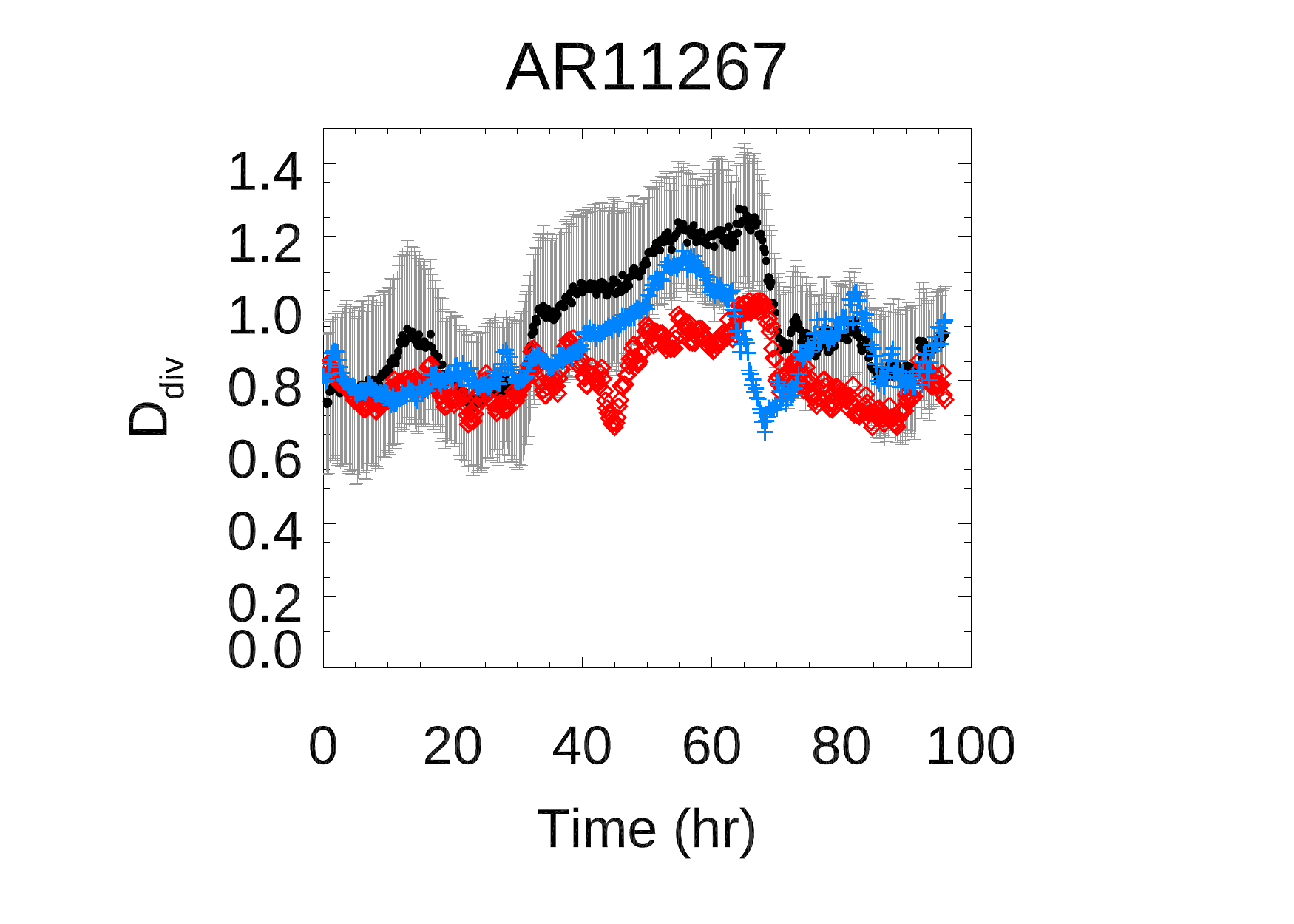}}
    \centerline{\includegraphics[trim={1.cm 2.cm  1.2cm  0.2cm},clip,width=6.1cm]{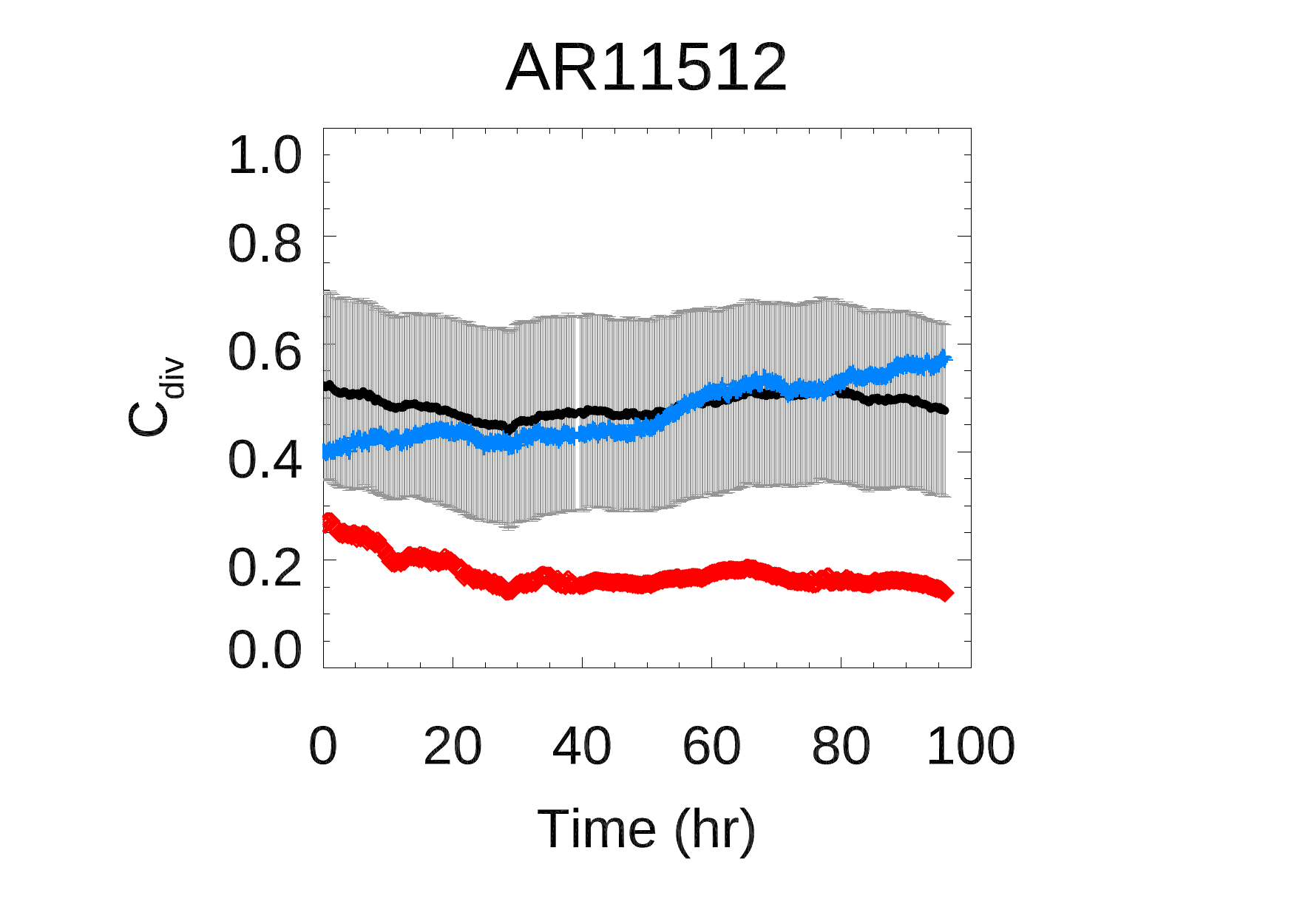}\includegraphics[trim={1.cm 2.cm  1.2cm  0.2cm},clip,width=6.1cm]{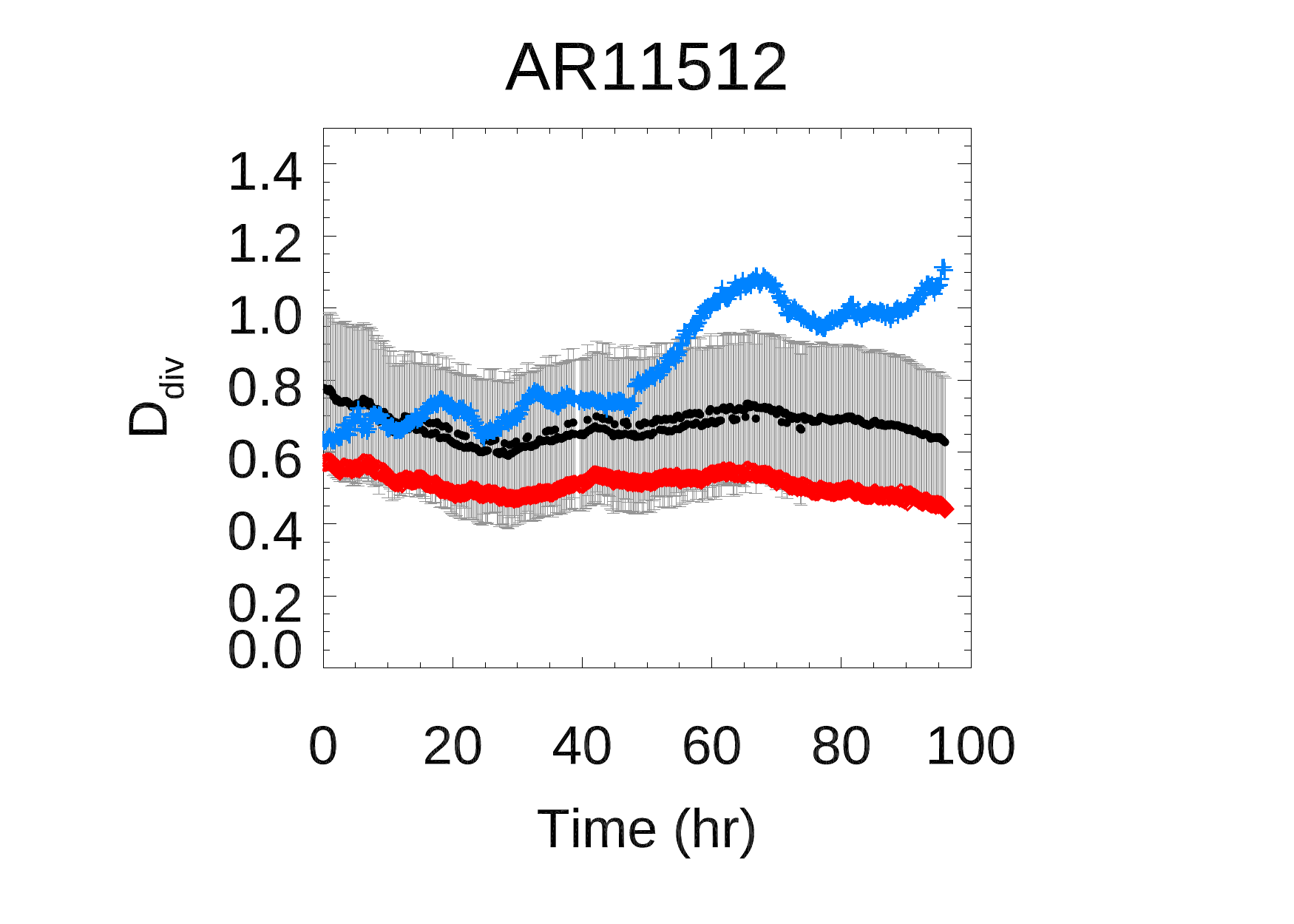}}
    \centerline{\includegraphics[trim={1.cm 2.cm  1.2cm  0.2cm},clip,width=6.1cm]{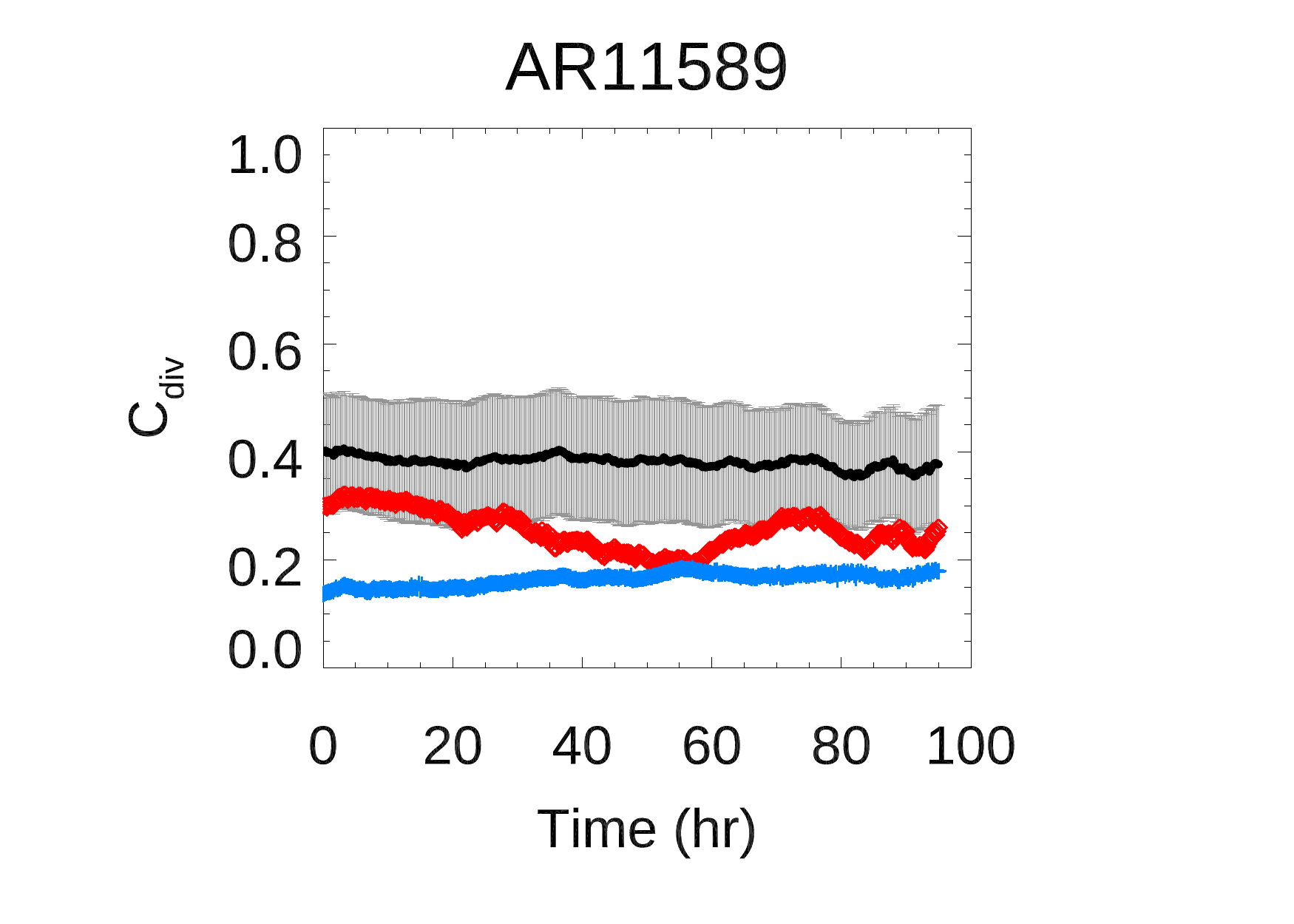}\includegraphics[trim={1.cm 2.cm  1.2cm  0.2cm},clip,width=6.1cm]{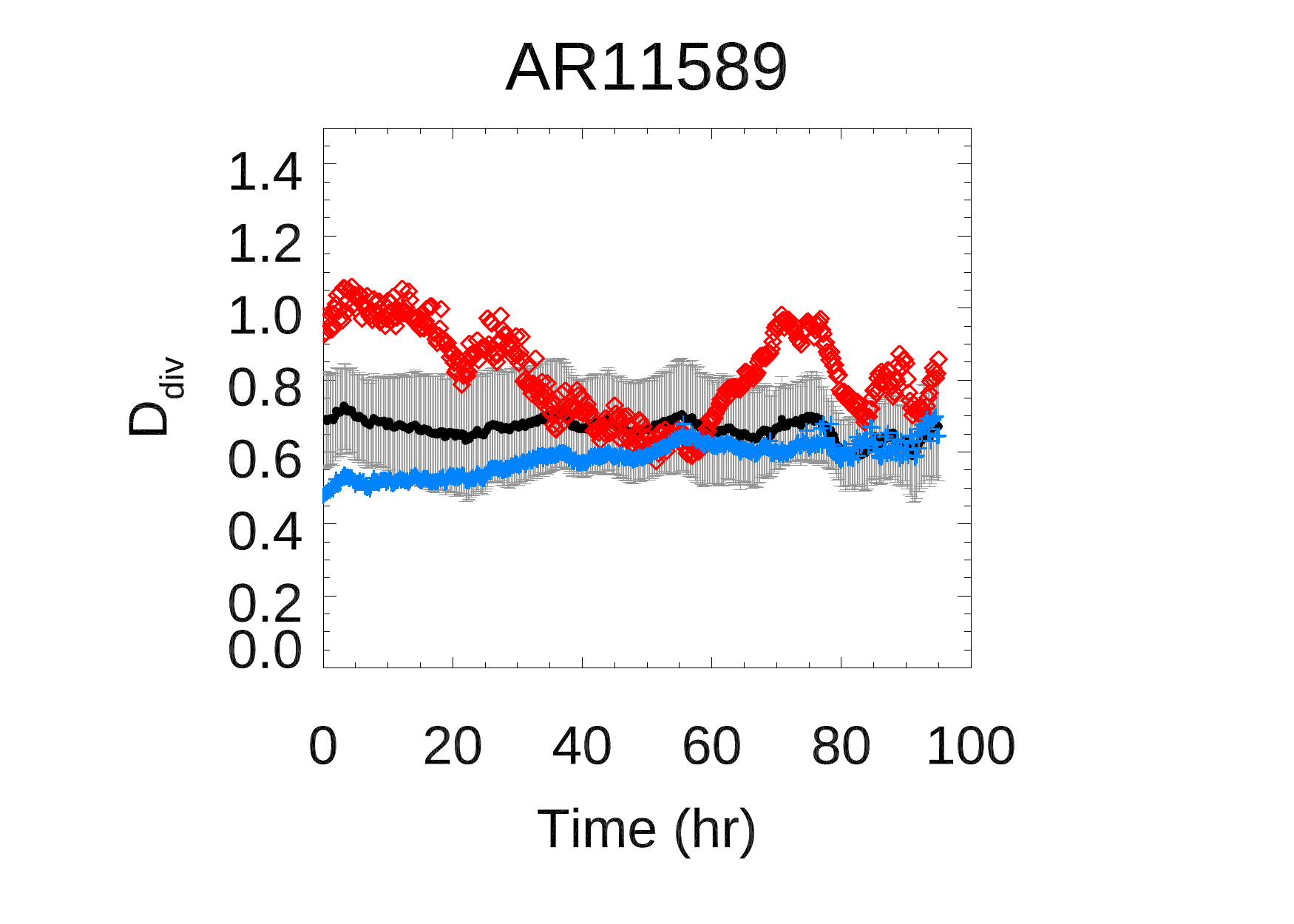}}
    \centerline{\includegraphics[trim={1.cm 2.cm  1.2cm  0.2cm},clip,width=6.1cm]{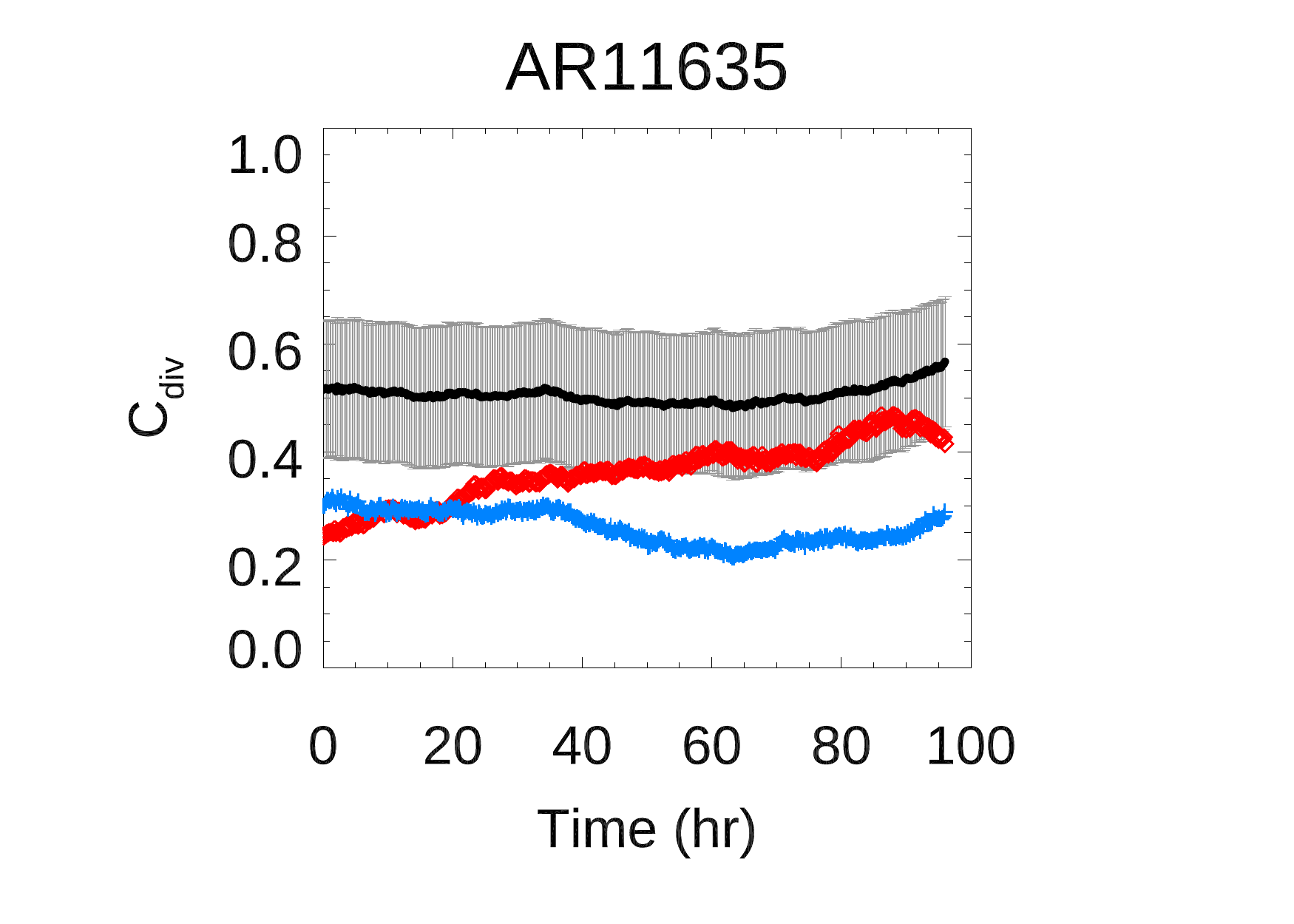}\includegraphics[trim={1.cm 2.cm  1.2cm  0.2cm},clip,width=6.1cm]{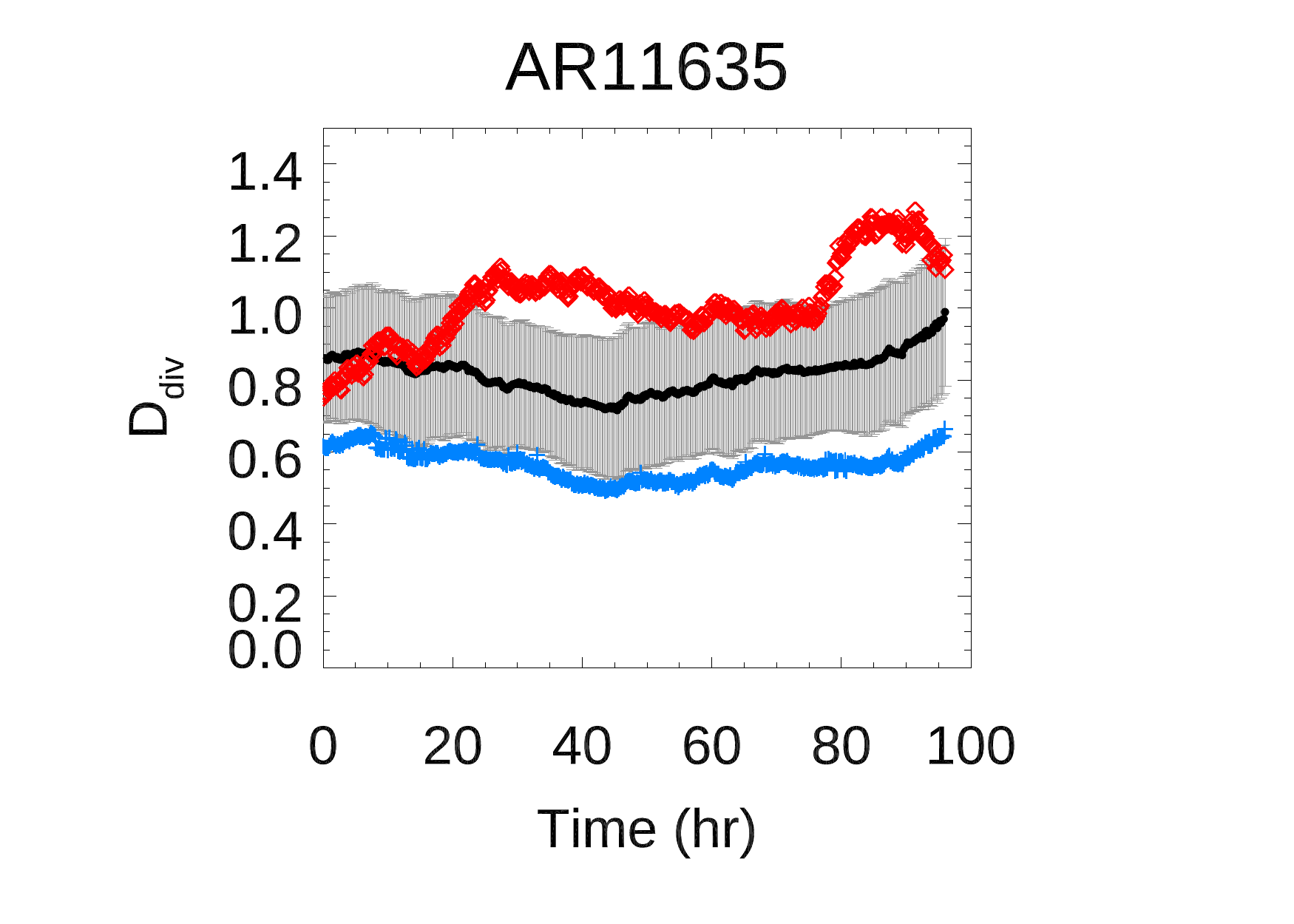}}
    \centerline{\includegraphics[trim={1.cm 0.5cm  1.2cm  0.2cm},clip,width=6.1cm]{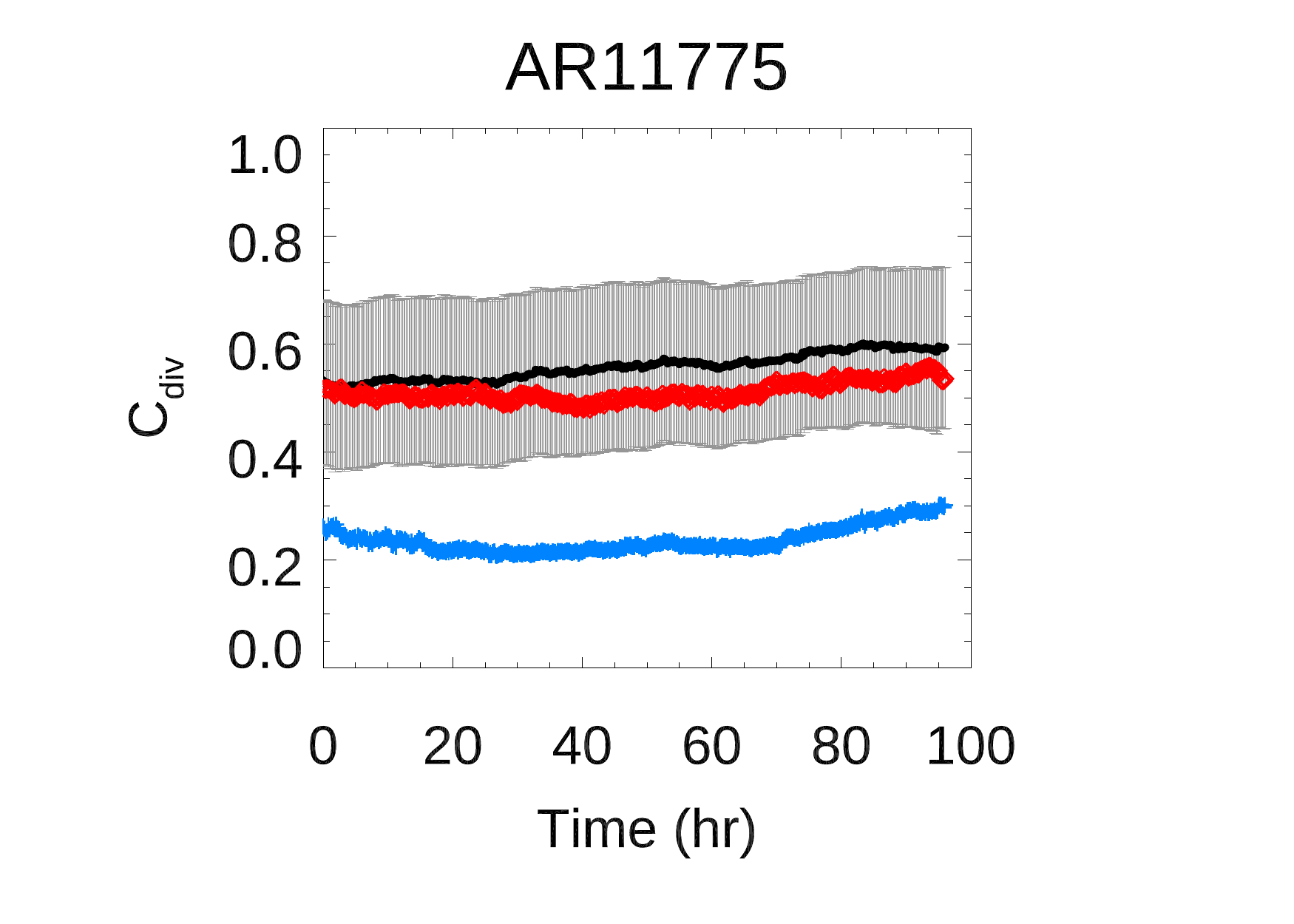}\includegraphics[trim={1.cm 0.5cm  1.2cm  0.2cm},clip,width=6.1cm]{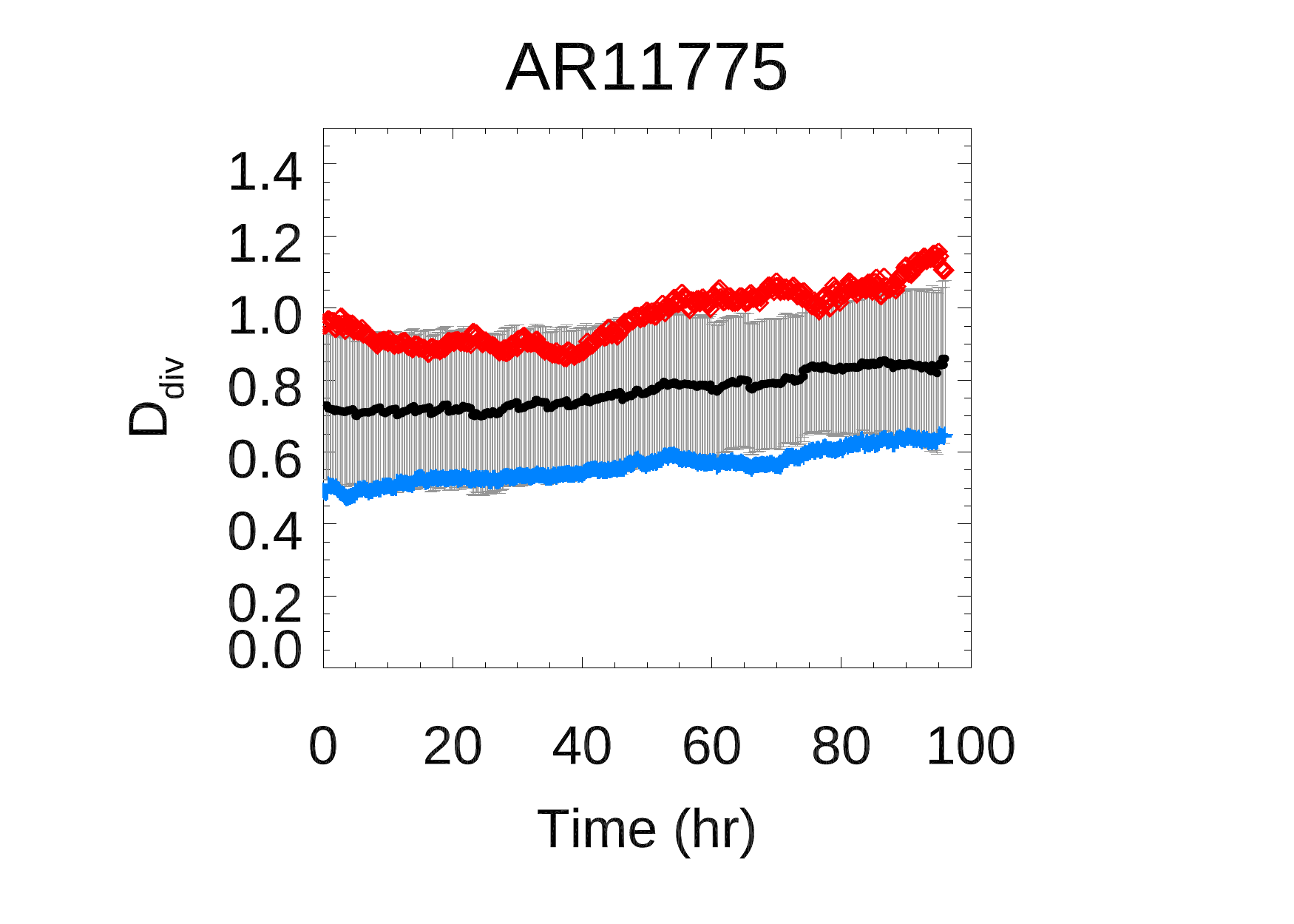}} 
              \caption{ \footnotesize {Same parameters as in Fig.~\ref{ff2} for ARs 11267, 11512, 11589, 11635, 11775 hosting only B and C flares. }}
   \label{ff2b}
   \end{figure*}
\clearpage
}

\section{Discussion}\label{5}

Comparisons that aimed to test the performances of different methods to assess the eruptive potential of ARs presented in the literature 
in specific contexts did not reveal clear outstanding performances of one method with respect to all others. Earlier findings  by, e.g. \citet{Barnes_Leka_2008}, \citet{Leka_2013}, confirmed by \citet{Barnes_2016}, reported  outcomes of considerable efforts devoted by several teams to test the ability  of a number of methods on common data sets. In their study, \citet{Barnes_2016} compared results from 11 algorithms on 13,000 magnetograms relevant to ARs that have hosted more than  3000 events during the period of  2000 to 2005, by applying standard verification statistics to determine the ability of the tested methods to identify flare signatures and predict eruptive events. The different algorithms were applied to sub-areas extracted from the full-disk line-of-sight magnetic field and continuum intensity images taken close to noon of each studied day observed by SOHO/MDI.  
\citet{Barnes_2016} showed that none among the tested methods clearly outperforms all others, and partially attributed this result  to the strong correlations among the parameters used by the various  methods to characterize the ARs. 
Furthermore, a workshop was held at Nagoya University in 2017 to quantitatively compare the performance of several operational solar flare-prediction methods. This led to the studies described by \citet{Leka2019a,Leka2019b} that present the compared methods and evaluation methodology applied, and describe the results from quantitative comparisons and method performance. 

\begin{table*}
\caption{Summary of the average value and standard deviation of the fractal ($D_0$ and $D_8$) and multifractal ($C_{\mathrm{div}}$ and $D_{\mathrm{div}}$) parameters measured  in the flare-productive and flare-quiet ARs samples, by considering unsigned  flux data in the analyzed ARs. FI and Max FI denote the Flare Index and its maximum value, respectively, as described in the main text. FI is computed during the transit of the analyzed region over the solar disk, while Max FI for the most intense event produced by the region. The FI values are given using $10^3 \,\mathrm{erg}\, \mathrm{cm}^{2}\,\mathrm{s}^{-1}$ units. Details are given in the text.}
\label{tblff}
\centering
\begin{tabular}{cccccccc}
\hline\hline
Class&Region & $D_0$  & $D_8$  &  $C_{\mathrm{div}}$ & $D_{\mathrm{div}}$ & FI & Max FI \\
\hline
productive &AR 11166 &1.79  $\pm$  0.02&1.50 $\pm$  0.03& 0.48 $\pm$  0.04 &  0.69 $\pm$ 0.12 & 303.6 & 150\\
&AR 11283 &1.79  $\pm$  0.02&1.56 $\pm$  0.03& 0.42 $\pm$  0.02 &  0.66 $\pm$ 0.05 & 602.7 & 210\\
&AR 11429 &1.82  $\pm$  0.01&1.55 $\pm$  0.01& 0.47 $\pm$  0.02 &  0.65 $\pm$ 0.04 & 1342.6 & 540\\
&AR 11515 &1.80  $\pm$  0.02&1.55 $\pm$  0.05& 0.44 $\pm$  0.05 &  0.66 $\pm$ 0.10 & 1064.1 & 69\\
&AR 11520 &1.904  $\pm$  0.003&1.65 $\pm$  0.01& 0.46 $\pm$  0.01 &  0.57 $\pm$ 0.02 & 444.3 & 140\\
& average & 1.82$\pm$0.01& 1.56$\pm$0.03 & 0.45$\pm$ 0.03&0.65$\pm$0.07&-&-\\
\hline
quiet&AR 11267 &1.64  $\pm$  0.01&1.31 $\pm$  0.04& 0.60 $\pm$  0.05 &  0.95 $\pm$ 0.15 & 8.7 & 4.1\\
&AR 11512 &1.73  $\pm$  0.01&1.46 $\pm$  0.02& 0.49 $\pm$  0.02 &  0.67 $\pm$ 0.04 & 12.4 & 4.2\\
&AR 11589 &1.824  $\pm$  0.006&1.63 $\pm$  0.01& 0.38 $\pm$  0.01 &  0.66 $\pm$ 0.03 & 6.7 & 3.3\\
&AR 11635 &1.807  $\pm$  0.003&1.51 $\pm$  0.01& 0.51 $\pm$  0.01 &  0.81 $\pm$ 0.05 & 29.4 & 4.1\\
&AR 11775 &1.792  $\pm$  0.003&1.45 $\pm$  0.02& 0.56 $\pm$  0.02 &  0.77 $\pm$ 0.05 & 6.3 & 1.3\\
&average & 1.76$\pm$0.01& 1.47$\pm$0.02 & 0.51$\pm$ 0.02&0.77$\pm$0.06&-&-\\
\hline
\end{tabular}
\vspace{0.2cm}
\end{table*}

Taking into account the conclusions of \citet{Barnes_2016} about the importance of combining the application of statistical analysis to the characterization of ARs by some parameters that are able to assess their eruptive potential, we used here two methods based on the study of the magnetic flux and helicity accumulation, and fractal and multifractal measurements. The detailed case study analysis was carried out on five flare-productive and five flare-quiet ARs. The methods applied in our study are based on different techniques. Following \citet{smyrli} and \citet{ermolli2014}, the two methods
were applied to LOS magnetogram time series of ARs within $\pm 30^{\circ}$ from the central meridian, because the  projection effects at longitudes greater than $\pm 30^{\circ}$ strongly affect the determination of the relevant parameters.
 The two methods have in common that they employ the signed flux and the absolute value of the magnetic flux.

\citet{smyrli} demonstrated that magnetic helicity accumulation in the ARs generating halo CMEs can exhibit significant changes when there are indications of newly emerging magnetic flux. The second result of their paper is that for 4/5 of the studied ARs helicity accumulation is in agreement with the hemispheric helicity rule \citep{See90}, but the remaining 1/5 part of the ARs, which produced impulsive CMEs, do not follow the hemispheric helicity rule. In the ARs considered in our study, three out of five of flare-productive ARs do not show an agreement with the helicity rule, while two ARs do. Moreover, the analysis of the helicity accumulation has shown two interesting results: the first one is related to the fact that flare-productive ARs show a persistent accumulation of higher magnitudes and senses of helicity, while the flare-quiet ARs are characterized by changes in the $H$ trend; the second result is related to lower values of the right-handed and left-handed magnetic helicity accumulation in flare-quiet ARs with respect to flare-productive ARs.

\citet{ermolli2014} and \citet{giorgi2014} investigated the temporal variation of the fractal and multi-fractal parameters of the total unsigned and signed flux of some ARs observed with the SDO/HMI. The main results obtained from their study are reflected back in the currently analyzed AR cases. Several of the solar flare events of the five flare-productive ARs occur during a decreasing phase of the $D_{\mathrm{div}}$ and, concurrently, during an increasing phase of the $D_{8}$ values. This result agrees with recent findings by \citet{Park2020} on that the prior flaring history of an AR  is an important factor to consider in development of robust flare-prediction methods.

The common feature of the two compared methods  is that they take into account the unsigned magnetic flux in their measurements. The variation  of the $D_{8}$ parameter \citep{ermolli2014} 
confirms the behavior of the positive/negative and unsigned flux in time, already investigated by \citet{smyrli}, in all the five flare-productive ARs analyzed in our study. We unveil an interesting common property in the two compared techniques: there is a prominent role of the decreasing phase of certain parameters. \citet{ermolli2014} found that several solar flare events occur during a decrease of $D_{\mathrm{div}}$. There is an interesting aspect in the magnetic flux and helicity accumulation investigation, i.e.,  that strong energetic flares (mainly X-class) with CMEs (see, e.g., Fig. 1, left panel, first, second and fifth rows) occur on the constant or decreasing evolution part of the unsigned magnetic flux, while a flare without CME appears anywhere as part of the temporal variation of the unsigned magnetic flux.

Our study aimed at exploiting the diverse information carried out by the individual methods on the pre-flare conditions. These comparative examinations necessarily differ from the studies on large samples. In their pioneering study, \citet{Bobra} examined 25 parameters deduced from vector magnetic maps of 2701 ARs and carried out true skill score analysis. The large size of the material obviously cannot allow scrutinizing the details of the pre-flare evolution, as they necessarily consider the individual cases like snapshots. In contrast, the present approach on a more limited sample allows us to compare the details of the pre-flare status and dynamics for the analyzed ARs.

\section{Conclusions} \label{6}

The large interest in the identification of physical parameters that carry information on the peak magnitudes and timescales of solar eruptive events has motivated many theoretical and observational studies of flaring ARs; see, e.g.,  recent works by \citet{Shibata_2011}, \citet{Harra_2016} and \citet{Toriumi_2017}, and references therein. Besides, the need to  refine our knowledge about the physical processes behind solar eruptive events and to mitigate their effects in the circumterrestrial environment has driven the development of several methods useful to assess the eruptive potential of ARs based on full-disk  observations of the ARs emerged into the solar atmosphere.

In this case study, here, we applied two different algorithms based on as many methods to assess the eruptive potential of ARs previously presented by \citet{smyrli}, and \citet{ermolli2014} on time series of SDO/HMI LOS magnetograms observations, i.e., present-day highest resolution full-disk photospheric data.
Based on the prediction capability of the two methods (see the summary Table \ref{SummaryTab}), we could conclude that the employed methods seem to complement each other in their ability to identify flaring ARs. 
By identifying suitable general characteristics of the AR, the methods based on helicity and fractal measurements allow discriminating regions that may host extreme class events right after their appearance on the solar disk, based on the reported  differences between the average values of the parameters measured in the flare-quiet and flaring ARs. 
Therefore, the two methods tested in our study could be employed as warning tools to identity ARs prone to flaring activity by following the emergence of any magnetic region in the solar photosphere.

\begin{figure}
   \centering
   \includegraphics[trim={0.2cm 0.2cm  0.4cm  0.2cm},clip,width=\textwidth]{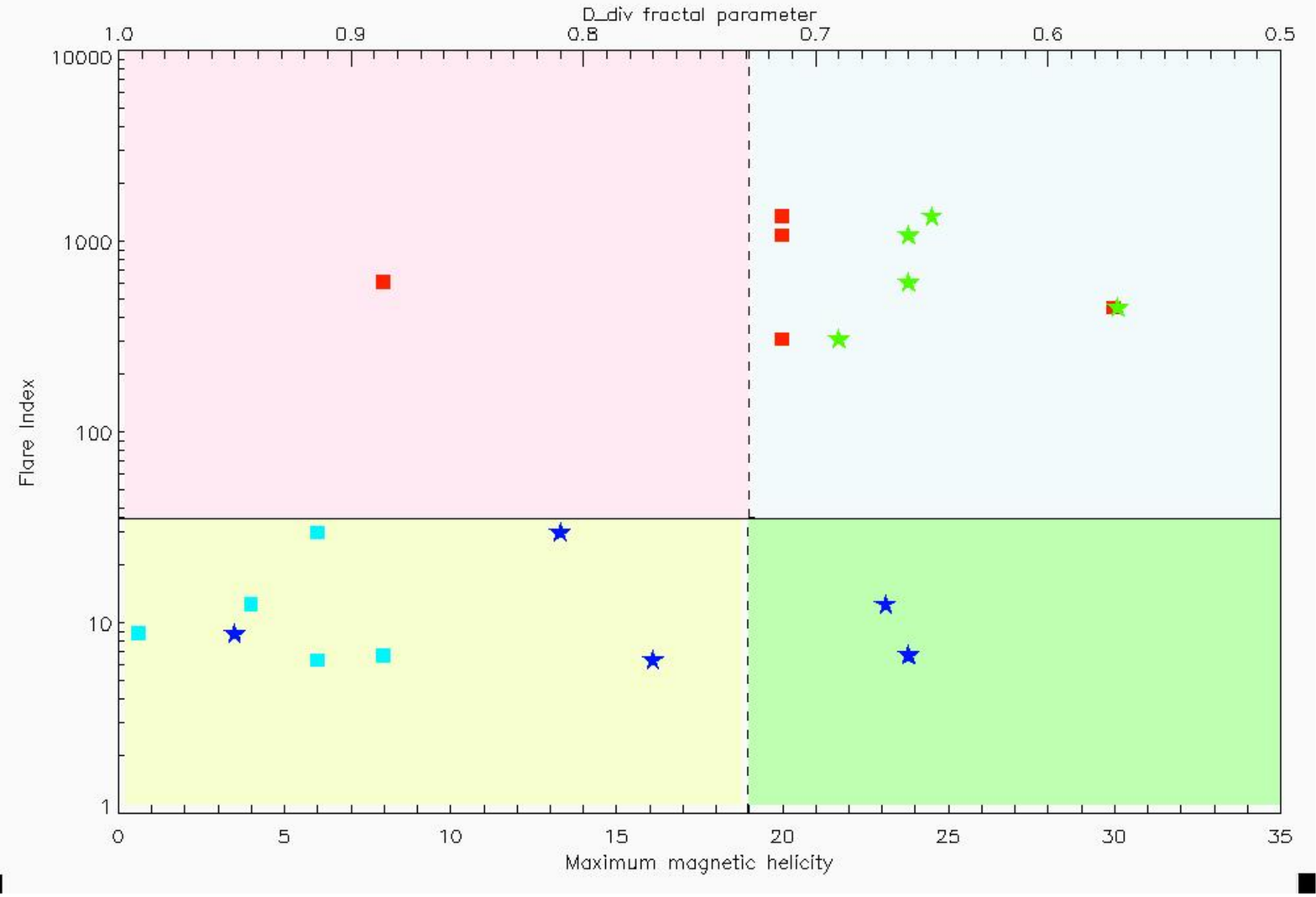}
   \caption{Distribution of data-points relevant to the relationship between the right- or left-handed magnetic helicity and the D$_{div}$ parameter as a function of FI. The horizontal axis at the bottom reports the absolute value of the right- or left-handed magnetic helicity (in units of $10^{42}$ Mx$^{2}$) for the analyzed ARs; the top horizontal axis reports D$_{div}$ values, while the vertical axis reports the FI. Red (light blue) squares indicate the absolute value of the right- or left-handed magnetic helicity for the flare-productive (flare-quiet) ARs; green (blue) stars indicate the D$_{div}$ parameter value for the flare-productive (flare-quiet) ARs.}
   \label{PlotFI}
   \end{figure}

In order to describe how the two methods could be employed to identify ARs prone to flaring activity, the Fig.\ref{PlotFI} was obtained by taking into account the relationship between the flare index FI and two parameters (left- or right- handed magnetic helicity $H$ and fractal parameter D$_{div}$), used in the analysis relevant to the first and the second method, respectively.

More precisely, as far as the first method is concerned, one of the results that we obtained was related to higher values of the left- or right- handed magnetic helicity in the flare-productive ARs with respect to the ones relevant to flare-quiet ARs. Therefore, we estimated, for each AR, the maximum value reached by the left- or right- handed magnetic helicity during the analyzed time interval and reported its absolute value as a function of the FI for the respective AR, as shown in the plot reported in Fig. \ref{PlotFI}. The results, indicated by the squares (red for the flare-productive ARs and light blue for the flare-quiet ARs) show that there are two classes, grouped in two different and well distinguished regions of the plot: pink and light blue color areas for flare-productive ARs and yellow area for the flare-quiet AR.

Similarly, the average value of the D$_{div}$ parameter obtained with the second method, is reported, for each AR as a function of the FI (green stars for the flare-productive ARs and blue stars for the flare-quiet ARs) in Fig. \ref{PlotFI}. In this case, we can see that the data-points are located in two different regions of the plot: light blue color area for flare-productive ARs and yellow plus green areas for the flare-quiet ARs.

Therefore, using a combination of these two methods, it should be possible to provide a quantitative estimation of the eruptive potential of an AR. Indeed, while the second method is particularly efficient in identifying productive ARs, the first one allows recognizing quiet regions very accurately. Thus, the suggested procedure could be the following: starting from the determination of the D$_{div}$ parameter, once this overcomes a certain value, which, based on the present analysis, we can set equal to $\sim$ 0.69, if the absolute value of the left- or right- handed magnetic helicity is greater than $20 \times 10^{42}$ Max and the FI becomes greater than $\sim$ 50, then the analyzed AR could be considered prone to flaring. More specifically, the plot quadrant which should be taken into account for this estimate is the one highlighted with the light blue color in Fig. \ref{PlotFI}. 

It is worth noting that none of the operational flare-prediction methods present in the literature are able to adequately respond to changes in flaring activity \citep{Park2020}. This shows that we need to specifically improve the performance of flare-prediction methods over short-term variations in flare activity.
Therefore, refining the capabilities of existing flare-prediction methods and deepening our knowledge of the physics behind the quantities that allow a more successful assessment of the eruptive potential of ARs be determined is important for solar and Space Weather research. In this respect, our next intention is to apply these two methods on a much bigger statistical sample, in order to extend our results. Moreover, we plan to investigate the potential of other parameters of flaring ARs, e.g. the ones based on properties of the horizontal gradient of the LOS component of the magnetic field \citep{Korsos2019,Korsos2020}, to further determining the energy class and onset time of upcoming flares.

\begin{acknowledgements}

The research leading to these results has received funding from the European Commission's Seventh Framework Programme under the grant agreements eHEROES (project No.~284461), F-Chroma (project No.~606862) and No.~312495 (SOLARNET project) and from the European Union's Horizon 2020 research and innovation programme under the grant agreements no.~739500 (PRE-EST project) and no.~824135 (SOLARNET project). FZ acknowledges that this research started thanks to a conversation with A.~Ludmany and the late T.~Baranyi during a eHEROES Meeting. Authors acknowledge support by the Universit\`{a} degli Studi di Catania (Piano per la Ricerca Universit\`{a} di Catania 2016-2018 -- Linea di intervento~1 ``Chance''; Linea di intervento~2 ``Dotazione ordinaria''; Fondi di Ateneo 2020-2022, Universit\`{a} di Catania, Linea Open Access), by the Istituto Nazionale di Astrofisica (INAF), by the Italian MIUR-PRIN grant 2017APKP7T on ``\textit{Circumterrestrial Environment: Impact of Sun-Earth Interaction}'', and by Space Weather Italian COmmunity (SWICO) Research Program. MBK is grateful to the Science and Technology Facilities Council (STFC), (UK, Aberystwyth University, grant number ST/S000518/1), for the support received while carrying out this research. RE is grateful to the STFC (UK), grant No. ST/M000826/1) for the support received. RE also acknowledges the support received by the Royal Society (grant nr. IE161153) and by the CAS President's International Fellowship Initiative grant No.2019VMA052.

\end{acknowledgements}

\newpage
\begin{landscape}
\begin{table*}
\caption{The Table summarises how the two applied prediction methods performed in the  selected 5 productive and 5 quiet AR cases. The performance of first (magnetic flux and magnetic helicity trend) and second (fractal and multi-fractal parameters) methods are based on their estimated threshold parameters. For the fractal and multifractal parameters, the threshold values considered here are the average values of the parameters derived by \citet{giorgi2014} from a large sample of ARs. The "Yes" ("No") means that the pre-flare signature of one of the method can (cannot) be seen. The "True" ("False") means that the pre-flare signature correctly indicated  (not) an upcoming event.}
\label{SummaryTab}
\resizebox{1.6\textwidth}{!}{%
\centering
\begin{tabular}{c|ccc|cccc}
\hline  
\hline
 &
  \multicolumn{3}{c}{Magnetic flux and magnetic helicity trend} &
  \multicolumn{4}{c}{Fractal and multi-fractal parameters}  \\
 \hline  
Region &
  $\Phi_{max}>2\cdot10^{22}$(Mx) &
  H   Continuous increasing &
  $\mid{}H\mid$max$>1.5\cdot10^{43}$ ($Mx^{2}$) &
  $D_{0} > 1.74 \pm 0.08$ &
  $D_{8}>1.42 \pm 0.11$ &
  $C_{div}<0.55 \pm 0.08$ &
  $D_{div}<0.81 \pm 0.15$  \\
  \hline  
Productive &     &  &     &    &    &    &    \\
11166      & Yes-True & Yes-True & Yes-True & Yes-True & Yes-True& Yes-True& Yes-True \\
11283      & No-False  & Yes-True    & No-False  & Yes-True& Yes-True& Yes-True& Yes-True \\
11429      & Yes-True & Yes-True     & Yes-True& Yes-True& Yes-True& Yes-True& Yes-True \\
11515      & Yes-True & Yes-True     & Yes-True & Yes-True& Yes-True& Yes-True& Yes-True \\
11520      & Yes-True & No-False (but change after CME) & Yes-True& Yes-True& Yes-True& Yes-True& Yes-True  \\
Quiet      &     &       &    &    &    &   &         \\
11267      & No-True  & No-True & No-True  & No-True & No-True & No-True &No-True \\
11512      & No-True  & No-True & No-True  & No-True & Yes-False  & Yes-False  & Yes-False\\
11589      & No-True  & No-True & No-True  & Yes-True & Yes-True & Yes-True & Yes-False  \\
11635      & No-True  & No-True      & No-True  & Yes-False  & Yes-False  & Yes-False  & No-True \\
11775      & No-True  & No-True      & No-True  & Yes-False  & Yes-False  & Yes-False  & Yes-False \\
\hline 
\end{tabular}

}

\end{table*}
\end{landscape}

\bibliography{raa_ms2021-0172.bib}

\end{document}